\documentclass[
aps,
pra,
showpacs
amsmath,
amssymb,
reprint,
floatfix,
lengthcheck,
superscriptaddress
]{revtex4-1}

\usepackage{graphicx}
\usepackage{mathtools}
\usepackage{braket}
\usepackage{hyperref}
\usepackage{bbold}
\usepackage{calrsfs}
\usepackage{dsfont}
\usepackage{xcolor}

\renewcommand{\i}{\mathrm{i}}
\renewcommand{\vec}[1]{\mathbf{#1}}
\renewcommand{\d}{\,\mathrm{d}}

\DeclareMathAlphabet{\pazocal}{OMS}{zplm}{m}{n}
\newcommand{\Ib}{\pazocal{I}}

\DeclareMathOperator*{\SumInt}{%
\mathchoice%
  {\ooalign{$\displaystyle\sum$\cr\hidewidth$\displaystyle\int$\hidewidth\cr}}
  {\ooalign{\raisebox{.14\height}{\scalebox{.7}{$\textstyle\sum$}}\cr\hidewidth$\textstyle\int$\hidewidth\cr}}
  {\ooalign{\raisebox{.2\height}{\scalebox{.6}{$\scriptstyle\sum$}}\cr$\scriptstyle\int$\cr}}
  {\ooalign{\raisebox{.2\height}{\scalebox{.6}{$\scriptstyle\sum$}}\cr$\scriptstyle\int$\cr}}
}

\begin{document}

\title{Light-Matter Response in Non-Relativistic Quantum Electrodynamics: Quantum Modifications of Maxwell's Equations}

\author{Johannes Flick}
  \email[Electronic address:\;]{flick@seas.harvard.edu}
  \affiliation{John A. Paulson School of Engineering and Applied Sciences, Harvard
University, Cambridge, MA 02138, USA}
  \affiliation{Max Planck Institute for the Structure and Dynamics of Matter and Center for Free-Electron Laser Science \& Department of Physics, Luruper Chaussee 149, 22761 Hamburg, Germany}
\author{Davis M. Welakuh}
  \email[Electronic address:\;]{davis.welakuh@mpsd.mpg.de}
  \affiliation{Max Planck Institute for the Structure and Dynamics of Matter and Center for Free-Electron Laser Science \& Department of Physics, Luruper Chaussee 149, 22761 Hamburg, Germany}
  \author{Michael Ruggenthaler}
  \email[Electronic address:\;]{michael.ruggenthaler@mpsd.mpg.de}
  \affiliation{Max Planck Institute for the Structure and Dynamics of Matter and Center for Free-Electron Laser Science \& Department of Physics, Luruper Chaussee 149, 22761 Hamburg, Germany}
  \author{Heiko Appel}
  \email[Electronic address:\;]{heiko.appel@mpsd.mpg.de}
  \affiliation{Max Planck Institute for the Structure and Dynamics of Matter and Center for Free-Electron Laser Science \& Department of Physics, Luruper Chaussee 149, 22761 Hamburg, Germany}
  \author{Angel Rubio}
  \email[Electronic address:\;]{angel.rubio@mpsd.mpg.de}
  \affiliation{Max Planck Institute for the Structure and Dynamics of Matter and Center for Free-Electron Laser Science \& Department of Physics, Luruper Chaussee 149, 22761 Hamburg, Germany}
  \affiliation{Center for Computational Quantum Physics, Flatiron Institute, 162 5th Avenue, New York, NY 10010, USA}

\date{\today}

\begin{abstract}
We derive the full linear-response theory for non-relativistic quantum electrodynamics in the long wavelength limit, {show quantum modifications of the well-known Maxwell's equation in matter} and provide a practical framework to solve the resulting equations by using quantum-electrodynamical density-functional theory. We highlight how the coupling between quantized light and matter changes the usual response functions and introduces new types of cross-correlated light-matter response functions. These cross-correlation responses lead to {measurable} changes in Maxwell's equations due to the quantum-matter-mediated photon-photon interactions. Key features of treating the combined matter-photon response are that natural lifetimes of excitations become directly accessible from first principles, changes in the electronic structure due to strong light-matter coupling are treated fully non-perturbatively, and for the first time self-consistent solutions of the back-reaction of matter onto the photon vacuum and vice versa are accounted for. By introducing a straightforward extension of the random-phase approximation for the coupled matter-photon problem, we {calculate} the first ab-initio spectra for a real molecular system that is coupled to the quantized electromagnetic field. Our approach can be solved numerically very efficiently. The presented framework leads to a shift in paradigm by highlighting {how electronically excited states arise as a modification of the photon field and} that experimentally observed effects are always due to a complex interplay between light and matter. At the same time the findings provide a new route to analyze as well as propose experiments at the interface between quantum chemistry, nanoplasmonics and quantum optics.
\end{abstract}

\date{\today}

\maketitle
 

\section{Introduction}
Recent years have seen tremendous experimental advances in the nascent
field of strongly-coupled light-matter systems~\cite{ebbesen2016, sukharev2017}. In particular, new experimental advances have been demonstrated in polaritonic
chemistry~\cite{george2015,hiura2018,thomas2019}, solid-state physics~\cite{riek2015}, biological
systems~\cite{coles2017}, nanoplasmonics~\cite{chikkaraddy2016,benz2016}, two-dimensional materials~\cite{kleemann2017, bisht2019} or optical waveguides~\cite{mirhosseini2018}, among others.

In this so-called strong-coupling regime, as a result of mixing matter and photon degrees-of-freedom~\cite{ruggenthaler2017b,flick2018b}, novel effects emerge such as changes in chemical pathways~\cite{galego2015,galego2016,kowalewski2016} ground-state electroluminescence~\cite{cirio2016}, cavity-controlled chemistry for molecular ensembles~\cite{herrera2016, galego2017}, or optomechanical coupling in optical cavities~\cite{roelli2015}, new topological phases of matter~\cite{shin2018}, superradiance~\cite{mazza2019} or superconductivity~\cite{sentef2018}.

Due to the inherent complexity of such coupled fermion-boson problems described in general by quantum electrodynamics (QED), the theoretical treatment is usually drastically simplified. One common approximation is to restrict the description of the system to simplified effective models that heavily rely on input parameters. Current state of the art in the theoretical description of strong light-matter coupling very often employs a few-level approximation. This approximation leading to the Rabi or Jaynes-Cummings model~\cite{braak2011, xie2016} in the single-emitter case, or the Dicke model~\cite{garraway2011} in the many-emitter case, is however often not sufficient~\cite{rabl2018,schaefer2018}, in particular when observables besides the energy are of interest~\cite{schaefer2018}, such as in experimental setups involving the modification of chemical reactivity~\cite{ebbesen2016}.

Alternatively, in linear spectroscopy, the current theoretical description is built on the \textit{semi-classical} approximation~\cite{gilbert2010}. Herein, the many-particle electronic system is treated quantum mechanically and the electromagnetic field appears as an external perturbation. As an external perturbation, the electromagnetic field probes the quantum system, but is not a dynamical variable of the complete system ({see also supplemental material ~\ref{sec:semi-classics}}). Since in the strong-coupling regime light and matter {must be} on the same level, a semi-classical approximation is not adequate and the feedback between light and matter has to be considered.

{It is, however, long known that the radiative lifetimes are finite.} {Furthermore, experimentally excited-state properties are usually inferred from (de)excitations of the photon field, which is in stark contrast to the usual semi-classical theoretical description based solely on the electronic subsystem.} 

In free-space, {this mismatch can be circumvented since excited-state properties such as} radiative lifetimes of atoms and molecules can be calculated {perturbatively} using the theory of Wigner-Weisskopf~\cite{weisskopf1930} employing the Markov approximation. {However, this perturbative treatment of the coupling of light and matter becomes insufficient in the case that strong light-matter coupling is achieved, e.g., due to many emitters or due to reducing the mode volume of a cavity. In such cases} the Markov approximation breaks down and the Wigner-Weisskopf theory is not applicable anymore~\cite{buzek1999}. Additionally it is not straightforward how to extend the original formulation of Wigner-Weisskopf to many electronic levels and hence {to an ab-initio treatment of electronic} systems.

As a consequence, the current literature shows a large gap for situations, where light and matter is strongly coupled and observables such as excited-state densities, radiative lifetimes, or electron-photon correlated observables of interest. {A good example is} the control of the radiative lifetimes of single molecules~\cite{lettow2007,wang2017} by changing the environment. In such cases the properties of the many-body system are changed, e.g., the excitation energies and lifetimes are {strongly} modified. This happens because certain modes of the photon vacuum field are enhanced which can lead to a strong coupling of light with matter. {Alternatively, increasing the number of particles leads to an enhancement of the coupling due to the self-consistent back-reaction of matter onto the photon field and vice versa. It is important to realize that such changes are non-perturbative for the photon field as well as for the matter subsystem and hence need a self-consistent { implementation}. This fact is most pronounced in the appearance of polaritonic states and their influence on chemical and physical properties of matter~\cite{ebbesen2016,ruggenthaler2017b}.}

In this paper, we close this gap by presenting a practical and general framework that subsumes electronic-structure theory, nanoplasmonics, and quantum optics. We present a {new description that challenges our conception of light and matter as distinct entities~\cite{ruggenthaler2017a}} {and that expresses the excited states as modifications of the photon field}. {We do so by introducing a linear-response formalism {for coupled matter-photon systems}. This {formalism} leads naturally to modifications of Maxwell's equations and the ability to calculate radiative lifetimes in arbitrary photon environments, including free-space, high-Q optical cavity or nanoplasmonic structures.} {We make this approach practical by introducing a linear-response framework for quantum-electrodynamical density-functional theory  (QEDFT)~\cite{ruggenthaler2017b,tokatly2013,ruggenthaler2014,ruggenthaler2015,flick2018b}. This development is specifically timely since QEDFT has now been successfully applied to real systems in equilibrium~\cite{flick2017c} -- which demonstrates the feasibility of ab-initio strong-coupling calculations -- yet an accurate and efficient approach to excited states within QEDFT has been missing. This work therefore furthermore closes a gap within the QEDFT framework.}

\section{Light-matter interaction in the long wavelength limit}
\label{sec:theory}
Our fundamental description of how the charged constituents of atoms, molecules and solid-state systems, i.e., electrons and positively charged nuclei, interact is based on QED~\cite{ryder1996, craig1998, spohn2004, ruggenthaler2017b}, thus the interaction is mediated via the exchange of photons. Adopting the Coulomb gauge for the photon field allows us to single out the longitudinal interaction among the particles which gives rise to the well-known Coulomb interaction and leaves the photon field purely transversal. Assuming then that the kinetic energies of the nuclei and electrons are relatively small, allows us to take the non-relativistic limit for the matter subsystem of the coupled photon-matter Hamiltonian, which gives rise to the so-called Pauli-Fierz Hamiltonian~\cite{spohn2004, ruggenthaler2014, ruggenthaler2017b} of non-relativistic QED. In a next step one then usually assumes that the combined matter-photon system is in its ground state such that the transversal charge currents are small and that the coupling to the (transversal) photon field is very weak. Besides the Coulomb interaction it is then only the physical mass of the charged constituents (bare plus electromagnetic mass~\cite{spohn2004}) that is a reminder of the photon field in the usual many-body Schr\"odinger Hamiltonian. In this work, however, we will not disregard the transversal photon field{, which makes the presented framework much more versatile and applicable to situation outside of standard quantum mechanics (see also appendix~\ref{app:lifetimes}).}

\subsection{Novel Spectroscopy from quantum description of light-matter interaction}
In the following, we consider cases, in which the semi-classical approximation breaks down, as outlined in the introduction. From the Pauli-Fierz Hamiltonian, we make the long-wavelength or dipole approximation in the length-gauge~\cite{rokaj2017} since the wavelength of the photon modes are usually much larger than the extend of the electronic subsystem which leads (in SI units) to~\cite{tokatly2013,ruggenthaler2014,pellegrini2015}~\footnote{In principle, QEDFT can be formulated for each level of theory of QED as presented in Ref.~\cite{ruggenthaler2014}. As a consequence, our formalism can be extended to more general formulations, including full minimal coupling, beyond the dipole approximation.}
\begin{align}
\label{eqn:h-dipole-2} 
\hat{H}(t)&={\hat{H}_e}+\sum_{\alpha=1}^{M}\frac{1}{2}\left[\hat{p}^2_{\alpha}+\omega^2_{\alpha}\left(\hat{q}_{\alpha}-\frac{\boldsymbol{\lambda}_{\alpha}}{\omega_{\alpha}} \cdot \textbf{R} \right)^2\right]+\frac{j_{\alpha}(t)}{\omega_\alpha}\hat{q}_\alpha,
\end{align} 
{where $\hat{H}_e$ is the standard many-body electronic Hamiltonian}~\cite{szabo1989}. We further restrict ourselves to arbitrarily many but a finite number $M$ of modes $\alpha\equiv (\textbf{k},s)$ with $s$ being the two transversal polarization directions that are perpendicular to the direction of propagation $\textbf{k}$. The frequency $\omega_\alpha$ and polarization ${\boldsymbol \epsilon_\alpha}$ that enter in ${\boldsymbol \lambda_\alpha} = \boldsymbol \epsilon_\alpha \lambda_\alpha$ with $\lambda_\alpha = S_\textbf{k} (\textbf{r})/\sqrt{\epsilon_0}$ and mode function $S_\textbf{k} (\textbf{r})$ define these electromagnetic modes. $S_\textbf{k} (\textbf{r})$ is normalized, has the unit $1/\sqrt{\text{V}}$ with the volume $V$ and we choose a reference point $\textbf{r}_0$ where we have placed the matter subsystem to determine the fundamental coupling strength~\footnote{All results presented in this paper are independent of $\textbf{r}_0$.}. These photon modes couple via the displacement coordinate $\hat{q}_{\alpha} = \sqrt{\frac{\hbar}{2\omega_{\alpha}}}(\hat{a}_{\alpha} + \hat{a}_{\alpha}^{\dagger})$, where $\hat{q}_\alpha$ is given in terms of photon annihilation $\hat{a}_{\alpha}$ and creation $\hat{a}_{\alpha}^{\dagger}$ operators, to the total dipole moment $\textbf{R} = \sum_{i=1}^{N} e\textbf{r}_i$~\footnote{{Throughout this paper, we use the implicit definition $e=-|e|$.}}. The $\hat{q}_{\alpha}$ {appears in the contribution of mode $\alpha$ to the displacement field} $\hat{\vec{D}}_{\alpha} = \epsilon_0 \omega_{\alpha} \boldsymbol{\lambda}_{\alpha} \hat{q}_{\alpha}$~\cite{rokaj2017}. Further, the conjugate momentum of the displacement coordinate is given by $\hat{p}_{\alpha} = -i\sqrt{\frac{\hbar\omega_{\alpha}}{2}}(\hat{a}_{\alpha} - \hat{a}_{\alpha}^{\dagger})$. Besides a time-dependent external potential $v(\textbf{r},t)$, we also have an external perturbation $j_{\alpha}(t)$ that acts directly on the mode $\alpha$ of the photon subsystem. Here $j_{\alpha}(t)$ is connected to a classical external charge current $\textbf{J}(\textbf{r},t)$ that acts as a source for the inhomogeneous Maxwell's equation.

Formally, however, due to the length-gauge transformations, the $j_{\alpha}(t)$ corresponds to the time-derivative of this (mode-resolved) classical external charge current~\cite{tokatly2013,ruggenthaler2014} {(see also appendix~\ref{app:Maxwell})}. Physically the static part $j_{\alpha,0}$ merely polarizes the vacuum of the photon field and leads to a static electric field~\cite{dimitrov2017, ruggenthaler2015}. The time-dependent part $\delta j_{\alpha}(t)$ then generates real photons in the mode $\alpha$. This term is also known as a source term in quantum field theory~\cite{ryder1996}, where it generates the particles (here the photons) that are studied. From this perspective it becomes obvious that instead of using $\delta j_{\alpha}(t)$ one could equivalently slightly change the initial state of the fully coupled system by adding incoming photons that then scatter off the coupled light-matter ground state~\cite{spohn2004}.
\subsection{Linear Response in the Length Gauge}%
With the Hamiltonian of Eq.~(\ref{eqn:h-dipole-2}) in length gauge we can then in principle solve the corresponding time-dependent Schr\"odinger equation (TDSE) for a given initial state of the coupled matter-photon system $\Psi_0(\vec{r}_1 \sigma_1,...,\vec{r}_N \sigma_N, q_{1},...,q_M)$
\begin{align}
\label{eqn:TDSE}
 \i \hbar \frac{\partial}{\partial t} \Psi(\vec{r}_1 \sigma_1,...,t) = \hat{H}(t)\Psi(\vec{r}_1 \sigma_1,...,t),
\end{align}
where $\sigma$ correspond to the spin degrees-of-freedom. However, instead of trying to solve for the infeasible time-dependent many-body wave function, we restrict ourselves to weak perturbations $\delta v(\vec{r},t)$ and $\delta j_{\alpha}(t)$ and assume that our system is in the ground state of the coupled matter-photon system initial time. In this case, first-order time-dependent perturbation theory can be used to approximate the dynamics of the coupled matter-photon system ({for details see supplemental material ~\ref{app:linresp}}). This framework gives us access to linear spectroscopy, e.g., the absorption spectrum of a molecule. Traditionally, if we made a decoupling of light and matter, i.e., we assumed $\Psi_0(\vec{r}_1 \sigma_1,...,\vec{r}_N \sigma_N, q_{1},...,q_M) \simeq \psi_0(\vec{r}_1 \sigma_1,...,\vec{r}_N \sigma_N) \otimes \varphi_0(q_{1},...,q_M)$, we would only consider the matter subsystem $\psi$ (the photonic part $\varphi$ would be completely disregarded). Physically, we would investigate the classical dipole field that the electrons induced due to a classical external perturbation $\delta v(\vec{r},t)$. To determine this induced dipole field we would only consider the linear response of the density operator $\hat{n}(\vec{r}) = \sum_{i=1}^{N}\delta(\vec{r} - \vec{r}_i)$ which would be given by the usual density-density response function in terms of the electronic wave function $\psi_0$ only~\footnote{In the following, we suppress the spin component of the wave function and focus exclusively on the spatial and mode dependence, i.e., $\Psi(\vec{r}_1 ,...,\vec{r}_N , q_{1},...,q_M;t)$.}

In this work however, since we do not assume the decoupling of light and matter, the full density-density response is taken with respect to the combined ground-state wave function $\Psi_0$ and is consequently different to the traditional density-density response. Further, since we can also perturb the photon field in the cavity by $\delta j_{\alpha}(t)$ which will subsequently induce density fluctuations, the density response $\delta n$ gets a further contribution leading to 
\begin{align}
\delta n(\textbf{r}t) = &\int dt' \int d\textbf{r}' \chi^n_n(\textbf{r}t,\textbf{r}'t') \delta v(\textbf{r}'t')\label{eq:delta_n1}\\
&+ \sum_{\alpha=1}^{M} \int dt' \chi^n_{q_\alpha}(\textbf{r}t,t') \delta j_\alpha(t')\nonumber.
\end{align}
Here the response function $\chi^n_n(\textbf{r}t,\textbf{r}'t')$ corresponds to the density-density response but with respect to the coupled light-matter ground state and $\chi^n_{q_\alpha}(\textbf{r}t,t')$ corresponds to the response induced by changing the photon field. In the standard linear-response formulation, due to the decoupling ansatz, changes in the transversal photon field would not induce any changes in the electronic subsystem. Since obviously we now have a cross-talk between light and matter, we accordingly have also a genuine linear-response of the quantized light field
\begin{align}
\delta q_\alpha(t) = &\int dt' \int d\textbf{r}' \chi^{q_\alpha}_{n}(t,\textbf{r}'t') \delta v(\textbf{r}'t')\label{eq:delta_q1}
\\
&+  \sum_{\alpha'=1}^{M}\int dt' \chi^{q_\alpha}_{q_{\alpha'}}(t,t')\delta j_{\alpha'}(t'),\nonumber
\end{align}
where $\chi^{q_\alpha}_{n}(t,\textbf{r}'t')$ is the full response of the photons due to perturbing the electronic degrees, and $\chi^{q_\alpha}_{q_\alpha'}(t,t')$ is the photon-photon response function. The response function $\chi^{q_\alpha}_{n}(t,\textbf{r}t')$ is in general not trivially connected to $\chi_{q_\alpha}^{n}(\textbf{r}t,t')$, due to the different time-ordering of $t$ and $t'$.  

The entire linear-response in non-relativistic QED for the density and photon coordinate can also be written in matrix form~\cite{hoffmann2016}. In this form we clearly see that the density response of the coupled matter-photon system depends on whether we use a classical field $\delta v(\vec{r},t)$, photons, which are created by $\delta j_{\alpha}(t)$, or combinations thereof for the perturbation. Furthermore, we can also decide to not consider the classical response of the coupled matter-photon system due to $\delta n(\vec{r},t)$, but rather directly monitor the quantized modes of the photon field $\delta q_{\alpha}(t)$. This response yet again depends on whether we choose to use a classical field $\delta v(\vec{r},t)$ that induces photons in mode $\alpha$ or whether we directly generate those photons by an external current $\delta j_{\alpha}(t)$. And we also see that the different modes are coupled, i.e., that photons interact. Similarly as charged particles interact via coupling to photons, also photons interact via coupling to the charged particles.
Keeping the coupling to the photon field explicitly therefore, on the one hand, changes the standard spectroscopic observables, and on the other hand also allows for many more spectroscopic observables than in the standard matter-only theory.

\subsection{Maxwell-Kohn-Sham linear-response theory}%
\begin{figure*}
\centerline{\includegraphics[width=\textwidth]{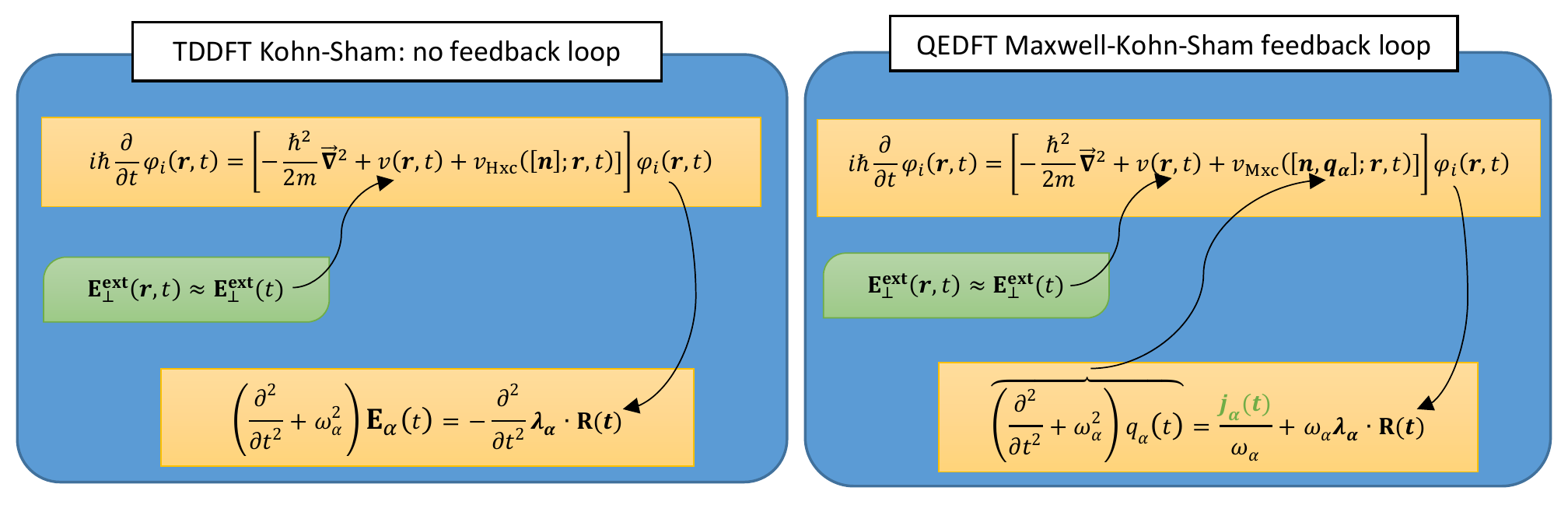}}
\caption{Schematics of the {Maxwell} KS approach contrasted with schematics of the usual semi-classical {KS} theory. While in the semi-classical approach the {KS} orbitals are used as fixed input into the {mode-resolved} inhomogeneous Maxwell's equation in vacuum {through the total dipole $\textbf{R}(t) =  \int d \textbf{r} \, {e}\vec{r} \, \sum_{i} |\varphi_i(\vec{r},t)|^2 $ (see also { appendix}~\ref{app:Maxwell})}, in the {Maxwell KS} framework the induced field acts back on the orbitals, which leads to an extra {self-consistency} cycle.}
\label{fig:change-maxwell}
\end{figure*}
The problem of this general framework in practice is that already in the simplified matter-only theory we usually cannot determine the exact response functions of a many-body system. The reason is that the many-body wave functions, which we use to define the response functions, are difficult, if not impossible to determine beyond simple model systems. So in practice we need a different approach that avoids the many-body wave functions. Several approaches exist that employ reduced quantities instead of wave functions~\cite{fetter2003, stefanucci2013, bonitz1998}. The workhorse of these many-body methods is DFT and its time-dependent formulation TDDFT~\cite{dreizler2012, engel2011, ullrich2011}. Both theories have been extended to general coupled matter-photon systems within the framework of QED~\cite{ruggenthaler2017b, tokatly2013, ruggenthaler2014, ruggenthaler2015, flick2018a}.

QEDFT allows us to solve instead of the TDSE equivalently a non-linear fluid equation for the charge density $n(\vec{r},t)$ coupled non-linearly to the mode-resolved inhomogeneous Maxwell's equation~\cite{ruggenthaler2011, tokatly2013, ruggenthaler2014, ruggenthaler2015}. While these equations are in principle easy to handle numerically, we do not know the forms of all the different terms explicitly in terms of the basic variables of QEDFT, i.e. $(n(\vec{r},t), q_{\alpha}(t))$. To find accurate approximations one then employs the Kohn-Sham (KS) scheme, where we model the unknown terms by a numerically easy to handle auxiliary system in terms of wave functions. The simplest approach is to use non-interacting fermions and bosons which lead to a similar set of equations, which are however uncoupled. Enforcing that both give the same density and displacement field dynamics gives rise to mean-field exchange-correlation (Mxc) potentials and currents~\cite{flick2015, ruggenthaler2015b, dimitrov2017}. Formally this Mxc potential and current is defined as the difference of the potential/current that generate a prescribed internal pair in the auxiliary non-interacting and uncoupled system $(v_{\rm s}([n], \vec{r},t), j_{\alpha}^{\rm s}([q_{\alpha}],t))$ and the potential/current that generates the same pair in the physical system defined by Eq.~(\ref{eqn:h-dipole-2}) which we denote by $(v([n, q_{\alpha}], \vec{r},t),j_{\alpha}([n, q_{\alpha}],t))$, i.e.,
\begin{align}
 v_{\rm Mxc}([n, q_{\alpha}], \vec{r},t) &= v_{\rm s}([n], \vec{r},t) - v([n, q_{\alpha}], \vec{r},t),
 \\
 j_{\alpha, {\rm M}}([n],t)& = j_{\alpha}^{\rm s}([q_{\alpha}],t) - j_{\alpha}([n,q_{\alpha}],t) 
 \label{m_current}\\
 &= - \omega_{\alpha}^2 \boldsymbol{\lambda}_{\alpha}\cdot \textbf{R}(t). \nonumber
\end{align}
In the time-dependent case we only have a mean-field contribution to the Mxc current~\cite{tokatly2013, ruggenthaler2015} where the total dipole moment is written as $\textbf{R}(t) =  \int d \textbf{r} \, {e}\vec{r} \, n(\vec{r},t) $. Further, we have ignored the so-called initial-state dependence because we assume (for notational simplicity and without loss of generality) in the following that we always start from a ground state~\cite{maitra2002, ruggenthaler2015b} of the matter-photon coupled system. In this way we can recast the coupled Maxwell-quantum-fluid equations in terms of coupled non-linear Maxwell-KS equations for auxiliary electronic orbitals, which sum to the total density $\sum_{i} |\varphi_i(\vec{r},t)|^2 = n(\vec{r},t)$, and the displacement fields $q_{\alpha}(t)$, i.e.,
\begin{align}
{i \hbar }\frac{\partial}{\partial t}\varphi_i(\textbf{r},t)=&\left[-\frac{\hbar^2}{2m_e}\vec{\boldsymbol\nabla}^2+ v_{\rm KS}([v, n, q_{\alpha}],\textbf{r},t)\right] \varphi_i(\textbf{r},t) ,
\label{KS-matter}
\\
\left(\frac{\partial^{2}}{\partial t^{2}} + \omega_{\alpha}^{2}\right) & q_{\alpha}(t) = -\frac{j_{\alpha}(t)}{\omega_{\alpha}} + \omega_{\alpha}\boldsymbol{\lambda}_{\alpha}\cdot \textbf{R}(t). \label{Max0}
\end{align}
Here we use the self-consistent KS potential $v_{\rm KS}([v, n, q_{\alpha}],\textbf{r},t) = v(\vec{r},t) + v_{\rm Mxc}([n, q_{\alpha}], \vec{r},t)$ that needs to depend on the fixed physical potential $v(\vec{r},t)$~\cite{ruggenthaler2015b}, and instead of the full bosonic KS equation for the modes $\alpha$ we just provide the Heisenberg equation for the displacement field. Although the auxiliary bosonic wave functions might be useful for further approximations it is only $q_{\alpha}(t)$ that is physically relevant and thus we get away with merely coupled classical harmonic oscillators, i.e., the mode resolved inhomogeneous Maxwell's equation. {To highlight the extra self-consistency due to coupling between light and matter we contrast the traditional electron-only KS theory with the Maxwell KS theory in Fig.~\ref{fig:change-maxwell}.} It is then useful to divide the Mxc potential into the usual Hartree-exchange-correlation (Hxc) potential that we know from electronic TDDFT and a correction term that we call photon-exchange-correlation {potential} (pxc), i.e.,
\begin{align*}
 v_{\rm Mxc}([n, q_{\alpha}], \vec{r},t) = v_{\rm Hxc}([n], \vec{r},t) + v_{\rm pxc}([n, q_{\alpha}], \vec{r},t).
\end{align*}
Clearly, the correction term $v_{\rm pxc}$ will vanish if we take the coupling $|\boldsymbol{\lambda}_{\alpha}|$ to zero and recover the purely electronic case. Since by construction the Maxwell KS system reproduces the exact dynamics, we also recover the exact linear-response of the interacting coupled system ({see also supplemental material ~\ref{app:linresp1}}). We can express this with the help of the Mxc kernels defined by the functional derivatives of the Mxc quantities 
\begin{align*}
f^{n}_\text{Mxc}(\textbf{r}t,\textbf{r}'t') &= \frac{\delta v_\text{Mxc}(\textbf{r}t)}{\delta n(\textbf{r}'t')}, \quad
f^{q_\alpha}_\text{Mxc}(\textbf{r}t,t') = \frac{\delta v_\text{Mxc}(\textbf{r}t)}{\delta q_\alpha(t')}, \\
g^{n_\alpha}_\text{M}(t,\textbf{r}'t') &= \frac{\delta j_{\alpha,\text{M}}(t)}{\delta n(\textbf{r}'t')}, \quad
g^{q_{\alpha'}}_\text{M}(t,t') = \frac{\delta j_{\alpha,\text{M}}(t)}{\delta q_{\alpha'}(t')} \equiv 0.
\end{align*}
and use the corresponding definitions for the Hxc kernel (that only for the variation with respect to $n$ has a non-zero contribution) and the pxc kernels. We note that using Eq.~(\ref{m_current}) we explicitly find
\begin{align}
 g^{n_{\alpha}}_\text{M}(t-t',\textbf{r}) = -\delta (t-t') \, \omega_{\alpha}^2 \boldsymbol{\lambda}_{\alpha}\cdot {e}\textbf{r}.
\label{eq:gnm}
\end{align}
and $g^{q_{\alpha'}}_\text{M}(t,t')$ vanishes, since $j_{\alpha,M}$ in Eq.~(\ref{m_current}) has no functional dependency on $q_\alpha$.
Via these kernels we find with $\chi^{n}_{n, {\rm s}}(\vec{r}t,\vec{r}' t')$ and $\chi^{q_{\alpha}}_{q_{\alpha'}, {\rm s}}(t,t')$, where $\chi^{q_{\alpha}}_{q_{\alpha'}, {\rm s}}(t,t') \equiv 0$ for $\alpha \neq \alpha'$, the uncoupled and non-interacting response functions that
 \begin{widetext}
\begin{align}
\label{eq:chi_nn}
\chi^n_n(\textbf{r}t,\textbf{r}'t') &=\chi^n_{n, {\rm s}}(\textbf{r}t,\textbf{r}'t') +\iint \text{d}\textbf{x}  \text{d}\tau \chi^n_{n, {\rm s}}(\textbf{r}t,\textbf{x}\tau)\left(\iint \text{d}\tau'\text{d}\textbf{y}f^n_\text{Mxc}{(\textbf{x}\tau,\textbf{y}\tau')}\chi^n_n{(\textbf{y}\tau',\textbf{r}'t')}\right.
\\
&\left.\quad + \sum_\alpha \int \text{d}\tau'f^{q_\alpha}_\text{Mxc}{ (\textbf{x}\tau,\tau')}\chi^{q_\alpha}_n{(\tau',\textbf{r}'t')}\right), \nonumber
\\
\label{eq:chi_qn}
\chi_{q_{\alpha'}}^{q_{\alpha}}(t,t') &= \chi_{q_{\alpha',s}}^{q_{\alpha}}(t,t') + \sum_{\beta}\iiint d\tau d\tau'd\textbf{x} \;  \chi_{q_{\beta,s}}^{q_{\alpha}}(t,\tau)  g_{M}^{n_{\beta}}(\tau,\textbf{x}\tau') \chi_{q_{\alpha'}}^{n}(\textbf{x}\tau',t') ,
\end{align}
 \end{widetext}
and accordingly for the mixed matter-photon response functions
 \begin{widetext}
\begin{align}
\label{eq:chi_nq}
\chi_{q_{\alpha}}^{n}(\textbf{r}t,t') &= \iint d\tau d\textbf{x} \; \chi_{n,s}^{n}(\textbf{r}t,\textbf{x}\tau) \left(\iint  d\tau'  d\textbf{y}f_{\text{Mxc}}^{n}(\textbf{x}\tau,\textbf{y}\tau')  \chi_{q_{\alpha}}^{n}(\textbf{y}\tau',t')  +  \sum_{\alpha'}\int  d\tau' f_{\text{Mxc}}^{q_{\alpha'}}(\textbf{x}\tau,\tau') \chi_{q_{\alpha}}^{q_{\alpha'}}(\tau',t')\right)  ,\\
\label{eq:chi_qq}
\chi_{n}^{q_{\alpha}}(t,\textbf{r}'t') &=  \sum_{\beta}\iiint d\tau d\tau' d\textbf{y} \; \chi_{q_{\beta,s}}^{q_{\alpha}}(t,\tau)  g_{M}^{n_{\beta}}(\tau,\textbf{y}\tau') \chi_{n}^{n}(\textbf{y}\tau',\textbf{r}'t') .
\end{align}
 \end{widetext}
Here we employed the formal connection between response functions and functional derivatives $\chi^{n}_{n}(\vec{r} t,\vec{r}' t') = \delta n(\vec{r},t)/ \delta v(\vec{r}',t')$ as well as $\chi^{q_{\alpha}}_{q_{\alpha'}}(t,t') = \delta q_{\alpha}(t)/ \delta j_{\alpha'}(t')$ and accordingly for the auxiliary system. The Mxc kernels correct the unphysical responses of the auxiliary system to match the linear response of the interacting and coupled problem. So in practice, instead of the full wave function, what we need are approximations to the unknown Mxc kernels. Later we will provide such approximations, show how accurate they perform for a model system and then apply them to real systems. If we decouple light and matter, i.e., $\Psi_0  \simeq \psi_{0} \otimes \varphi_{0}$, and disregard the photon part $\varphi_0$ (as is usually done in many-body physics), we recover the response function of Eq.~(\ref{eq:chi_nn}) with $f^{q_{\alpha}}_{\text{Mxc}} \equiv 0$, and $f^{n}_{\text{Mxc}}\rightarrow f^{n}_{\text{Hxc}}$. The response function, which is calculated with the bare matter initial state $\psi_0$, then obeys the usual  Dyson-type equation relating the noninteracting and interacting response in TDDFT~\cite{petersilka1996, casida1996} with $v_{\rm Mxc}([n, q_{\alpha}],\vec{r},t) \rightarrow v_{\rm Hxc}([n],\vec{r},t)$.

\subsection{{Excited states as properties of the photon field}}%
~\label{sec:Observables}
\begin{figure*}
\centerline{\includegraphics[width=\textwidth]{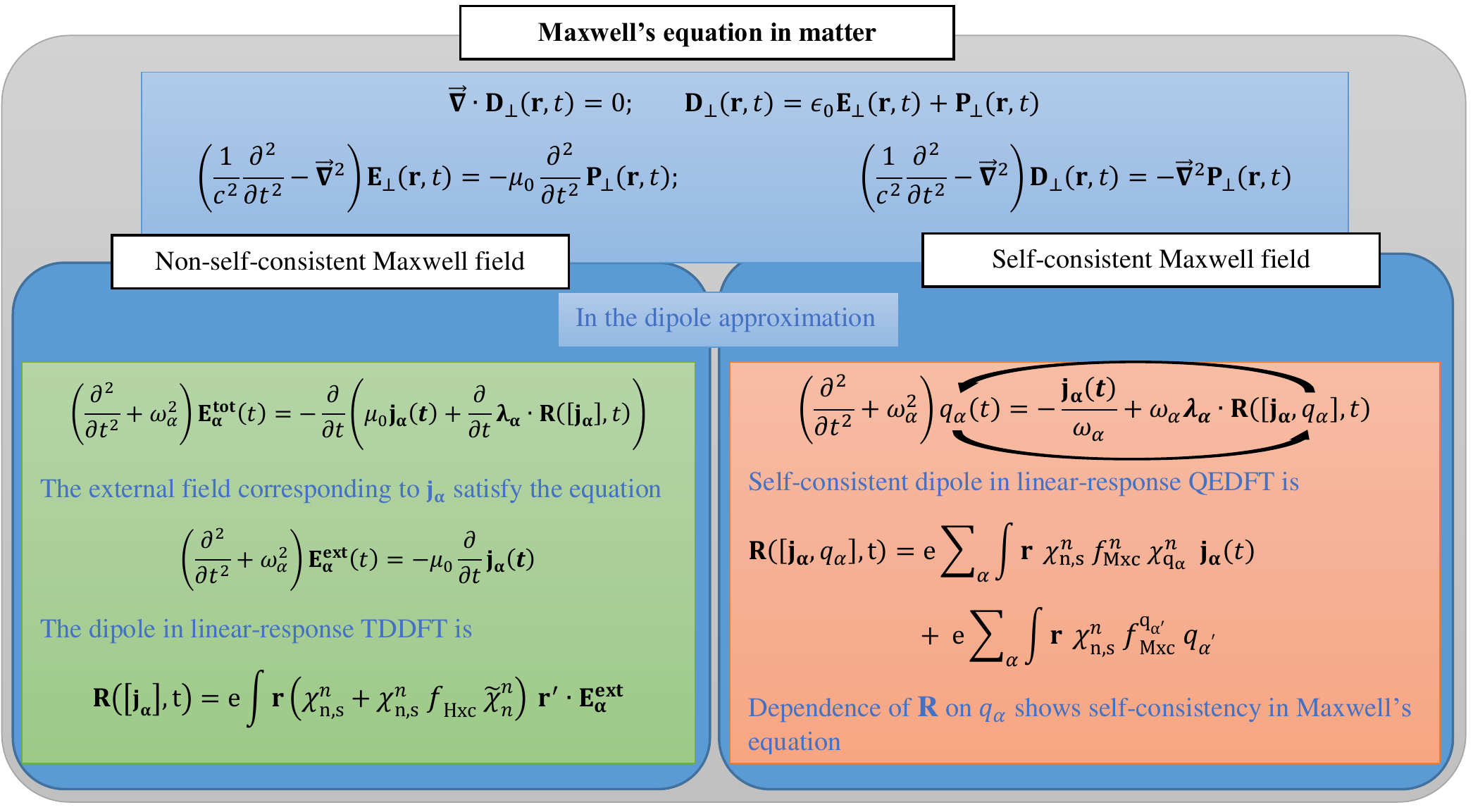}}
\caption{
{Schematics that contrasts the usual Maxwell's equation (left) with the fully self-consistent Maxwell's equation (right). Top: The induced transversal electric field $\textbf{E}_{\perp}$ as a consequence of the induced polarization $\textbf{P}_{\perp}$, which can be equivalently expressed in terms of the auxiliary displacement field $\textbf{D}_{\perp}$. Left: mode-resolved non-self-consistent Maxwell's equation with no backreaction. The external charge current $\textbf{j}_{\alpha}$ induces the external electric field in $\textbf{E}_{\alpha}^{\textrm{tot}}=\textbf{E}_{\alpha}+\textbf{E}_{\alpha}^{\textrm{ext}}$ which acts as an external perturbation through the dipole. Since the constituents of $\tilde{\chi}^n_n$ expressed in TDDFT are purely electronic, the induced field does not couple back to the Maxwell field. Right: self-consistent Maxwell's equation in which $\textbf{j}_{\alpha}$ induces the internal field $q_{\alpha}(t)$ through the electron-photon correlated dipole which has an explicit dependence as seen in the QEDFT form of $\chi_{q_{\alpha}}^{n}$. The self-consistency of the induced field through the dipole introduces nonlinearities in the coupled system thus changes the Maxwell field at the level of linear-response.} }
\label{fig:maxwell}
\end{figure*}
Following the above discussion, the usual response functions will change and novel response functions are introduced if we keep the matter-photon coupling explicitly. {This leads to many exciting consequences. Firstly, we get the completely self-consistent response of the system including all screening, retardation and other effects that become important when either the matter subsystem is becoming large~\cite{ehrenreich1966, mochan1985, maki1991,luppi2010} or when strong-coupling situations are considered. Since light and matter influence each other non-perturbatively the usual simplified approximations that only treat one part of the system accurately become unreliable~\cite{schaefer2018,rabl2018} (see also discussion in Sec.~\ref{sec:beyond}). Secondly, due to the matter-mediated photon-photon interactions (see appendix~\ref{app:Maxwell} and Fig.~\ref{fig:maxwell}) the usual Maxwell's equations are changed. A very interesting consequence is that in contrast to a purely classical theory we can theoretically distinguish whether a system is perturbed by a free current (that in turn would generate a classical electromagnetic field) or by a free electromagnetic field, e.g., a classical laser pulse. Thirdly, we rectify fundamental failings of standard quantum mechanics, such as the prediction of infinitely-lived excited states. The inclusion of the photon modes introduces the missing photon bath that leads to finite lifetimes (see appendix~\ref{app:lifetimes} and Sec.~\ref{sec:num-lifetime}). In connection to this it becomes important that we suddenly have access to a wealth of new observables that describe the photon field. Most importantly this implies the possibility to completely change our perspective of excited states of atoms and molecules. Indeed, in line with the experimental situation where changes in the photon field give us information on the excited states, we can view excited-state properties as arising from quantum modifications of the Maxwell's equations in matter}
\begin{align*}
\left(\frac{\partial^{2}}{\partial t^{2}} + \omega_{\alpha}^{2}\right)\delta q_{\alpha}(t) = -\frac{\delta j_{\alpha}(t)}{\omega_{\alpha}} + \omega_{\alpha}\boldsymbol{\lambda}_{\alpha}\cdot\int d\textbf{r} \;  {e}\textbf{r} \delta n(\textbf{r},t). 
\end{align*}
The response of the density is then found with help of the response functions Eqs.~(\ref{eq:chi_nn})-(\ref{eq:chi_qq}). In the usual case of an external classical field $\delta v(\vec{r},t)$ and $\delta j_\alpha(t) = 0$ we then find the induced field by (suppressing detailed dependencies with $\int d\textbf{r}\rightarrow \int $ and $\int d\textbf{r} \sum_\alpha \rightarrow\SumInt $)
\begin{widetext}
\begin{align}
\left(\frac{\partial^{2}}{\partial t^{2}} + \omega_{\alpha}^{2}\right)\delta q_{\alpha}(t) =   \omega_{\alpha} \boldsymbol{\lambda}_{\alpha} \cdot \int {e}\vec{r} \; \chi^n_{n, {\rm s}} \delta v  + \omega_{\alpha} \boldsymbol{\lambda}_{\alpha} \cdot \int {e}\vec{r} \; \chi^n_{n, {\rm s}} f^{n}_\text{Mxc}\chi^{n}_{n}\delta v + \omega_{\alpha} \boldsymbol{\lambda}_{\alpha} \cdot \SumInt {e}\vec{r} \;  \chi^n_{n, {\rm s}} f^{q_{\alpha'}}_\text{Mxc} \delta q_{\alpha'}  . \label{ v(r,t)}
\end{align}
\end{widetext}
Here the first term on the right-hand side corresponds to the non-interacting matter-response. However, due to the electron-electron interaction we need to take into account also the self-polarization of interacting matter (second term). Finally, the third term describes the matter-mediated photon-photon response. {The excited states of the coupled light-matter system are in this description changes in the photon field. That this perspective is actually quite natural becomes apparent if one considers the nature of the emerging resonances for a real system (see Fig.~\ref{fig:azulene-lifetime}). These resonances are mainly photonic in nature, as they describe the emission/absorption of photons (see appendix~\ref{app:lifetimes}).} Let us consider {now in more detail} what the terms on the right-hand side {of the modified Maxwell's equations} mean physically. First of all, in a matter-only theory the self-consistent solution of the Maxwell's equations together with the response of the bare matter-system would correspond approximately to the first two terms on the right-hand side {(see appendix~\ref{app:Maxwell})}. The photon-photon interaction would not be captured in such an approximate approach. Secondly, to highlight the physical content of the different terms we can make the mean-field contributions due to 
\begin{align}
v_{M}(\textbf{r}t) &= \sum_{\alpha} \left(\int d\textbf{r}' \boldsymbol{\lambda}_{\alpha} \cdot {e}\textbf{r}'n(\textbf{r}'t) - \omega_{\alpha} q_{\alpha}(t) \right)\boldsymbol{\lambda}_{\alpha} \cdot {e}\textbf{r} \label{eq:vm}
\\
&\quad + \int \d \textbf{r}'\frac{e^2 n(\vec{r}'t)}{4 \pi \epsilon_0 |\textbf{r} - \textbf{r}'|} \nonumber
\end{align}
explicit
 \begin{widetext}
\begin{align*}
\left(\frac{\partial^{2}}{\partial t^{2}} + \omega_{\alpha}^{2}\right)\delta q_{\alpha}(t) &= \omega_{\alpha} \boldsymbol{\lambda}_{\alpha} \cdot \int {e}\vec{r} \; \chi^n_{n, {\rm s}} \delta v +   \omega_{\alpha} \boldsymbol{\lambda}_{\alpha} \cdot \int {e}\vec{r} \; \chi^n_{n, {\rm s}} \left[ \frac{e^2}{4 \pi \epsilon_0 |\textbf{r}' - \textbf{r}''| } + \sum_{\alpha'} \left(\boldsymbol{\lambda}_{\alpha'} \cdot {e}\textbf{r}''\right)\boldsymbol{\lambda}_{\alpha'} \cdot {e}\textbf{r}' \right]\chi^{n}_{n}\delta v  \nonumber
\\
& - \omega_{\alpha} \boldsymbol{\lambda}_{\alpha} \cdot \SumInt {e}\vec{r} \;  \chi^n_{n, {\rm s}} \left(\omega_{\alpha'} \boldsymbol{\lambda}_{\alpha'} \cdot {e}\vec{r}' \right)\delta q_{\alpha'}  + \omega_{\alpha} \boldsymbol{\lambda}_{\alpha} \cdot \int {e}\vec{r} \; \chi^n_{n, {\rm s}} f^{n}_\text{xc}\chi^{n}_{n}\delta v \nonumber\\
&+ \omega_{\alpha} \boldsymbol{\lambda}_{\alpha} \cdot \SumInt {e}\vec{r} \;  \chi^n_{n, {\rm s}} f^{q_{\alpha'}}_\text{xc}\delta q_{\alpha'} . 
\end{align*}
 \end{widetext}
The second term on the right-hand side then corresponds to the random-phase approximation (RPA) to the instantaneous matter-matter polarization. Here a new term that corresponds to the dipole self-energy induced by the coupling to the photons arises. The third term on the right hand side is the RPA approximation to the dipole-dipole mediated photon interaction. {To give these terms further physical meaning note that in the usual perturbative derivation of the van-der-Waals interaction~\cite{craig1998} the first two terms would cancel and leave the photonic dipole-dipole interaction that gives rise to the $R^{-6}$ for small distances and the $R^{-7}$ for larger distances.} The rest are exchange-correlation (xc) contributions that arise due to more complicated interactions among the electrons and photons. The last term effectively describe photon-photon interactions mediated by matter. {In addition, we want to highlight that xc contributions are directly responsible for multi-photon effects, such as two-photon or three-photon processes (see Fig.~\ref{mixed}).} If we only keep the mean-field contributions of the coupled problem, we will denote the resulting approximation in the following as photon RPA (pRPA) to distinguish it from the bare RPA of only the Coulomb interaction. We see how the Maxwell's equations in matter change for bound charges, i.e., fields due to the polarization of matter, only. A new term, the photon-photon interaction, appears. For free charges, i.e., due to an external charge current $\delta j_{\alpha}(t)$, we see similar changes. Clearly, if we would not have a coupling to matter, then there would be no induced density change and we just find the vacuum Maxwell's equations coupled to an external current for the electric field. In other terms, the displacement field trivially corresponds to the electric field {(see appendix~\ref{app:Maxwell})}.

\section{Examples for the coupled matter-photon response}
~\label{sec:rabi}
\begin{figure}
\centerline{\includegraphics[width=0.5\textwidth]{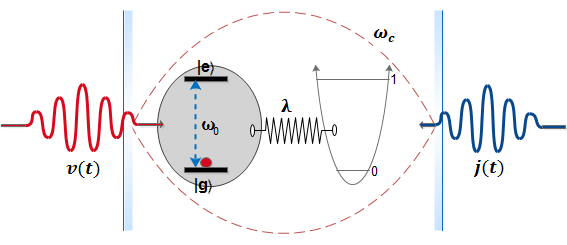}}
\caption{Two-level system {(with excitation $\omega_0$)} coupled to one mode of the radiation field {(with frequency $\omega_c$)}. The matter subsystem is driven by an external classical field $v(t)$ and the photon mode is driven by an external classical current $j(t)$ {and both subsystems are coupled with a coupling strength $\lambda$}.}
\label{Fig:TwoLevel}
\end{figure}

\begin{figure*}
\begin{minipage}[c]{0.32\textwidth}
\includegraphics[width=2.2in,height=2.2in]{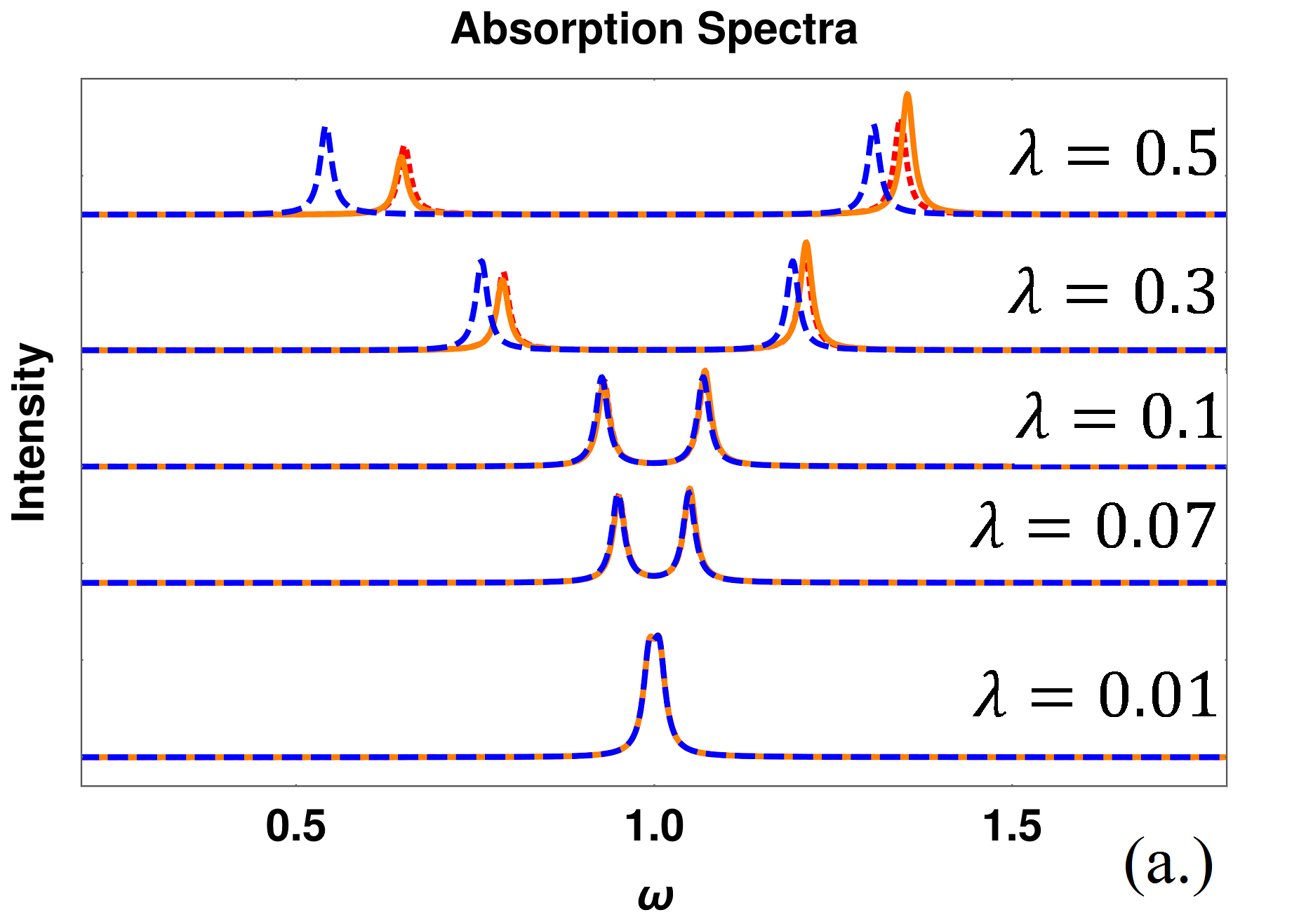}
\end{minipage}%
\begin{minipage}[c]{0.32\textwidth}
\includegraphics[width=2.2in,height=2.2in]{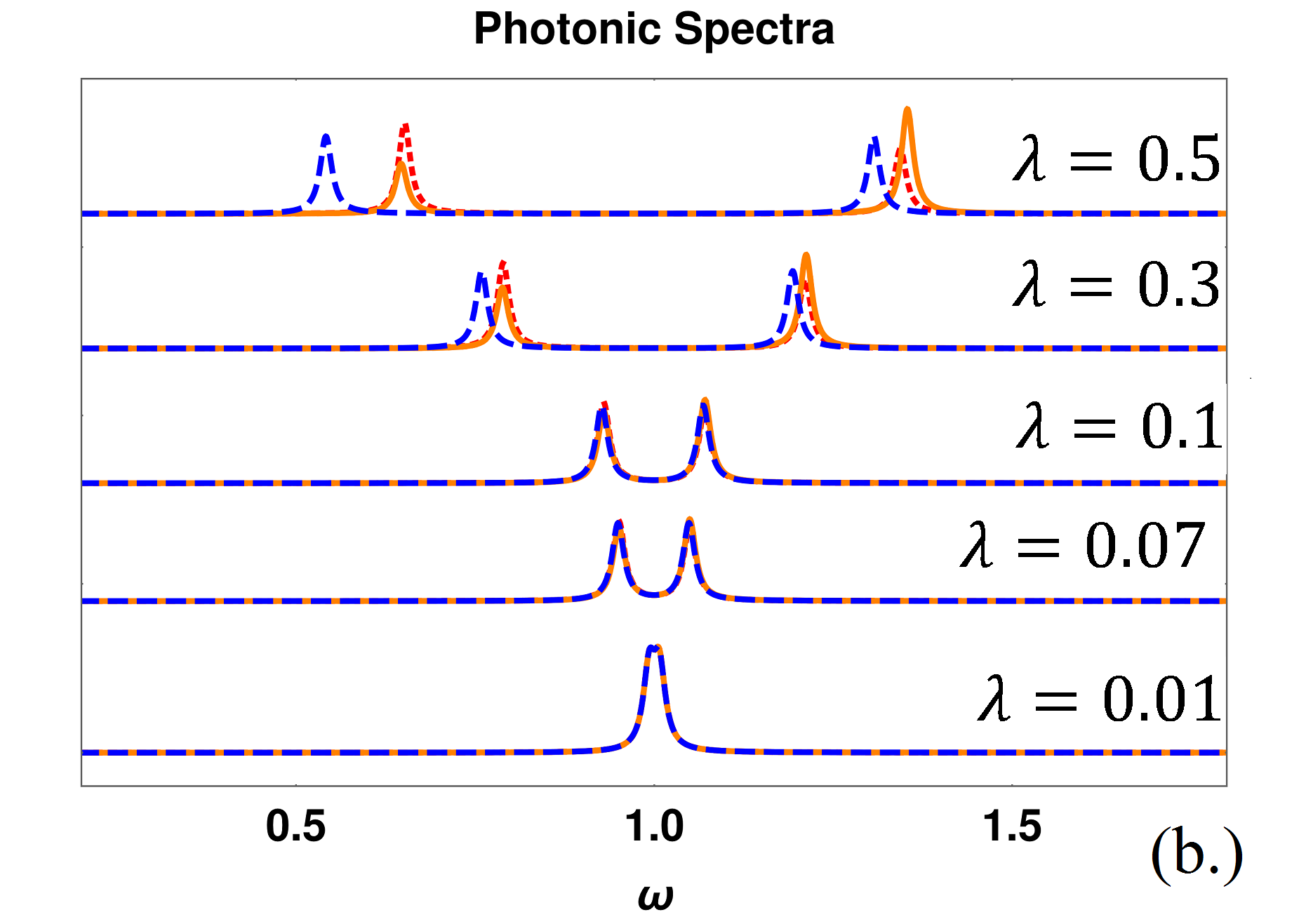}
\end{minipage}%
\begin{minipage}[c]{0.32\textwidth}
\includegraphics[width=2.2in,height=2.2in]{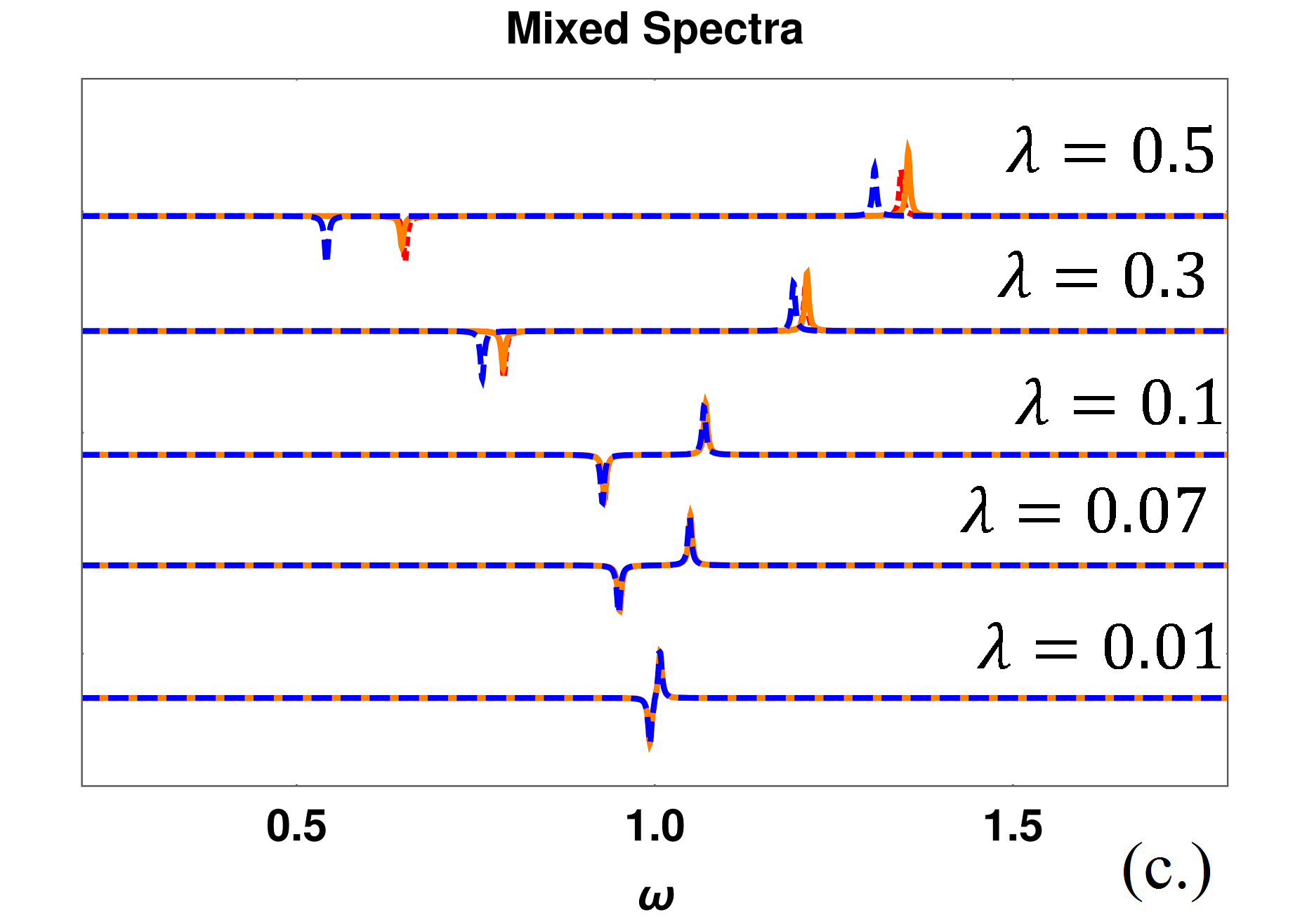}
\end{minipage}\\ 
\begin{minipage}[c]{0.39\textwidth}
\includegraphics[width=2.5in,height=2.2in]{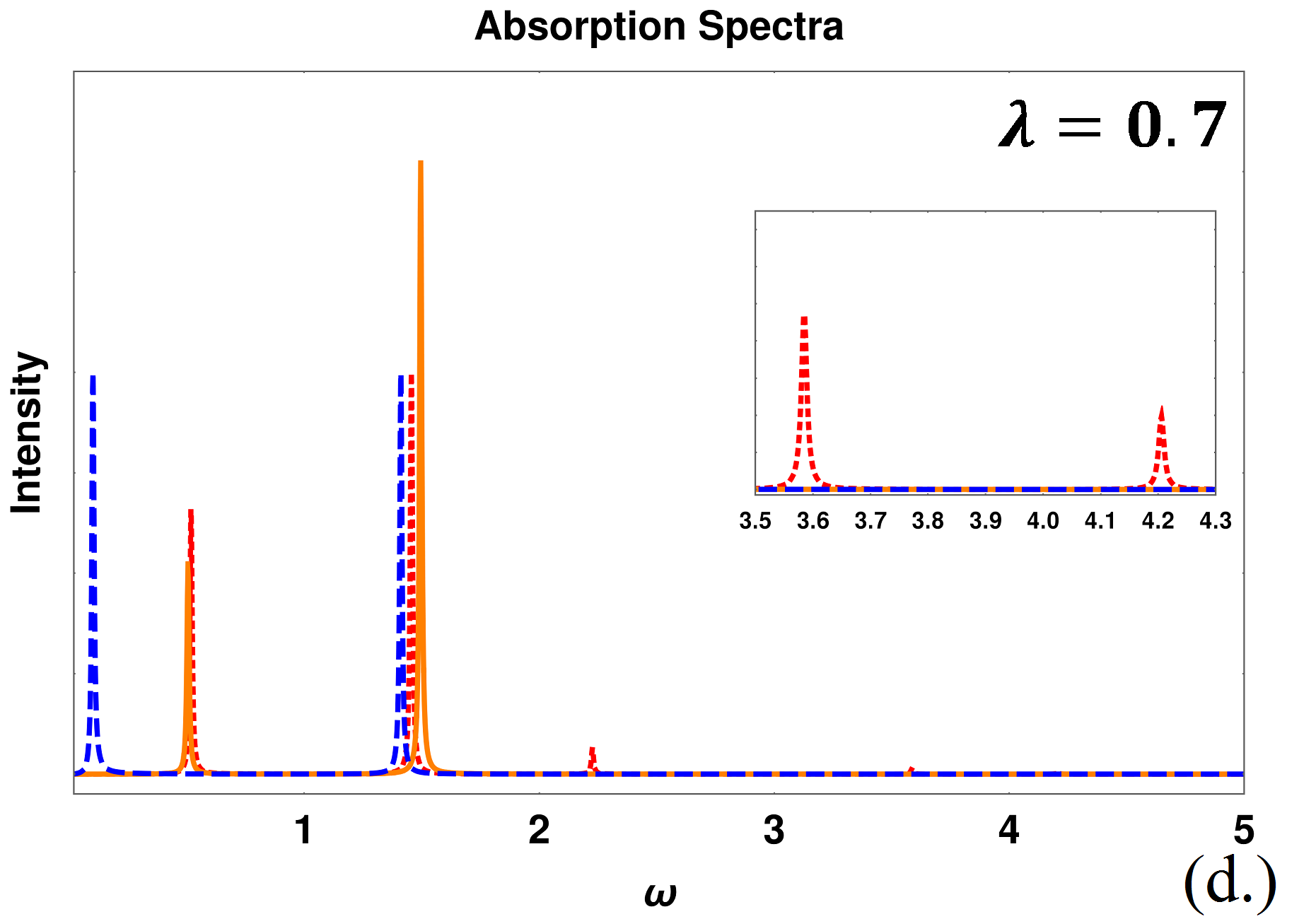}
\end{minipage}%
\begin{minipage}[c]{0.32\textwidth}
\includegraphics[width=2.5in,height=2.2in]{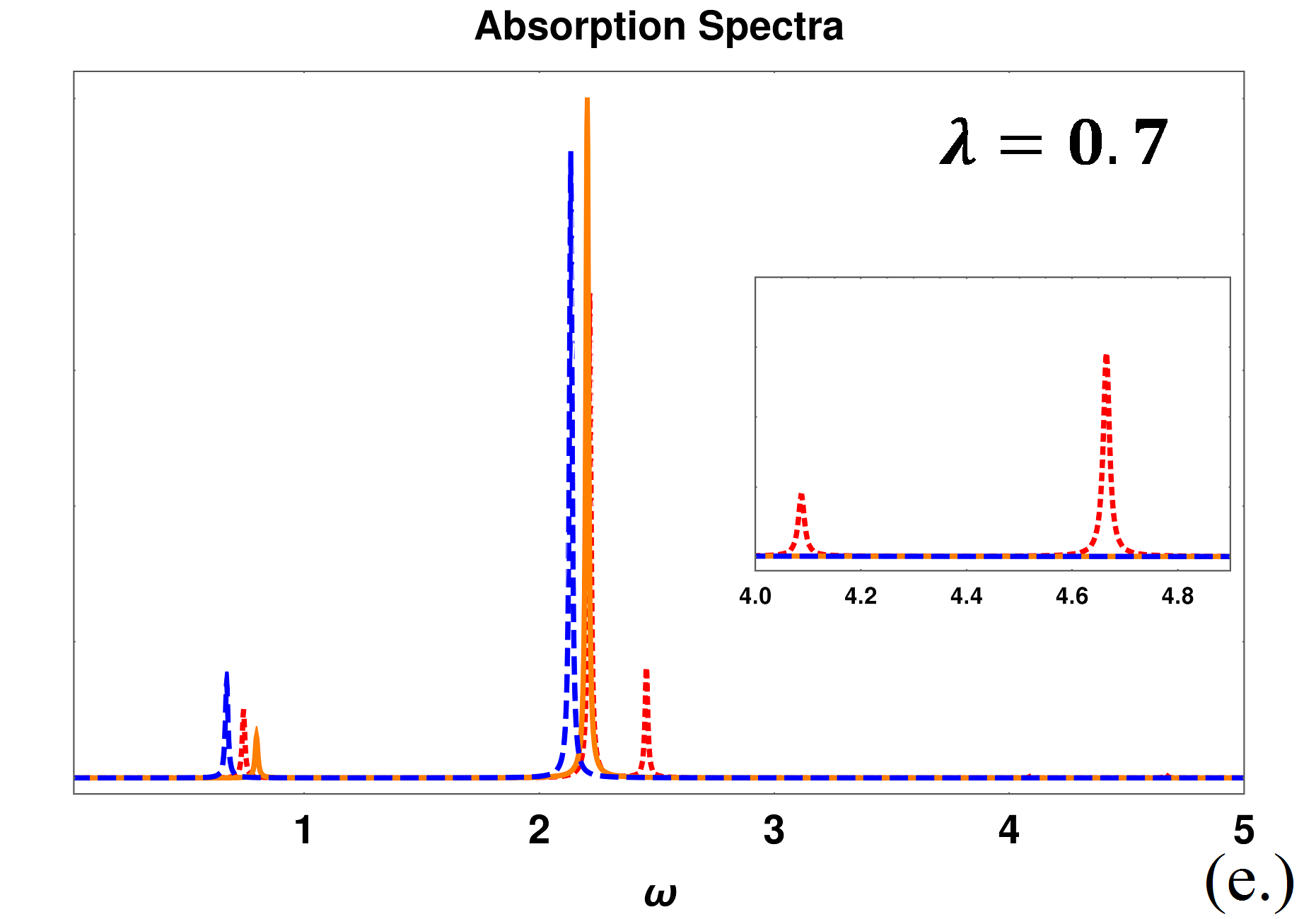}
\end{minipage}%
\hspace{2em}\begin{minipage}[c]{0.22\textwidth}
\includegraphics[width=0.5in,height=2.0in]{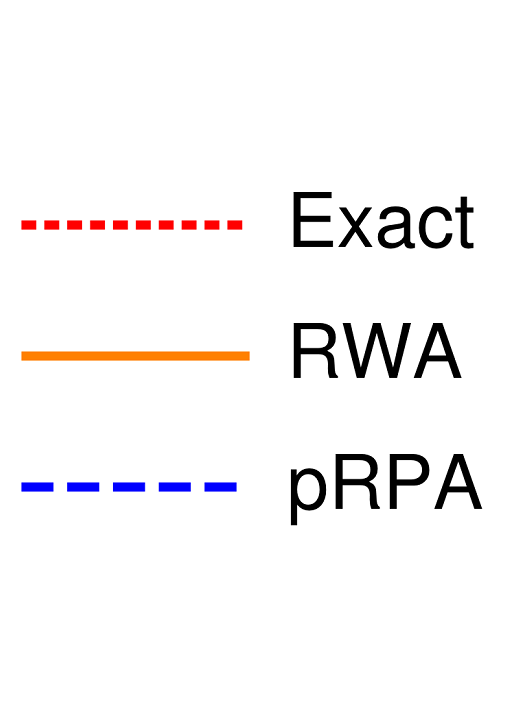}
\end{minipage} 
\caption{Linear-response spectra for the extended Rabi model (dotted-red) compared to the pRPA (dashed-blue) and RWA (full-orange) approximations and for different coupling strengths $\lambda$. (a.) Absorption spectra due to matter-matter response, (b.) spectra due to photon-photon response, (c.) spectra due to matter-photon or photon-matter response. {(d.) The case for $\lambda=0.7$ shows all excitations that arise in strong coupling. {(a.) through (d.) describes resonant coupling.}  In (e.) the field is half-way detuned from atomic resonance, i.e., $\omega_{0}=2$ and $\omega_{c}=1$ with strength and energies shifted to frequencies favoring 2-photon processes. The insets in (d.) and (e.) zoom into the frequency axis showing many-photon process.}}  \label{mixed}
\end{figure*}

{In this section, we discuss the new perspective enabled by the linear response formalism of QEDFT in more detail for a simple {and illustrative} model system. {We discuss} a slight generalization of the Rabi model~\cite{rabi1936,rabi1937}, which is the standard model of quantum optics. The Rabi model describes a single electron on two lattice sites{/energy levels} interacting with a single photon mode. We schematically depict the system in Fig~\ref{Fig:TwoLevel} and present all further details of this system in appendix~\ref{sec:app:rabi}.}

{First,} let us analyze the optical spectra for such a system and scrutinize the different approximations to the Mxc kernels. {We will compare the numerical exact results, with the mean-field (pRPA) and the rotating-wave approximation (RWA).} In Fig.~\ref{mixed}~(a), (b) and (c) we see how the optical spectra {of the resonantly coupled system (i.e. $\delta = \omega_{0} - \omega_{c} = 0$)} change for an increasing electron-photon coupling strength $\lambda$. Already for small coupling, the splitting of the electronic state into an upper and lower polariton becomes apparent. Approximately these states are given in terms of the RWA as $|+, 0\rangle$ and $|-,0\rangle$. The difference in energy between the lower and upper polariton is called the Rabi splitting $\Omega_R$ and is used to indicate the strength of the matter-photon coupling. In molecular experiments values of up to $\Omega_R/\omega_c \simeq 0.25$ have been measured~\cite{shalabney2015,george2016}. {Up to $\lambda = 0.1$ the different spectra for the exact (dotted-red), the pRPA (dashed-blue) as well as the RWA (full-orange) are in close agreement before they start to differ.} Already the mean-field treatment is enough to recover the quantized matter-photon responses, even for the coupled matter-photon spectra in Fig~\ref{mixed}~c. Consequently the pRPA seems a reasonable approximation for linear-response spectra even for relatively strong coupling situations. Only upon increasing the coupling strength further and thus going into the ultra-strong coupling regime, {the} discrepancies becomes large. For ultra-strong coupling (for $\lambda = 0.3$ the Rabi splitting is already of the order of $0.5{\omega_c}$) the approximations do not recover the exact results. Increasing further leads then to not only a disagreement in transition frequencies but also the weights of the transitions become increasingly different.

Besides a simple check for the approximations to the Mxc kernels, the extended Rabi model also allows us to get some understanding of the novel response functions $\chi^{\sigma_x}_q$, $\chi_{\sigma_x}^q$ and $\chi^{q}_q${, where $\sigma_{x}$ is the expectation-value of the corresponding Pauli matrix and describes the density/occupation changes between the two sites/energy levels}.
{This means,} we consider mixed spectroscopic observables where we perturb one subsystem and then consider the response in the other. We analogously employ $ \chi_{\sigma_{x}}^{q}(\omega)$ and $\chi_{q}^{\sigma_{x}}(\omega)$, respectively, to determine a ``mixed polarizability'' {(see supplemental material ~\ref{app:linresp3})}. If we plot this mixed spectrum (see Fig.~\ref{mixed}~(c) displayed in dotted-red for the numerically exact case), we find that we have positive and negative peaks. Indeed, this highlights that excitations due to external perturbations can be exchanged between subsystems, i.e., energy absorbed in the electronic subsystem can excite the photonic subsystem and vice versa. The oscillator strength of the photonic spectrum (based on $\chi^{q}_q$) in Fig.~\ref{mixed}~(b) provides us with a measure of how strong the displacement field (and with this also the electric field) reacts to an external classical charge current with frequency $\omega$. Similarly, the mixed spectrum (based on $\chi^{\sigma_x}_q$ or $\chi_{\sigma_x}^q$) in Fig.~\ref{mixed}~(c) provides us with information of how strong one subsystem of the coupled system reacts upon perturbing the other one. The oscillator strength here is not necessarily positive. What is absorbed by one subsystem can be transferred to the other. 

{In Fig.~\ref{mixed} (d) and (e), we show specifically the absorption spectra of the Rabi model for ultra-strong coupling, i.e., $\lambda = 0.7$. In this regime, three new peaks arise for the exact case accounting for high-lying excited states with non-vanishing dipole moments due to the strong electron-photon coupling. The new absorption peaks in Fig.~\ref{mixed}~(d), also shown in the inset, describes the resonant coupling case which the RWA and pRPA fail to capture in strong coupling, since processes beyond one-photon are involved. Similarly, Fig.~\ref{mixed}~(e) depicts the case were the field is half-detuned from the electronic resonance indicating a two-photon process. Clearly in ultra-strong coupling the absorption peaks are merely shifted close to the bare frequencies of the individual subsystems, but remain dressed by the photon field as new peaks arise due to the coupling.} {The pRPA and RWA capture the first of the two peaks around $\omega=2$, which is also the frequency of the atom, but fail to capture higher lying non-vanishing contributions to the spectra. These higher-lying peaks correspond to multi-photon processes.} {With more accurate approximation for the xc potential results closer to the exact ones can be obtained.} We note at this point that the peaks in Fig.~\ref{mixed} are artificially broadened and in reality correspond to sharp transitions due to excited states with infinite lifetimes. How to get lifetimes {quantitatively} will be discussed in the next section. 

\section{Coupled matter-photon response: real systems}
\label{sec:general} 
\begin{figure}[t] 
\centerline{\includegraphics[width=0.5\textwidth]{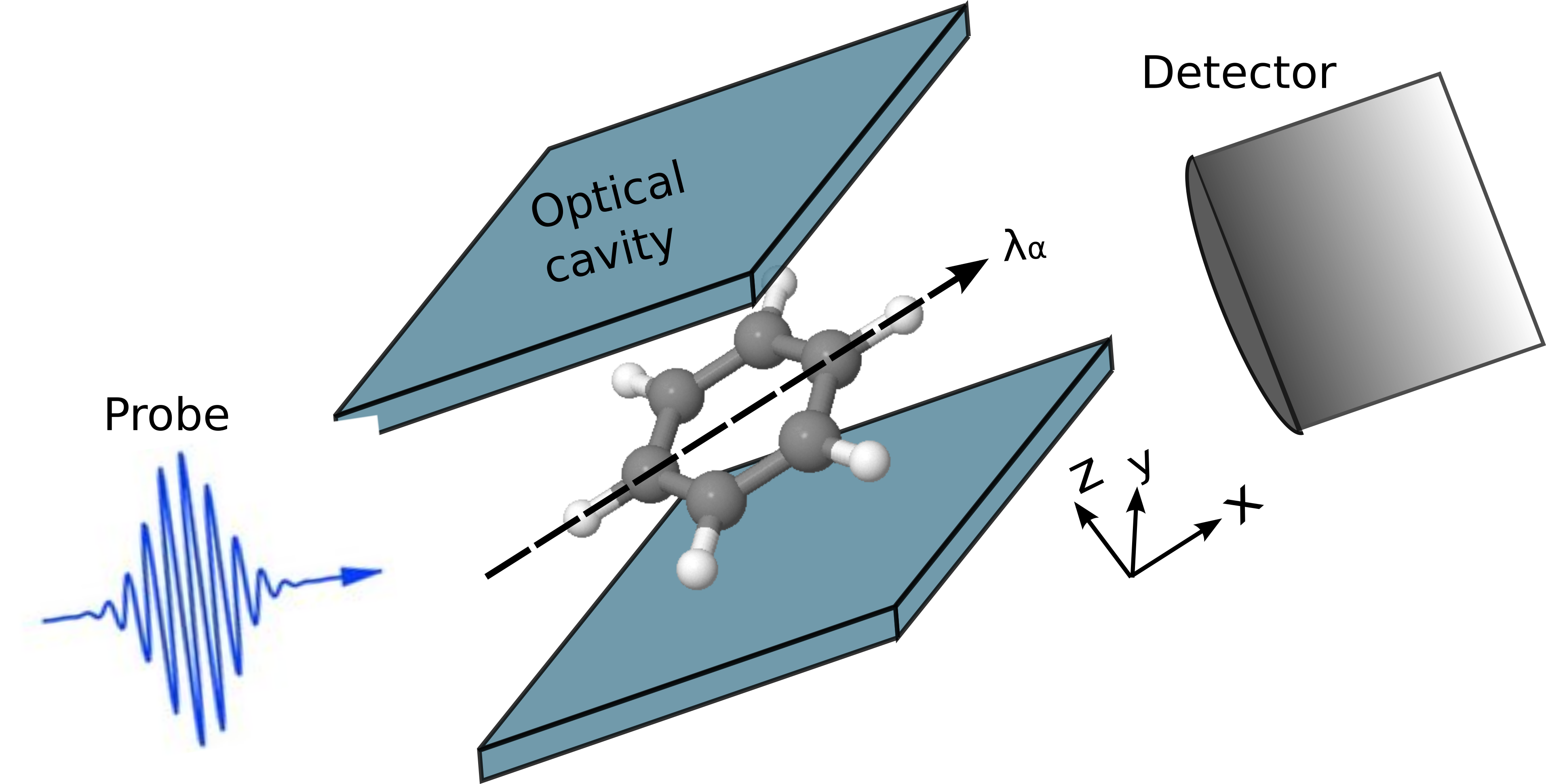}}
\caption{Schematic of absorption spectroscopy in optical cavities: Benzene (C$_{6}$H$_6$) molecule and ${\boldsymbol\lambda_\alpha}$ denotes the polarization direction of the photon field.}
\label{fig:azulene}
\end{figure}
{In this section, we apply the introduced formalism in pRPA approximation to real systems. We make the linear-response formulation practical by reformulating the problem as an eigenvalue equation in the frequency-domain. For electron-only problems this formulation is known as the \textit{Casida equation}~\cite{casida1996}. We refer the reader to appendix~\ref{sec:app:casida} for a derivation of our extension of the Casida equation, which includes transverse photon fields.}

For the following discussion, we consider benzene molecules in an optical cavity. In Fig.~\ref{fig:azulene} we schematically depict the experimental setup for a photoabsorption experiment under strong light-matter coupling {for a single molecule}. First we study the prototypical cavity QED setup where {a} molecule is strongly coupled to a single cavity mode of a high-Q cavity. In the second setup, we lift the restriction of only one mode and instead couple the benzene molecule to many modes that sample the electromagnetic vacuum field without enhancing the coupling to a specific mode by hand. In the third setup, we study the behavior of two molecules in an optical cavity, as well as a dissipative situation, where only {a few} modes {are} strongly coupled, embedded in a quasi-continuum of modes. {In the last example, we analyze the strong coupling of a single molecule to a continuum of modes. We find a transition from Lorentzian lineshape to a Fano lineshape~\cite{Ott2013} for increasing electron-photon coupling strength.} {These different setups provide us with the first ab-initio calculation for the spectrum of a real molecule in a high-Q cavity, the first ab-initio determination of intrinsic lifetimes and the first ab-initio calculation of the non-perturbative interplay between electronic structure, lifetime and strong-coupling. The {two} last situations need a self-consistent treatment of photons and matter alike and cannot be captured by any available electronic-structure or quantum-optical method.} All of those examples highlight the novel possibilities and perspectives that the QEDFT framework provides.
\subsection{Strong light-matter coupling}
The first results we discuss are a set of calculations, where a benzene molecule is strongly coupled to a single photon mode in an optical high-Q cavity. {We have implemented the linear-response pseudo-eigenvalue equation of Eq.~(\ref{eq:casida-deltav1}) into the real-space code OCTOPUS~\cite{marques2003,andrade2014} and details of the numerical parameters {are given in appendix~}\ref{sec:numerics}}~\footnote{{The routines used to perform all calculations in this work will be made publicly available. They can be easily transported to any other first principles code that has the matter linear-response equations implemented to make them ready to describe the complete QED response, i.e. joint matter-photon response, as described in this work.}}.

In the first calculation, we include a single cavity mode in resonance to the $\Pi$-$\Pi^*$ transition of the benzene molecule~\cite{yabana1999,marques2003}, i.e., $\omega_\alpha=6.88$~eV. For the light-matter coupling strength $\lambda_\alpha= |\boldsymbol\lambda_\alpha|$, we choose five different values, i.e. $\lambda_\alpha = (0, 2.77, 5.55, 8.32, 11.09)$ eV$^{1/2}$/nm that correspond to a transition from the weak to the strong-coupling limit and the cavity mode is assumed to be polarized along the x-direction.
\begin{figure}[t] 
\centerline{\includegraphics[width=0.5\textwidth]{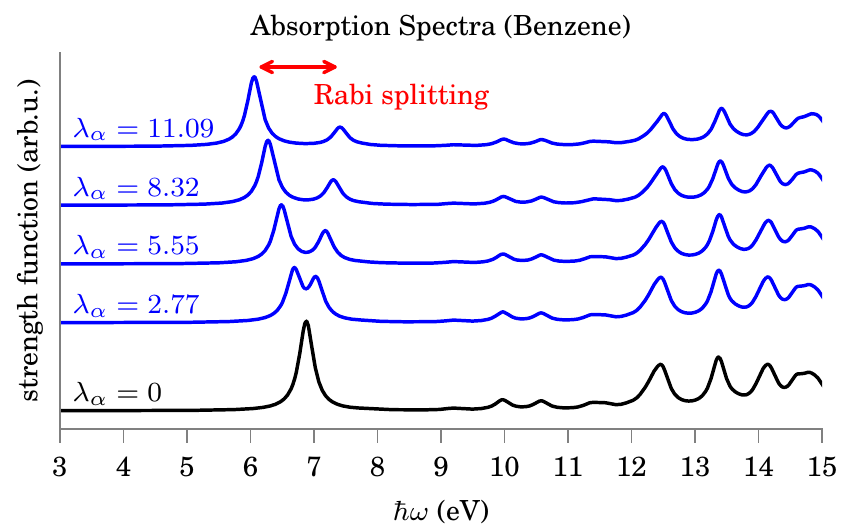}}
\caption{Absorption spectra for the benzene molecule in free space (black) and under strong light-matter coupling in an optical cavity to ultra-strong coupling (blue). The value for $\lambda_\alpha$ is given in units of $[$eV$^{1/2}$/nm$]$.}
\label{fig:azulene-results}
\end{figure}
In Fig.~\ref{fig:azulene-results}, we show the absorption spectra for these different values of $\lambda_\alpha$. We start by discussing the $\lambda_\alpha=0$ case that is shown in black. This spectrum corresponds to a calculation of the benzene molecule in free space and the spectrum is within the numerical capabilities identical to Ref.~\cite{marques2003}~\footnote{The spectrum in Ref.~\cite{marques2003} has been obtained using an explicit time-propagation with finite time. In the limit of zero broadening and including all unoccupied states, we would find identical spectra with very long propagated spectra.}. {We stress that here the broadening of the peaks is only done artificially since the photon bath is not included in the calculation. In the examples of Sec.~\ref{sec:num-lifetime} and \ref{sec:novel} we include many modes and hence sample the photon bath non-perturbatively.} We tune the electron-photon coupling strength $\lambda_\alpha$ in Fig.~\ref{fig:azulene-results}. We find for increasing coupling strength a Rabi splitting of the $\Pi$-$\Pi^*$ peak into two polaritonic branches. The lower polaritonic branch has higher intensity, compared to the upper polaritonic peak. Numerical values for the excitation energy $E_I$, the transition dipole moment $x_I$ and the oscillator strength $f_I$ are given in Tab.~\ref{tab:casida} {in the appendix}. This demonstrates that ab-initio theory is able to describe excited-state properties of strong light-matter coupling situations and captures the hybrid character of the combined matter-photon states. Thus predictive theoretical first-principle calculations for excited-states properties of real systems strongly coupled to the quantized electromagnetic field are now available. This will allow unprecedented insights into coupled light-matter systems, {since we have access to many observables that are not (or not well~\cite{schaefer2018}) captured by quantum-optical models}.
\subsection{Lifetimes of excitations from first principles}
\label{sec:num-lifetime}
Next we consider how {to} obtain lifetimes from QEDFT linear-response theory. In this example, we explicitly couple the benzene molecule to a wide range of photon modes similar as in the spontaneous emission calculation of Ref.~\cite{flick2017}. While in Ref.~\cite{flick2017}, {the system was simulated} with 200 photon modes, we choose here now 80.000 photon modes. The energies of the sampled photon modes cover densely a range from $0.19$~meV, for the smallest energy up to 30.51~eV for the largest one with a spacing of $\Delta \omega = 0.38$~meV. However, we do not sample the full three-dimensional mode space together with the two polarization possibilities per mode but rather consider a one-dimensional slice in mode space. This one-dimensional sampling of mode frequencies will change the actual three-dimensional lifetimes, but for demonstrating the possibilities of obtaining lifetimes this is sufficient~\footnote{A detailed analysis of real lifetimes would besides a proper sampling of the mode space also include considerations with respect to the bare mass of the particles.}. The sampling of the photon modes corresponds to the modes of a quasi-one dimensional cavity. We choose a cavity of length $L_x$~\cite{flick2017} in $x$-direction with a finite width in the other two directions that are much more confined. Thus we employ $\omega_\alpha = \alpha c\pi /L_x$ and $\boldsymbol\lambda_\alpha = \sqrt{\frac{2}{\hbar\epsilon_0 L_xL_y L_z}} \text{sin}( \omega_\alpha/c  \, x_0)\textbf{e}_x$, where $x_0 = L_x/2$ is the position of the molecule in $x$-direction. While we have a sine mode function in the $x$-direction, we assume a constant mode function in the other directions. For this example, we choose a cavity of length $L_x= 3250\mu \text{m}$ in $x$-direction, $L_y=10.58 \AA$ in $y$-direction and $L_z = 2.65\AA$ in $z$-direction.\\
\begin{figure}[ht] 
\centerline{\includegraphics[width=0.5\textwidth]{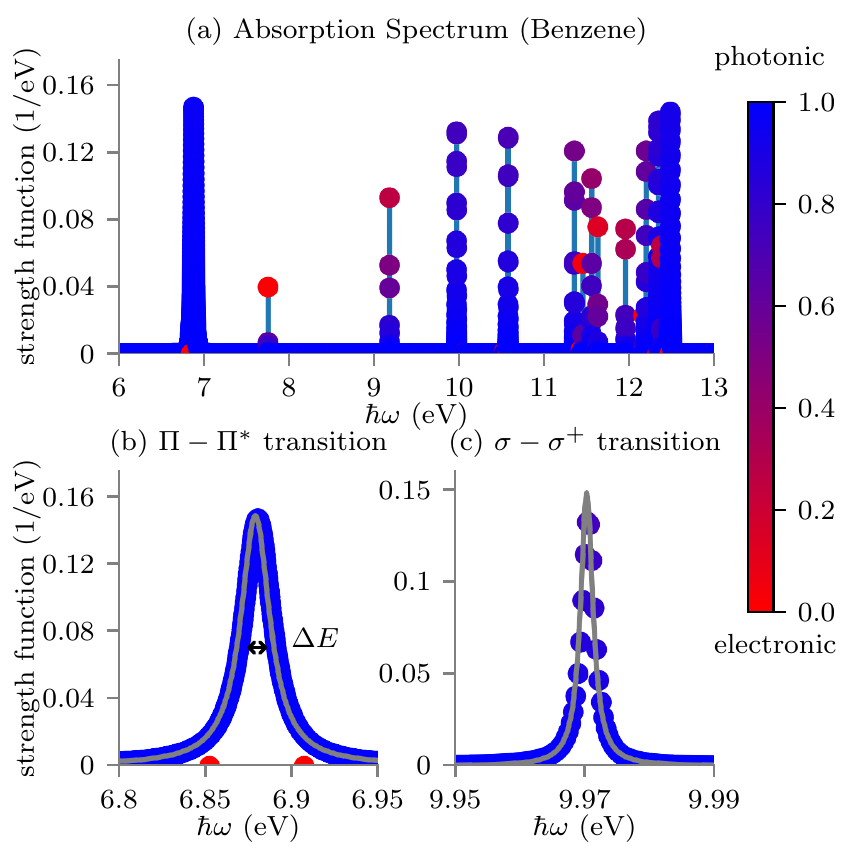}}
\caption{{First principles} lifetime calculation of the electronic excitation spectrum of the benzene molecule in an quasi one-dimensional cavity: (a) Full spectrum of the benzene molecule, (b) zoom to the $\Pi-\Pi^*$ transition{, where the black arrow indicates the full width at half maximum (FWHM) $\Delta E$}, (c) zoom to a peak contributing to the $\sigma-\sigma^+$ transition. The gray spectrum is obtained by Wigner-Weisskopf theory~\cite{weisskopf1930}. The dotted spectral data points correspond to many coupled electron-photon excitation energies which together comprise the natural lineshape of the excitation. Blue color refers to a more photonic nature of the excitations, vs. red color to a more electronic nature.}
\label{fig:azulene-lifetime}
\end{figure}
The results of this calculation are shown in Fig.~\ref{fig:azulene-lifetime}. In Fig.~\ref{fig:azulene-lifetime}~(a) we show the full spectrum. The electron-photon absorption function that has been obtained by coupling the benzene molecule to the quasi one-dimensional cavity with 80.000 cavity modes is plotted in blue. Since we have sampled the photon part densely, we do not need to artificially broaden the peaks anymore. Formulated differently, we can directly plot the oscillator strength and the excitation energies of our resulting eigenvalue equation and do not need anymore to employ the Lorentzian broadening. In Fig.~\ref{fig:azulene-lifetime} from blue (more photonic) to red (more electronic) for the electron-photon absorption spectrum we plot the different contributions of each pole in the response function. {These results confirm our intuition that resonances are mainly photonic in nature and that a Maxwell's perspective of excited states is quite natural.} In (b) we zoom to the $\Pi$-$\Pi^*$ transition. Due to quasi one-dimensional nature of the quantization volume, we find a broadening of the peak that is larger than it is for the case of a three-dimensional cavity due to the sampling of the electromagnetic vacuum. This is similar to changing the vacuum of the electromagnetic field. Accordingly the lifetimes of the electronic states are shorter if the electromagnetic field is confined to one dimension and we will discuss this in the next section.
\subsection{Connection to {standard} Wigner-Weisskopf theory}
\label{sec:beyond}

{{If the coupling between light and matter is very weak and neither subsystem gets appreciably modified due to the other, in contrast to the previous strong light-matter coupling case,} the radiative lifetimes of atoms and molecules can be calculated using {the perturbative} Wigner-Weisskopf theory~\cite{weisskopf1930} in single excitation approximation, as well as under the assumption of the Markov approximation. {These approximations are justified in the usual free-space case, where the results of Wigner and Weisskopf reproduce the prior results of Einstein based on the ad-hoc $\textit A$ and $\textit B$ coefficients.} {However it does not include the treatment of ensembles of molecules that effectively enhance the matter-photon coupling strength, as shown below.
}{Under the assumption of Wigner-Weisskopf theory, the radiative decay rate is given by}
\begin{align}
\Gamma_{3D} = \frac{\omega_0^3|\textbf{d}|^2}{3 \pi \epsilon_0\hbar c^3}.
\label{eq:gamma-ww3}
\end{align}
{For a one-dimensional cavity in x-dimension the results change to~\cite{buzek1999}}
\begin{align}
\Gamma_{1D} = \frac{\omega_0|\textbf{d}|^2}{L_yL_z \epsilon_0\hbar c}
\label{eq:gamma-ww}
\end{align}
{For comparison, we show in Fig.~\ref{fig:azulene-lifetime} in grey the peaks that are predicted by Wigner-Weisskopf theory. Since our sampling is very dense, we find for both peaks shown in the bottom a good agreement with Eq.~\ref{eq:gamma-ww}.\\
In fact, if we take the continuum limit for the photon modes, we recover {in our framework} the lifetimes predicted by Wigner-Weisskopf theory including the diverging energy shifts~\cite{milonni1976}, i.e. the Lamb shift. Due to the Lamb shift, our resulting peaks are slightly shifted, due to the divergencies. These divergencies can be handled by renormalization theory.} The lifetimes can now be obtained the following way: We measure the full width at half maximum (FWHM), indicated by the black arrow in (b). In this case, we find $\Delta E_\text{FWHM} = 0.0204$~eV and the corresponding lifetime $\tau_{\Pi-\Pi^*}$ follows by $\tau_{\Pi-\Pi^*}=\hbar/\Delta E_\text{FWHM} = 32.27$~fs. {Using the Wigner-Weisskopf formula from Eq.~\ref{eq:gamma-ww}, and the dipole moments and energies from the LDA calculation without a photon field, we find a lifetime of $32.21$~fs. As a side remark, the same transition using Eq.~\ref{eq:gamma-ww3} has a free-space lifetime of $0.89$~ns, roughly in the range of the 2p-1s lifetime of the Hydrogen atom of $1.6$~ns.}\\
In Fig.~\ref{fig:azulene-lifetime}~c we finally show the ab-initio peak of the $\sigma-\sigma^+$ transition. We find a narrow ab-initio peak {that is not as well sampled as the $\Pi-\Pi^*$. We note in passing that we find a ionization energy of $9.30$~eV using $\Delta$-SCF in the benzene molecule with the LDA exchange-correlation functional. We note in passing that we find a ionization energy of $9.30$~eV using $\Delta$-SCF in the benzene molecule with the LDA exchange-correlation functional. In our simulation, coupling to peaks higher than the ionization energy are broadened by continuum (box) states.}

\subsection{Beyond the single molecule limit and dissipation in QEDFT}
\label{sec:novel}

\begin{figure}[ht] 
\centerline{\includegraphics[width=0.5\textwidth]{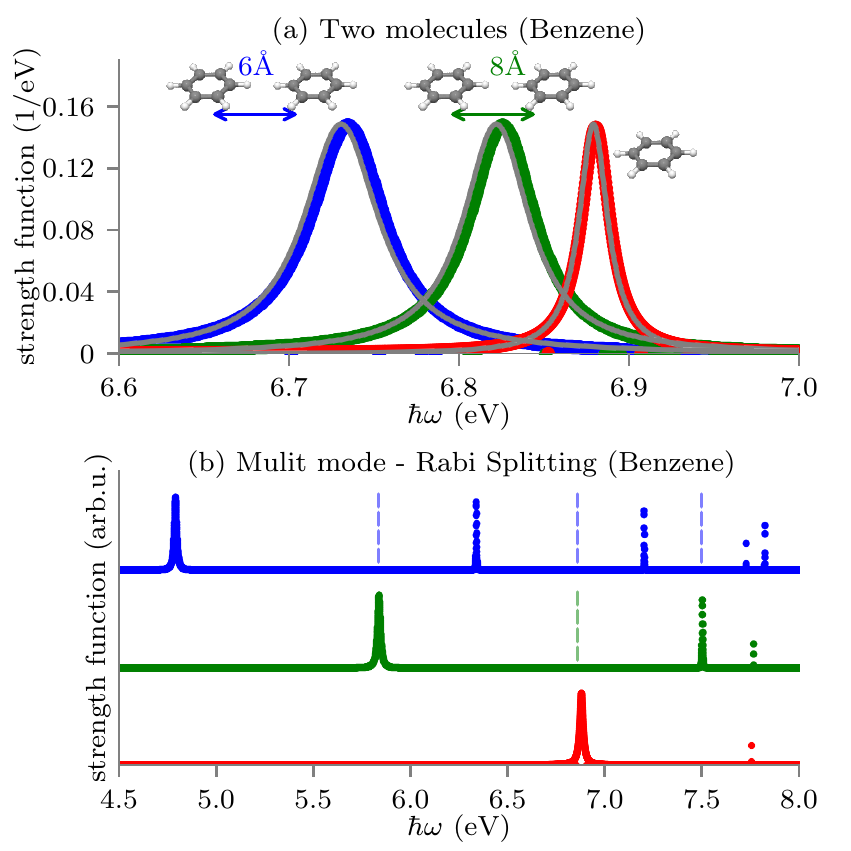}}
\caption{{(a) Two molecules of benzene strongly coupled to 80.000 cavity modes of an one-dimensional cavity. The further apart the molecules are, the closer the peak gets to the single molecule peak. Also we notice the doubled peak broadening (shorter lifetime). The gray spectrum is obtained by Wigner-Weisskopf theory~\cite{weisskopf1930}. (b) We show the Rabi splitting in a situation of a single strongly coupled mode with 80.000 cavity modes (green), and three strongly coupled modes with 80.000 cavity modes (blue). The red lines correspond to the same setup as in (a). The dashed lines refer to the frequency of the cavity modes. The peaks become broadened due to the interaction with the continuum.}}
\label{fig:azulene-two}
\end{figure}

{In contrast to the free-space result, where weak coupling as well as the assumption of a dilute gas of molecules are implied, in the case of single-molecule strong coupling~\cite{chikkaraddy2016} or when nearby molecules {or an ensemble of interacting molecules} modify the vacuum, the usual perturbative theories break down. Changes in the electronic and the photonic subsystem become self-consistent and the usual distinction of light and matter becomes less clear. In such situations the linear-response formulation of QEDFT as well as the Maxwell's perspective of excited-state properties becomes most powerful. Consider, for instance, two benzene molecules weakly coupled to a one-dimensional continuum of photon modes. If the molecules are far apart we just find the usual Wigner-Weisskopf result. But if we bring the molecules closer (see Fig.~\ref{fig:azulene-two} (a)), we see that the combined resonance shifts and the combined linewidth becomes broader, implying a shortened lifetime. In Fig.~\ref{fig:azulene-two} (b),  we consider the case of single-molecule strong coupling, where {a few} out of the 80.000 modes ha{ve} an enhanced coupling strength.} {In red, we show the spectrum where the molecule is coupled to the continuum, as is also shown in Fig.~\ref{fig:azulene-lifetime}. We then introduce a single strongly coupled mode at the $\Pi-\Pi^*$ transition energy and the resulting spectra is shown in green. We note that in the figure, the cavity frequencies are plotted in dashed lines. The single mode introduces the expected Rabi splitting into the upper and lower polariton} {and the peaks of the upper and lower polariton become broadened due to the interaction with the continuum. Interestingly, we find a different line broadening for the lower and the upper polaritonic peak, since only the sum of both has to be conserved. The smaller broadening for these two lower polaritonic states implies that the radiative lifetime of the lower and upper polaritonic state is longer than the lifetime of the excitation in weakly-coupled free-space.} {In blue, we show the spectra, where we have introduced three strongly coupled modes in addition to the cavity 80.000 modes of the continuum. We tune the two additional cavity modes in resonance to the lower and upper polariton peak of the green plot. We find additional peak splitting, but also a shifting of peak positions at 7.8~eV.}

{In the last numerical example, we study the strong coupling to the continuum for the case of a single molecule. The results are shown in Fig.~\ref{fig:azulene-fano}. Here, we effectively enhance the light-matter coupling strength by reducing the volume of the cavity along the $y$ and $z$ direction. For comparison, we show in red the setup that is also shown in Fig.~\ref{fig:azulene-lifetime}, where the excitations have Lorentzian lineshape consistent with Wigner-Weisskopf theory as discussed in the previous section. By gradually reducing the dimensions along the $y$ and $z$ direction, we find drastic changes in the lineshape of the excitations. These changes lead to the transition of the lineshape from a Lorentzian to a Fano lineshape, as becomes clearly visible for $L_xL_z=0.28\AA$.}

\begin{figure}[t] 
\centerline{\includegraphics[width=0.5\textwidth]{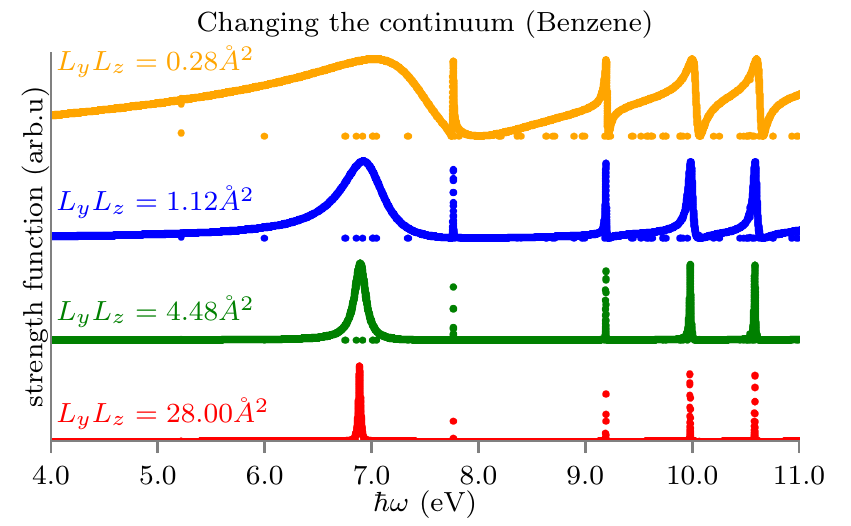}}
\caption{{Ab-intio lifetime calculation of the electronic excitation spectrum of the benzene molecule in an one-dimensional cavity along $x$-direction with different length in $L_y$ and $L_z$ direction. The red spectra refer to the same setup as in Fig.~\ref{fig:azulene-lifetime}. Effectively the electron-photon strength increases with smaller $L_y$ and $L_z$ length leading to a transition from a Lorentzian lineshape to a Fano lineshape.}}
\label{fig:azulene-fano}
\end{figure}

As a summary, {we have presented in this section, that lineshapes, as well as lifetimes can be inferred directly from first principle calculations. In case of Lorentzian lineshape, we find that the width of the} {calculated peaks (no need to introduce any artificial broadening as commonly done)} correspond to the lifetimes. These calculations demonstrate that ab-initio theory is able to capture the true nature of excitations, i.e., resonances with finite intrinsic lifetimes, without the need of an artificial bath or post-processing. This allows a new perspective { of well-known results. Furthermore, we find} that the excitations measured in absorption/emission experiments are mainly photonic in nature, and it is only the peak position that is dominated by the matter constituents. This is of course very physical, since what we see is the absorption/emission of a photon, not of the matter constituents. Further, since we describe the photon vacuum on the same theoretical footing as the matter subsystem, we have full control over the photon field making it straightforward to simulate very intricate changes, e.g., changing the character of a specific mode out of basically arbitrarily many, and investigating its influence on excited-states properties such as the radiative lifetime. This allows predictive first-principle calculations for intricate experimental situations similar to the ones encountered in Ref.~\cite{lettow2007,wang2017}.

\section{Summary and Outlook}
\label{sec:summary}
In this work we have introduced linear-response theory for non-relativistic quantum-electrodynamics in the long wavelength limit. Compared to the conventional matter-only response approaches, we have highlighted how in the coupled matter-photon case the usual response functions change, how novel photon-photon and matter-photon response functions are introduced, {how these novel response functions provide a photonic perspective on excited state properties, how the results lead to changes in the usual Maxwell's equation in matter} and how we can efficiently calculate {all} these response functions in the framework of QEDFT. By investigating a simple model system, we have shown how the spectrum of the matter subsystem is changed upon coupling to the photon field. Further we have demonstrated the range of validity of a simple yet reliable approximation to the in general unknown mean-field exchange-correlation kernels. Using this approximation we have presented the first ab-initio calculations of the spectrum of real system{s} (benzene {molecules}) coupled to the modes of the quantized electromagnetic field. In one example we have calculated the change upon strong coupling to a single mode of a high-Q cavity, which leads to a large Rabi splitting. In the second example we have calculated from first principles the natural linewidths of benzene coupled to a specific sampling of the vacuum field. {In the last examples, we demonstrated the abilities to calculate many-molecule systems, as well as dissipative strong-coupling situations, {as well as strong coupling to the continuum, where we find a transition from Lorentzian lineshape to Fano lineshape,} {where the usual (perturbative) approaches to light-matter coupling fail}.}
These results demonstrate the versatility and possibilities of QEDFT, where light and matter are treated on equal quantized footing. In the context of strong light-matter coupling, e.g., in polaritonic chemistry, the presented linear-response formulation allows now to determine polaritonically modified spectra from first principles. Together with ab-initio ground-state calculations~\cite{flick2017c} QEDFT now provides a workable first-principle description to analyze and predict photon-dressed chemistry and material sciences. In particular, our novel approach provides {a unique} practical computational scheme to compute photon-dressed excited-state potential-energy surfaces and non-adiabatic coupling elements that are required for ab-initio calculations in the emerging field of polaritonic chemistry. Further, in the context of standard ab-initio theory, the linear-response formulation of QEDFT now allows the calculation of intrinsic lifetimes {and provides access to quantum-optical observables. Specifically, due to the non-perturbative nature of the approach, quantum-optical problems where the self-consistent feedback between light and matter has to be taken into account, e.g., that many molecules change the photon vacuum and hence the Markov approximation breaks down, become feasible}. For optical physics, the presented linear-response framework presents an interesting opportunity to study the modifications of the Maxwell's equations in matter from first principles.
Finally we want to highlight that although the QEDFT linear-response framework is new, its similarity to the usual matter-only linear-response formulation in terms of an pseudo-eigenvalue problem makes it very easy to include in already existing first-principle codes. This, together with the above discussed novel possibilities in different fields of physics, shows that there are many interesting cases that can be studied with the presented method. 
\section{Acknowledgements}
We would like to thank Christian Sch\"afer and Norah Hoffmann for insightful discussions, and Sebastian Ohlmann for the help with the efficient massive parallel implementation. JF acknowledges financial support from the Deutsche Forschungsgemeinschaft (DFG) under Contract No. FL 997/1-1 and all of us acknowledge financial support from the European Research Council (ERC-2015-AdG-694097).


\appendix

\section{Modification of the Maxwell's equation}%
\label{app:Maxwell}
{In this section, we give more details on the modifications of the Maxwell's equations}.
{The semi-classical description of light-matter interaction is limited as a result of the transverse field being treated as an external perturbation. This approximation breaks the feedback loop between light and matter that leads to apparent changes in the Maxwell's equation. {Let us start from the classical description and assume that we are interested in the induced fields due to an external perturbation. If everything is perfectly classical there is no difference whether we perturb by an external transversal field $\vec{a}_{\perp}$ or an external classical current $\vec{j}_{\perp}$ due to the inhomogeneous Maxwell's equation in vacuum
\begin{align}
\label{eq:classicalMaxwell}
\biggl(\frac{1}{c^{2}}\frac{\partial^{2}}{\partial t^{2}} - {\boldsymbol\nabla}^2\biggl) \vec{a}_{\perp}(\textbf{r},t) = \mu_0 {c} \textbf{j}_{\perp}(\textbf{r},t). 
\end{align}
Now, if we have some theory to relate these external perturbation to the induced current $\textbf{J}_{\perp}[\vec{a}_{\perp}]$, the induced field reads
\begin{align}
\biggl(\frac{1}{c^{2}}\frac{\partial^{2}}{\partial t^{2}} - {\boldsymbol\nabla}^2\biggl) \vec{A}_{\perp}(\textbf{r},t) = \mu_0 {c} \textbf{J}_{\perp}([\vec{a}_{\perp}],\textbf{r},t), 
\end{align}
from which we can calculate the induced physical fields, e.g., the transversal electric field in Coulomb gauge is $\vec{E}_{\perp}(\vec{r},t) = -\frac{1}{c}\partial_{t} \vec{A}_{\perp}(\vec{r},t)$~\footnote{Some textbooks~\cite{craig1998} define the connection of the electric field to the vector potential without the prefactor $\frac{1}{c}$. We use the current notation to be consistent with relativistic literature and Ref.~\cite{ruggenthaler2014}.}. We can again combine these two results and look at the total field $\vec{A}_{\perp}^{\rm{tot}} = \vec{a}_{\perp} + \vec{A}_{\perp}$, which obeys
\begin{align}
\biggl(\frac{1}{c^{2}}\frac{\partial^{2}}{\partial t^{2}} - {\boldsymbol\nabla}^2\biggl) \vec{A}_{\perp}^{\rm tot}(\textbf{r},t) = \mu_0 {c} \left( \textbf{j}_{\perp}(\textbf{r},t) + \textbf{J}_{\perp}([\vec{j}_{\perp}],\textbf{r},t) \right). 
\end{align}
Using the Maxwell relations once more we can equivalently find for, e.g., the induced electric field
\begin{align}
\label{eq:vacEfield-full}
\biggl(\frac{1}{c^2} \frac{\partial^2}{\partial t ^2} - {\boldsymbol\nabla}^2\biggl)\vec{E}_{\perp}(\vec{r},t) = -\mu_0\tfrac{\partial}{\partial t} \vec{J}_{\perp}([\vec{a}_{\perp}],\vec{r},t).
\end{align}
We can now make a connection to the Maxwell's equation in matter, where the $\textbf{j}_{\perp}$ is called the free current and $\textbf{J}_{\perp}$ the bound current. Assuming that we can express the transversal induced current locally around the center of charge as $\vec{J}_{\perp}(\vec{r},t)\approx \tfrac{\partial}{\partial t} \textbf{P}_{\perp}(\vec{r}, t)$, where we use the polarization
\begin{align}
\vec{P}_{\perp}(\textbf{r},t) =  \epsilon_0 e\sum_{\alpha=1}^M{\boldsymbol\lambda_\alpha(\textbf{r})\int \text{d} \vec{r}' \, {\boldsymbol\lambda_\alpha(\textbf{r}')}}\cdot\vec{r}' n([\vec{a}_{\perp}],\vec{r}',t),\nonumber
\end{align}
and expand the electric field in the modes $\boldsymbol\lambda_\alpha(\textbf{r})$ as
\begin{align}
\vec{E}_{\perp}(\textbf{r},t) = \sum_{\alpha=1}^{M}{\boldsymbol \lambda_\alpha(\textbf{r})}E_\alpha(t),
\end{align}
we can rewrite the above equation at the center of charge, i.e., ${\boldsymbol\lambda_\alpha(\textbf{r}})\rightarrow \boldsymbol\lambda_\alpha$, as
\begin{align} 
\label{eq:vacEfield}
\biggl( \frac{\partial^2}{\partial t ^2} +\omega_\alpha^2\biggl) {E}_\alpha(t) = -\frac{\partial^2}{\partial t^2} {\boldsymbol\lambda_\alpha}\cdot\textbf{R}([\vec{a}_{\perp}],t) . 
\end{align}
Using this kind of approach we can connect $\delta n(\vec{r},t)$ of Eq.~(\ref{eq:delta_n1}) to the induced electric field $\delta\vec{E}_{\perp}(\textbf{r},t)$, where we employ a spatially homogeneous vector potential $\vec{a}_{\perp}(t)$ that gives rise to the external electric field $\vec{E}^{\rm ext}_{\perp}(t) = - \tfrac{1}{c} \tfrac{\partial}{\partial t} \vec{a}_{\perp}(t)$. In a final step, to avoid solving the above mode-resolved Maxwell's equations, one often even ignores the spatial dependence of the induced field and merely uses $E_{\alpha}(t) = - {\boldsymbol\lambda_\alpha}\cdot\textbf{R}([\vec{a}_{\perp}],t)$. If we now determine in linear response $\textbf{R}([\vec{a}_{\perp}],t)$ we immediately see that when $\chi^{n}_{n}$ is changed due to strong light-matter coupling also the induced field is changed.} {Furthermore, the reformulation of the linear-response kernel in Eq.~(\ref{eq:chi_nn}) shows that we get a feedback from the induced photon field onto the matter. Such intrinsic back-reaction (screening) effects are very important for large systems, as is well known from solid-state physics, where the bare (vacuum) electric field as determined by Eq.~(\ref{eq:vacEfield}) does not agree with the measured spectrum. One needs to include the self-consistent polarization of the system that counter-acts the external perturbing field. This can be done {approximately} in linear response by solving self-consistently a Maxwell's equation with the matter response as input~\cite{ehrenreich1966, mochan1985, maki1991,luppi2010}.}}
{In the theory of classical electrodynamics, a convenient way to do so is to switch to the Maxwell's equations in matter. In the above considerations this means we introduce the displacement field $\vec{D}_{\perp} = \epsilon_0 \vec{E}_{\perp} + \vec{P}_{\perp}$, where now all the knowledge about how the system reacts to an external perturbation is encoded again in $\vec{P}_{\perp}$ such that we find
\begin{align}
\label{eq:MaxwellinMatter}
\biggl(\frac{1}{c^2} \frac{\partial^2}{\partial t ^2} - {\boldsymbol\nabla}^2\biggl)\vec{D}_{\perp}(\vec{r},t) =  - \nabla^2 \vec{P}_{\perp}([\vec{a}_{\perp}],\vec{r},t).
\end{align}
After expanding $\vec{D}_{\perp}(\vec{r},t) = \epsilon_0 \sum_{\alpha}\omega_{\alpha} {\boldsymbol \lambda}_{\alpha}(\vec{r}) q_{\alpha}(t)$ and then performing the long wave-length limit we arrive at
\begin{align}
\label{eq:displacement}
\left(\frac{\partial^2}{\partial t^2} + \omega_{\alpha}^2\right) q_{\alpha}(t) = \omega_{\alpha} {\boldsymbol \lambda}_{\alpha} \cdot \vec{R}([\vec{a}_{\perp}],t),
\end{align}
which is the classical analogue of Eq.~\ref{Max0}. In the usual decoupled light-matter description without self-consistency we then simply determine $\vec{R}([\vec{a}_{\perp}],t)$ from the electric permittivity and ignore any feedback that describes how the matter system affects (screens) the field. Approximate self-consistency is found once the induced field $\vec{E}_{\perp}$ is taken into account to screen the perturbing field $\vec{E}^{\rm ext}_{\perp}$. But in our case we want to go beyond this simple approximate self-consistency which will break down once the coupling between light and matter is strong. Note that in the macroscopic Maxwell's equation the electric field becomes $E_{\alpha}(t) = \omega_{\alpha} q_{\alpha}(t) - {\boldsymbol \lambda}_{\alpha} \cdot \vec{R}([\vec{a}_{\perp}],t)$, and we see that if we ignore the spatial dependence in determining $\vec{E}_{\perp}$ we basically assume $\vec{D}_{\perp} = \vec{E}_{\perp}$.} 

{In our description we keep the photon field as a dynamical variable of the system such that the Maxwell field couples to the electronic system}{, leading to a fully self-consistent description of the light-matter response. Besides the changes in $\chi^{n}_{n}$, which when used as an input into Eqs.~\eqref{eq:vacEfield-full} or \ref{eq:displacement}, captures the self-consistent response of the light-matter system, we can now also directly access the induced electric field by considering the response of the displacement field due to $\chi^{q}_{n}$ and the use of
\begin{align*}
\hat{\textbf{E}}_{\perp} = \sum_{\alpha=1}^M \boldsymbol{\lambda}_{\alpha}\omega_{\alpha}   \left( \hat{q}_{\alpha} - \frac{\boldsymbol{\lambda}_{\alpha}}{\omega_{\alpha}} \cdot\textbf{R} \right) .
\end{align*}	
As discussed in Sec.~\ref{sec:Observables}, this leads to a complete change of perspective, since it highlights that the excited states of the coupled light-matter system can be viewed as changes in the quantized Maxwell field in accordance to the usual experimental situation. On the other hand, we can now also investigate what the quantum description of the coupled light-matter system does to the Maxwell's equations.
We therefore consider the case where the free (time-derivative of the) current $\delta j_{\alpha}(t)$ is non-zero while the external classical field is zero, i.e., $\delta v(\textbf{r},t)=0$. In this case, we find
\begin{widetext}
\begin{align}
\left(\frac{\partial^{2}}{\partial t^{2}} + \omega_{\alpha}^{2}\right)\delta q_{\alpha}(t) = -\frac{\delta j_{\alpha}(t)}{\omega_{\alpha}} +   \omega_{\alpha} \boldsymbol{\lambda}_{\alpha} \cdot \int {e}\vec{r} \; \chi^n_{n, {\rm s}} f^{n}_\text{Mxc}\chi^{n}_{q_\alpha}\delta j_{\alpha} + \omega_{\alpha} \boldsymbol{\lambda}_{\alpha} \cdot \SumInt {e}\vec{r} \;  \chi^n_{n, {\rm s}} f^{q_{\alpha'}}_\text{Mxc}\delta q_{\alpha'}  . \label{ j(t)}
\end{align}
\end{widetext}
If we contrast this to the classical Maxwell's equation in matter 
\begin{align}
\left(\frac{\partial^2}{\partial t^2} + \omega_{\alpha}^2\right) \delta q_{\alpha}(t) = - \frac{\delta j_{\alpha}(t)}{\omega_{\alpha}} + \omega_{\alpha} {\boldsymbol \lambda}_{\alpha} \cdot \delta\vec{R}([\vec{j}_{\perp}],t),
\end{align}
where $\vec{R}([\vec{j}_{\perp}],t)$ would be determined from the response of the matter system due to the corresponding external field $\vec{a}_{\perp}$, we see that besides the self-consistent response of the matter system (second term on the right hand side) also a genuine new (matter-mediated) photon-photon interaction term (third term on the right hand side) appears.} Making again the mean-field explicit leads to
\begin{widetext}
\begin{align*}
\left(\frac{\partial^{2}}{\partial t^{2}} + \omega_{\alpha}^{2}\right)\delta q_{\alpha}(t) &= -\frac{\delta j_{\alpha}(t)}{\omega_{\alpha}} +   \omega_{\alpha} \boldsymbol{\lambda}_{\alpha} \cdot \int {e}\vec{r} \; \chi^n_{n, {\rm s}} \left[ \frac{e^2}{4 \pi \epsilon_0 |\textbf{r}' - \textbf{r}''| } + \sum_{\alpha'} \left(\boldsymbol{\lambda}_{\alpha'} \cdot {e}\textbf{r}''\right)\boldsymbol{\lambda}_{\alpha'} \cdot {e}\textbf{r}' \right]\chi^{n}_{q_\alpha}\delta j_{\alpha} \nonumber \\
& - \omega_{\alpha} \boldsymbol{\lambda}_{\alpha} \cdot \SumInt {e}\vec{r} \;  \chi^n_{n, {\rm s}} \left(  \omega_{\alpha'} \boldsymbol{\lambda}_{\alpha'} \cdot {e}\vec{r}' \right)\delta q_{\alpha'} 
+ \omega_{\alpha} \boldsymbol{\lambda}_{\alpha} \cdot \int {e}\vec{r} \; \chi^n_{n, {\rm s}} f^{n}_\text{xc}\chi^{n}_{q_\alpha}\delta j_{\alpha} \nonumber\\
&+ \omega_{\alpha} \boldsymbol{\lambda}_{\alpha} \cdot \SumInt {e}\vec{r} \;  \chi^n_{n, {\rm s}} f^{q_{\alpha'}}_\text{xc}\delta q_{\alpha'} . 
\end{align*}
\end{widetext}
If we ignore the xc contributions to the matter-photon and photon-photon response we get the pRPA approximation to the Maxwell's equation in matter. In this pRPA form we clearly see how the Maxwell's equation becomes non-linear because of the feedback between light and matter. Such non-linearities of the Maxwell's equations are investigated in great detail in high-energy physics in the context of strong-field QED~\cite{diPiazza2012}. In that case the strong fields lead to particle creation and thus a matter-mediated photon-photon interaction. In our case, we do not need these high energies because we consider the photon-photon interaction due to condensed matter in form of atoms, molecules or solids and use, e.g., a cavity to enhance the coupling. That the changes in the Maxwell's equations are not purely theoretical concepts but lead to observable effects can be seen in many physical situations. As mentioned before, the most well-known effect are polarization effects in solid-state systems~\cite{luppi2010}, but more strikingly are effects due to the quantum-matter-mediated photon-photon interactions, see e.g. Ref.~\cite{firstenberg2013}. In this context, the presented ab-initio method allows to theoretically investigate the photon-photon interactions and possibly predict systems with very strong photon-photon correlations. In such cases the strong photon correlations could be used to give complementary insights into molecular systems or to imprint the photonic correlations on the matter subsystem. {Besides these differences we highlight that the quantized Maxwell's equation in matter, if we allow for both, a free external current and a free external field, can indeed discriminate between these two sorts of perturbations. In a purely classical theory, due to Eq.~\ref{eq:classicalMaxwell}, there can be no difference. This provides a completely new playground to investigate the difference between classical and quantum physics.}

\section{Novel photonic observables and radiative lifetimes}%
\label{app:lifetimes}

In the presented framework, besides the above highlighted changes in, e.g., the Maxwell's equations, novel observables become accessible. For instance, one can monitor the response of the matter system due to a perturbation of the photonic subsystem by an external current. This allows to investigate directly the cross-correlation between the matter and the photon subsystem induced by $\chi^{n}_{q_{\alpha}}$. Also note, that this cross-correlation observable allows to distinguish between the response due to a purely classical field $\delta v(\vec{r},t)$ or due to a quantized field, since $\delta j_{\alpha}(t)$ generates photons (which is equivalent to just use a slightly different initial state with an incoming photon pulse) that then perturb the correlated matter-photon system. This makes the presented framework applicable to also determine observables due to novel spectroscopies that use quantum light~\cite{dorfman2016}. This area of spectroscopy is so far not accessible with common first-principle methods. One further important observable that can be captured in this approach is the intrinsic lifetimes of excited states, which is not accessible in standard matter-only quantum mechanics. Let us briefly explain what we mean by this. 
In standard quantum mechanics we find besides the ground state also other eigenstates, i.e., excited states. Hereby an eigenstate is a square-integrable eigenfunction of a self-adjoint, usually unbounded Hamiltonian. If we excite a matter system from its ground state into such an excited state, it will remain in this state as long as we do not perturb it. In quantum mechanics we then also have generalized eigenstates, so-called scattering states, which are not square-integrable and that constitute the continuous spectrum of such a Hamiltonian~\cite{blanchard2015}. The simplest example is the free electronic Hamiltonian $\hat{T} = \sum_{i=1}^{N} -\tfrac{\hbar^2}{2 m_e} {\boldsymbol\nabla}^{2}_i$ which in infinite space has a purely continuous spectrum consisting of non-normalizable plane-waves~\cite{teschl2014}. The physical interpretation of such scattering states - as already the name indicates - is that particles propagate to infinity and do not stay bound anywhere. Thus exciting a matter system from its ground state into such a generalized eigenstate corresponds to the physical process of ionization. Ionization, however, is something completely different than the process of spontaneous emission. That is, if we put an atom or molecule into an ``excited state'', even without a further perturbation it will relax to the ground state by emitting radiation. The time the system stays in this ``excited state'' before emitting a photon is called the lifetime. The process of spontaneous emission clearly cannot be captured by standard quantum mechanics where matter and light are decoupled. Non-relativistic QED, however, does capture this process~\cite{spohn2004} by coupling the matter system to the quantized electromagnetic field which consists of infinitely many harmonic oscillators. In this way the excited states of the bare matter system turn into resonances and the ground state (usually) remains the only eigenstate of the combined matter-photon system. While formally these resonances are indeed scattering states of the combined matter-photon system, it is only the photonic part that shows a scattering behavior, i.e., a photon leaves the vicinity of the matter subsystem. The matter subsystem just relaxes to the only stable state, its ground state~\cite{spohn2004}. In linear response such relaxation processes express themselves as finite linewidths of excitations, where the linewidth can be associated with the lifetimes of the different resonances. In our slightly simplified treatment based on Eq.~(\ref{eqn:h-dipole-2}) we only consider a finite number of photon modes, and hence we do not have genuine resonances. However, by including enough modes we sample the influence of the vacuum and instead of one sharp transition peak (which numerically is usually artificially broadened) we get many that approximate the resonance. In this way linear-response theory for non-relativistic QED in the long wavelength limit can determine lifetimes of real systems. We show an example for such an ab-initio lifetime calculation in Sec.~\ref{sec:num-lifetime}. This provides a further new field of research that the presented framework makes accessible for ab-initio theory. While it is conceptually very interesting to revisit well-known results for intrinsic radiative lifetimes of gas phase molecules, since we can now study the nature of resonances in detail (see, e.g., the discussion on the photonic nature of resonances in Sec.~\ref{sec:num-lifetime}), we have now access to even more exciting experimental situations. By changing the environment, e.g., putting the molecule inside a cavity and thus enhance certain modes while suppressing others, one can change and control the radiative lifetimes of single molecules~\cite{lettow2007,wang2017} {(see also Sec.~\ref{sec:novel})}. We can thus theoretically study and predict realistic experimental situations where non-trivial changes in the photonic vacuum, e.g., due to nearby surfaces or other physical entities, directly influence intrinsic lifetimes and properties of resonances. 

\section{Linear-response theory as a pseudo-eigenvalue problem}%
\label{sec:app:casida}
{In this section, we reformulate the linear-response theory of coupled electron-photon systems as a pseudo-eigenvalue problem. The entire linear-response in non-relativistic QED for the density and photon coordinate can be written in matrix form as}
\begin{align}
\begin{pmatrix}
\delta n\\
\delta q_1\\
\delta q_2\\
\vdots\\
\delta q_{M}
\end{pmatrix} = 
\begin{pmatrix}
\chi^n_n &\chi^n_{q_1} & \chi^n_{q_2} &\hdots &\chi^n_{q_{M}}\\
\chi^{q_1}_n &\chi^{q_1}_{q_1} & \chi^{q_1}_{q_2} &\hdots &\chi^{q_1}_{q_M}\\
\chi^{q_2}_n &\chi^{q_2}_{q_1} & \chi^{q_2}_{q_2} &\hdots &\chi^{q_2}_{q_M}\\
\vdots & \vdots & \vdots &\ddots & \vdots \\
\chi^{q_{M}}_n &\chi^{q_M}_{q_1} & \chi^{q_M}_{q_2} &\hdots &\chi^{q_{M}}_{q_{M}}\\
\end{pmatrix}
\begin{pmatrix}
\delta v\\
\delta j_{1}\\
\delta j_{2}\\
\vdots\\
\delta j_{M}
\end{pmatrix} \label{Matrix eq}
\end{align}
{where we imply integration over time and space when appropriate. In this form we clearly see that the density response of the coupled matter-photon system depends on whether we use a classical field $\delta v(\vec{r},t)$, photons, which are created by $\delta j_{\alpha}(t)$, or combinations thereof for the perturbation.} The explicit coupling between the subsystems (i.e. matter and photons) demonstrates changes in the subsystems as a result of the back-reaction between matter and photons. The cross-talk between the respective coupled subsystems shows up in the cross-correlation response functions which {leads to} changes in the respective observables $(n(\vec{r},t), q_{\alpha}(t))$. This becomes evident by considering an external perturbation of the coupled system with the external potential $\delta v(\textbf{r}t)$ reduces to the coupled set of responses
\begin{align}
\begin{cases}
\delta n(\textbf{r}t) = \iint dt'd\textbf{r}' \chi_{n}^{n}(\textbf{r}t,\textbf{r}'t') \delta v(\textbf{r}'t'), \\
\delta q_{\alpha}(t) = \iint dt'd\textbf{r}' \chi^{q_\alpha}_{n}(t,\textbf{r}'t') \delta v(\textbf{r}'t'). \label{delta_v}
\end{cases}
\end{align} 
Here, the cross-correlation response function $\chi^{q_\alpha}_{n}(t,\textbf{r}'t')$ accounts for the action of the matter subsystem on the photon field which gives rise to a response of the photon field as a result of perturbing the matter. In the semi-classical approach in which TDDFT is based on, the cross-correlation response function do not show up but rather just {a simplified form (since there the wave function describes only the matter subsystem) of }the $\chi^{n}_{n}(\textbf{r}t,\textbf{r}'t')$. Similarly, a perturbation of the coupled system with the external charge current $\delta j_{\alpha}(t)$ results in
\begin{align}
\begin{cases}
\delta n(\textbf{r}t) = \sum_{\alpha=1}^{M}\int dt' \chi_{q_{\alpha}}^{n}(\textbf{r}t,t') \delta j_{\alpha}(t'), \\
\delta q_{\alpha}(t) = \sum_{\alpha'=1}^{M}\int dt'  \chi^{q_\alpha}_{q_{\alpha'}}(t,t') \delta j_{\alpha'}(t'). \label{delta_j}
\end{cases}
\end{align}
The cross-correlation response function $\chi^{n}_{q_\alpha}(\textbf{r}t,t')$ accounts for the action of the photon field on the matter thus specifying the response of the density by perturbing the photon field and $\chi^{q_\alpha}_{q_{\alpha'}}(t,t')$ describes how photon interacts via matter thus specifying changes in response of the photon field.

Next, we need to find an efficient way to solve these linear-response equations in terms of the Maxwell KS system. First, performing a Fourier transformation from time $t$ and $t'$ to frequency space $\omega$ and using the Hxc and pxc kernels, we write the response functions of Eqs.~(\ref{eq:chi_nn})-(\ref{eq:chi_qq}) in the following compact notation
\begin{align}
\chi^n_n&=\chi^n_{n, {\rm s}} \!+\! \chi^n_{n, {\rm s}}\left[\left(f^n_\text{pxc}+f^n_\text{Hxc}\right) \chi^n_n + \sum_\alpha f^{q_\alpha}_\text{pxc}\chi^{q_\alpha}_{n}\right] , \label{FT1}
\\
\chi^{q_\alpha}_{n}&=  \sum_{\beta} \chi^{q_{\alpha}}_{q_{\beta,{\rm s}}} \; g^{n_{\beta}}_\text{M}\;\chi^{n}_{n} , \label{FT2}
\\
\chi^n_{q_\alpha} &= \chi^n_{n, {\rm s}} \left[ \sum_{\alpha'}f^{q_{\alpha'}}_\text{pxc} \; \chi^{q_{\alpha'}}_{q_{\alpha}} + \left(f^{n}_\text{pxc}+f^{n}_\text{Hxc}\right) \;\chi^{n}_{q_\alpha}\right] , \label{FT3}
\\
\chi^{q_\alpha}_{q_{\alpha'}} & = \chi^{q_\alpha}_{q_{\alpha',{\rm s}}} + \sum_{\beta}\chi^{q_\alpha}_{q_{\beta},{\rm s}} \; g^{n_{\beta}}_\text{M} \;\chi^n_{q_{\alpha'}} . \label{FT4}
\end{align} 
Those equations are coupled with respect to the external perturbations as seen in Eqs.~(\ref{delta_v})-(\ref{delta_j}). The perturbation with respect to the external potential $\delta v(\textbf{r}t)$ results in a coupled set of response functions $\left\{\chi^{n}_{n}(\textbf{r}t,\textbf{r}'t'),\chi^{q_{\alpha}}_{n}(t,\textbf{r}'t')\right\}$ and for the external current $\delta j_{\alpha}(t)$ gives the coupled set $\left\{\chi^{n}_{q_{\alpha}}(\textbf{r}t,t'), \chi^{q_{\alpha}}_{q_{\alpha'}}(t,t')\right\}$. These pairs of coupled response functions have to be solved in a self-consistent way to obtain the exact interacting response functions.
The response functions of Eqs.~(\ref{delta_v}) and (\ref{delta_j}) can be expressed in frequency space through a Fourier transform that yields
\begin{eqnarray}
\delta n_{v}(\textbf{r},\omega) &=& \int d\textbf{r}' \chi_{n}^{n}(\textbf{r},\textbf{r}',\omega) \delta v(\textbf{r}',\omega), \label{App13a} \\
\delta q_{\alpha,v}(\omega) &=& \int d\textbf{r}'\chi_{n}^{q_{\alpha}}(\omega,\textbf{r}')\delta v(\textbf{r}',\omega), \label{App13b} \\
\delta n_{j}(\textbf{r},\omega) &=& \sum_{\alpha}\chi_{q_{\alpha}}^{n}(\textbf{r},\omega)\delta j_{\alpha}(\omega), \label{App13c} \\ 
\delta  q_{\alpha,j}(\omega) &=& \sum_{\alpha'}\chi_{q_{\alpha'}}^{q_{\alpha}}(\omega) \delta j_{\alpha'}(\omega) . \label{App13d}
\end{eqnarray}
A perturbation with the external potential $\delta v(\textbf{r},\omega)$ induces the responses $\delta n_{v}(\textbf{r},\omega)$ and $\delta q_{\alpha,v}(\omega)$. Making a substitution of Eqs.~(\ref{FT1}) and (\ref{FT2}) into the density and displacement field response due to an external potential $\delta v(\textbf{r},\omega)$ yields after some algebra the following eigenvalue problem (for a detailed derivation, we refer the reader to appendix~\ref{app:linresp2})
\begin{widetext}
\begin{align}
&\left.
\begin{pmatrix}
L(\Omega_q) & K(\Omega_q) &  M(\Omega_q) &M(\Omega_q)  \\
K^*(\Omega_q) & L(\Omega_q) &  M^*(\Omega_q) &M^*(\Omega_q)  \\
N & N^* &\omega_\alpha & 0  \\
N & N^* & 0 & \omega_\alpha 
\end{pmatrix}\begin{pmatrix}
\textbf{X}_{1}(\Omega_q) \\
\textbf{Y}_{1}(\Omega_q) \\
\textbf{A}_{1}(\Omega_q) \\
\textbf{B}_{1}(\Omega_q) 
\end{pmatrix}
=\Omega_q
\begin{pmatrix}
1 & 0 & 0 & 0\\
0 & -1 & 0 & 0  \\
0 & 0 & 1 & 0 \\
0 & 0 & 0 & -1 
\end{pmatrix}
\right.
\begin{pmatrix}
\textbf{X}_{1}(\Omega_q) \\
\textbf{Y}_{1}(\Omega_q) \\
\textbf{A}_{1}(\Omega_q) \\
\textbf{B}_{1}(\Omega_q) 
\end{pmatrix}.\label{eq:casida-deltav}
\end{align}
\end{widetext}
In this equation, $\textbf{X}_1$ and $\textbf{Y}_1$ are the contributions to the full solution in the matter part of the equation, while $\textbf{A}_1$ and $\textbf{B}_1$ are the contributions to the solution in the photon part of the equation. Further, $\Omega_q$ refers to the many-body electron-photon excitation energies. {In comparison to the standard linear-response formulation of TDDFT, new $2\times2$ blocks arises, the $M$-block accounts for the explicit electron-photon interaction, the $N$-block accounts for the dipole coupling of the electronic system to the photon field and the $\omega_{\alpha}$-block are the frequencies of the photon field.} The quantity $L_{ai,jb}(\Omega_q) =\delta_{ab}\delta_{ij}\left(\epsilon_a -\epsilon_i \right) + K_{ai,jb}(\Omega_q)$ contains the difference of two Kohn-Sham energies $\epsilon_a$ and $\epsilon_i$, where $i$ refers to occupied orbitals and the index $a$ to unoccupied orbitals. The coupling-matrix $K$ is given by
\begin{align}
K_{ai,jb}(\Omega_q) &= \iint d\textbf{r} d\textbf{y}\varphi_i(\textbf{r})\varphi_a^*(\textbf{r})  f^n_{Mxc}{(\textbf{r},\textbf{y},\Omega_q)}\varphi_b(\textbf{y})\varphi^{*}_j(\textbf{y}). \label{52a}
\end{align}
The quantity $K_{ai,jb}(\Omega_q)$ differs from the electron-only case since $f^n_{Mxc} = f^n_{Hxc} +f^n_{pxc}$. Treating the photon field only externally reduces this matrix to the standard coupling matrix in TDDFT linear response with $f^n_{Mxc} = f^n_{Hxc}$. The two new coupling functions appearing, $M$ and $N$ that couple the matter block are given explicitly as
\begin{eqnarray}
M_{\alpha,ai}(\Omega_q)  &=&\int d\textbf{r} \varphi_i(\textbf{r})\varphi_a^*(\textbf{r}) f^{q_\alpha}_{Mxc}{ (\textbf{r},\Omega_q)} , \label{52b} \\
N_{\alpha,ia} &=& \frac{1}{2\omega_\alpha^2} \int d\textbf{r} \varphi^*_i(\textbf{r})\varphi_a(\textbf{r}) g^{n_{\alpha}}_{M}(\textbf{r}) . \label{52c}
\end{eqnarray}
We emphasize here, that the exact coupling matrix $N_{\alpha,ia}$ has no frequency dependence since the exact kernel for Eq.~(\ref{52c}) is equivalent to just the mean-field kernel of the photon modes as can be seen from Eq.~(\ref{eq:gnm}).
Given the exact kernels, the nonlinear pseudo-eigenvalue problem in Eq.~(\ref{eq:casida-deltav}) allows to compute the exact excitation energies of the coupled matter-photon system.
Of course, in practice, approximations have to be employed for the matter-photon response kernels as is also required in the matter-only response formalism. Since the explicitly known mean-field kernel $g^{n_{\alpha}}_{M}(\textbf{r})$ is already exact, only $f^{q_\alpha}_{Mxc}{ (\textbf{r},\Omega_q)}$ and $f^{n}_{Mxc}{ (\textbf{r},\textbf{y},\Omega_q)}$ are left to be approximated. \\
The above matrix equation of Eq.~(\ref{eq:casida-deltav}) can be cast into a Hermitian eigenvalue form following the same transformations as, e.g., in Ref.~\cite{bauernschmitt1996}, where we assume real-valued orbitals, i.e., $K=K^{*}$, $M=M^{*}$ and $N=N^{*}$. Further, we drop the dependency on $\Omega_q$ for brevity. Then we find the pseudo-eigenvalue equation, reminiscent to the equations found for excitation energies in Hartree-Fock theory and TDDFT~\cite{casida1996}. The eigenvalue problem of Eq.~(\ref{eq:casida-deltav}) is now written in a compact Hermitian form as
\begin{eqnarray}
\left(
\begin{array}{ c c  }
U &   V     \\
W &  \omega_{\alpha}^{2}
\end{array}
\right)
\left(
\begin{array}{ c }
\textbf{E}_{1}  \\
\textbf{P}_{1}  
\end{array}	
\right) 
&=& 
\Omega_{q}^{2}
\left(
\begin{array}{ c }
\textbf{E}_{1} \\
\textbf{P}_{1} 
\end{array}	
\right), \label{eq:casida-deltav1}
\end{eqnarray}
where the matrices $U$, $V$ and $W$ are given by $U = (L-K)^{1/2}(L + K)(L-K)^{1/2}$, $V = 2(L-K)^{1/2}M^{1/2}N^{1/2}\omega_{\alpha}^{1/2}$ and $W = 2\omega_{\alpha}^{1/2}N^{1/2}M^{1/2}(L-K)^{1/2} $. The matrices are given explicitly by
\begin{eqnarray}
U_{qq'} &=& \delta_{qq'}\omega_{q}^{2} + 2\sqrt{\omega_{q}\omega_{q'}}K_{qq'}(\Omega_{q}), \label{matrix1}\\
V_{q\alpha} &=& 2\sqrt{\omega_{q}M_{\alpha q}(\Omega_{q})N_{\alpha q}\omega_{\alpha}  }     ,\\
W_{\alpha q} &=& 2\sqrt{\omega_{\alpha}N_{\alpha q}M_{\alpha q}(\Omega_{q}) \omega_{q}}    ,
\end{eqnarray}
where the off-diagonal matrices $V_{q\alpha}$ and $W_{\alpha q}$ are transpose of each other, i.e., $V_{q\alpha} = W_{\alpha q}^{\top}$. The index $q {= (a,i)}$ describes transitions from the electronic occupied {$(i)$} to unoccupied states {$(a)$} and thus the difference of Kohn-Sham energies is given by $\omega_q=\epsilon_a-\epsilon_i$. With $\alpha$ we denote the photon modes. The eigenvectors $\textbf{E}_{1}$ and $\textbf{P}_{1}$ can be used to compute oscillator strengths of the coupled matter-photon system (see appendix~\ref{app:linresp3}). In the decoupling limit of light-matter interaction, Eq.~(\ref{eq:casida-deltav1}) reduces to the well-known \textit{Casida equation} Eq.(\ref{eq:casida-alternate})~\cite{casida1996}.
So far we did not solve anything but have just rewritten the problem in terms of unknown Mxc kernels that correct the uncoupled and non-interacting auxiliary response functions. To actually solve this problem we need to provide approximations to these unknown quantities. Here it becomes advantageous to have divided the full Mxc kernels in Hxc and pxc terms, such that we can use well-established approximations from electronic TDDFT for the Hxc and specifically developed approximations for the pxc terms (see Sec.~\ref{sec:rabi} for more details). In the following, we will employ the above introduced pRPA approximation, which is a straightforward generalization of the standard RPA of electronic-structure theory and yields the following kernels
\begin{eqnarray}
\!\!\!\!f^{n}_\text{H}(\textbf{r},\textbf{r}') &=& \frac{e^2}{4 \pi \epsilon_0 |\textbf{r} - \textbf{r}'|} , \quad f^{q_{\alpha}}_\text{p}(\textbf{r}) =  -\omega_{\alpha} \boldsymbol{\lambda}_{\alpha}\cdot {e}\textbf{r}  , \nonumber\\
f^{n_{\alpha}}_\text{p}(\textbf{r},\textbf{r}') &=& \sum_{\alpha} \left(\boldsymbol{\lambda}_{\alpha} \cdot {e}\textbf{r}'\right)\boldsymbol{\lambda}_{\alpha} \cdot {e}\textbf{r} \;, \quad g^{n_{\alpha}}_\text{M}(\textbf{r}) =  -\omega_{\alpha}^2 \boldsymbol{\lambda}_{\alpha}\cdot {e}\textbf{r} \!.\nonumber
\end{eqnarray}
We note that at the pRPA level the matter-photon coupling mediated via $g^{n_{\alpha}}_{\rm M}$ is exact. The influence of the photon-matter xc contributions $f^{q_{\alpha}}_\text{xc}$ and $f^{n}_\text{xc}$ will be highlighted in the next section.
{By connecting to the eigenstates  $\textbf{E}_1$ and $\textbf{P}_1$, we can assign to each of the individual poles of the response function, i.e. the excitation energies, the amount of  photonic and electronic contribution to that excitation by using
\begin{align}
\sigma_e = \sum_{i=1}^{N_\text{pairs}} {\left|E_{1,i}\right|^2},\\
\sigma_p = \sum_{\alpha=1}^{M} {\left|P_{1,\alpha}\right|^2},
\end{align}
where $N_\text{pairs}$ corresponds to the number of occupied-unoccupied pairs of KS orbitals, in our case $30\times500$. The sum of $\sigma_e$ and $\sigma_p$ is normalized to one, i.e. $\sigma_e+\sigma_p=1$.}

In the pRPA approximation all the frequency dependence that we suppressed at times for brevity now genuinely vanishes (an adiabatic approximation) which allows us to express $M$ and $N$ of Eqs.~(\ref{52b}) and (\ref{52c}) as
\begin{align}
M_{\alpha,ai} &=-\omega_\alpha\int dr \varphi_i(\textbf{r})\varphi_a^*(\textbf{r}) {\boldsymbol\lambda_\alpha}\cdot {e}\textbf{r},\\
N_{\alpha,ia} &=-\frac{1}{2}\int dr \varphi^*_i(\textbf{r})\varphi_a(\textbf{r}){\boldsymbol\lambda_\alpha}\cdot {e}\textbf{r}.
\end{align}
Next we want to connect to the standard matter-only linear-response framework~\cite{ullrich2011}. In defining the oscillator strength for the density-density response function, we make use of the relationship between the polarizability tensor and susceptibility. The first-order dipole polarizability is given by 
\begin{equation}
\delta \textbf{R}(t) = \int d\textbf{r}\; {e}\textbf{r} \; \delta n(\textbf{r},t), \label{os6}
\end{equation}
and in frequency space  $\textbf{R}(\omega) = \overleftrightarrow{\alpha}(\omega) \textbf{E}(\omega)$. The dynamic polarizability tensor can then be written as
\begin{equation}
\overleftrightarrow{\alpha}_{\mu\nu}(\omega)  =  \int d\textbf{r}\; {e}{r}_{\mu} \frac{\delta n(\textbf{r},\omega)}{\delta E_{\nu}(\omega)}, \label{os7}
\end{equation}
with $\mu,\nu = (1,2,3)$ denoting all three spatial directions. Connecting to the QEDFT linear-response theory, we find
\begin{eqnarray}
\overleftrightarrow{\alpha}_{\mu\nu}(\omega) &=& \sum_{I}\frac{2\textbf{r}^{\dagger}_{\mu} S^{1/2}\textbf{Z}_{I}\textbf{Z}_{I}^{\dagger}S^{1/2} \textbf{r}_{\nu}}{\omega^2 - \Omega_{I}^{2} } , \label{atom-pol}
\end{eqnarray}
where $\textbf{r}_{\mu}^{I} = \int d\textbf{r}\;{e}{r}_{\mu} \sum_{i,a}\Phi_{ia}(\textbf{r})$ is the Kohn-Sham transition dipole matrix element of the many-body transition $I$. Further, we have used $S = (L-K)$ and the transition density is defined as $\Phi_{ia}(\textbf{r}) = \varphi^*_{i}(\textbf{r}) \varphi_{a}(\textbf{r})$ in terms of Kohn-Sham orbitals.
This then allows to obtain the full photoabsorption cross section from the trace of the polarizability tensor through
\begin{align}
\sigma(\omega) = \frac{4 \pi \omega}{c} \Ib m \ \text{Tr}\overleftrightarrow{\alpha}(\omega)/3 . \label{photo}
\end{align}
For the oscillator strength~\cite{casida1996, ullrich2011}, we find
\begin{align}
f_I = \frac{2}{3}\sum_{\mu=1}^3\left|\textbf{Z}^\dagger_{I} S^{1/2}\textbf{r}^{I}_{\mu}\right|^2 = \frac{2}{3}\omega_{I}\sum_{\mu=1}^3\left|\langle \Psi_{0}|{e}{r}_{\mu}|\Psi_{I}\rangle\right|^2
\label{eq:oscillator-strength}
\end{align}
and also in the case of QEDFT, the oscillator strength satisfy the Thomas-Reiche-Kuhn sum rule (also known as $f$-sum rule), i.e. $\sum_{I} f_I = N$, where $N$ is the total number of electrons in the system. At this point, we also want to introduce the dipole strength function ${S}(\omega)$~\cite{ullrich2011} that is defined as
\begin{align}
{S}(\omega) = \sum_I f_I \delta(\omega- \Omega_I) 
\label{eq:dipole-strength}
\end{align}
and integrates according to the $f$-sum rule to the total number of electrons.
For the non-standard part of our response theory, i.e., matter-photon and photon-photon perturbations, we use similar constructions to display the results. Their derivations and definitions are given in appendix~\ref{app:linresp3}. We will discuss their physical meaning in the next section where we employ a simple yet illuminating model system. This will not only allow us to explain many of the so far abstract ideas in a straightforward manner, but we can also test the accuracy of the pRPA. 

\section{Examples for the coupled matter-photon response: Details on the Rabi Model}
~\label{sec:app:rabi}
{In this section, we give more details on the model system that have been employed in Sec.~\ref{sec:rabi}. }
The model Hamiltonian we consider is given by (in this section we switch for simplicity to atomic units)
\begin{equation}
\hat{H}_{R}(t) = \frac{\omega_{0}}{2}\hat{\sigma}_{z} + \omega_{c}\hat{a}^{\dagger}\hat{a} + \lambda\hat{\sigma}_{x}\hat{q} + j(t)\hat{q}  + v(t)\hat{{\sigma}}_{x},\label{eqn1}
\end{equation}
where $\omega_{0}$ is the transition frequency between the ground state $\ket{g}$ and excited state $\ket{e}$ and $\hat{\sigma}_{x}$ as well as $\hat{\sigma}_{z}$ are the usual Pauli matrices. We only keep one photon mode with frequency $\omega_{c}$ and use the usual photon creation and annihilation operators to represent the harmonic oscillator of this mode. By further compressing the notation, we then describe the coupling between matter and light by a coupling strength $\lambda$ and the displacement coordinate $\hat{q} = \frac{1}{\sqrt{2\omega_{c}}} \left(\hat{a} + \hat{a}^{\dagger} \right)$. Finally, we couple the matter system to a classical external perturbation $v(t)$ and the photon system to a classical external current $j(t)$ (a pictorial representation of the coupled system is given in Fig.~\ref{Fig:TwoLevel}). We note for consistency with respect to other works~\cite{ruggenthaler2014, pellegrini2015, dimitrov2017} that in the above Rabi model we can perform a unitary transformation that allows us to exchange $\hat{\sigma}_x$ and $\hat{\sigma}_z$. Both forms of the extended Rabi model are therefore equivalent. We further note that with respect to the full non-relativistic QED problem in the long wavelength approximation of Eq.~(\ref{eqn:h-dipole-2}) the Rabi model does not include the dipole self-energy term proportional to $(\boldsymbol{\lambda}\cdot \mathbf{R})^2$. This is because the analogous term in this model is just a constant energy shift, i.e., it is proportional to $\sigma_x^2 = \hat{\mathds{1}}$~\cite{rokaj2017}. For more levels this is no longer the case~\cite{dimitrov2017} and this term has to be taken into account, else the resulting eigenstates do not have a proper continuum limit~\cite{rokaj2017}. The responses that we want to consider in the following are those observables that couple to the external perturbations. In our case this is $\sigma_{x}(t)= \langle \Psi(t)|\hat{\sigma}_{x} |\Psi(t)\rangle$ (in essence the atomic dipole) and the displacement field $q(t)=\langle \Psi(t)|\hat{q}| \Psi(t)\rangle$. \\
{The response of these observables $(\delta\sigma_{x}(t),\delta q(t))$ to perturbations by the external pair $\left(\delta v(t),\delta j(t)\right)$ can be written similarly as Eq.(\ref{Matrix eq}) in the collective form}
\begin{equation}
\left(
\begin{array}{ c }
\delta \sigma_{x}(t)  \\
\delta q(t)
\end{array}	
\right)
= \int dt'
\left(
\begin{array}{ c c }
\chi_{\sigma_{x}}^{\sigma_{x}}(t,t') & \chi_{q}^{\sigma_{x}}(t,t')  \\
\chi_{\sigma_{x}}^{q}(t,t') & \chi_{q}^{q}(t,t')   
\end{array}
\right)
\left(
\begin{array}{ c }
\delta v(t') \\
\delta j(t') 
\end{array}	
\right).  \label{RabiResp}
\end{equation}
Again we find besides the usual matter-matter response $\chi^{\sigma_x}_{\sigma_x}$ also matter-photon responses $\chi_{q}^{\sigma_{x}}$ and $\chi^{q}_{\sigma_{x}}$, respectively, as well as a photon-photon response function $\chi_{q}^{q}$.
Next, in analogy to Sec.~\ref{sec:theory}, we reformulate the coupled matter-photon problem in form of a {Maxwell} KS auxiliary problem. Using by now well-established results of QEDFT for the extended Rabi model systems~\cite{ruggenthaler2014, pellegrini2015} we can introduce two effective fields
\begin{eqnarray}
v_\text{Mxc}([\sigma_{x},q];t) &=& v_{s}([\sigma_{x}];t) - v([\sigma_{x},q];t)  ,  \\
j_\text{M}([\sigma_{x}];t) &=& j_{s}([q];t) - j([\sigma_{x},q];t)  ,  
\end{eqnarray}
that force the auxiliary uncoupled, yet non-linear {Maxwell} KS system to generate the same dynamics of the internal pair $(\sigma_{x}(t), q(t))$ as the corresponding coupled reference system. For an uncoupled initial {Maxwell} state $|\Psi_{0} \rangle = |\psi_0\rangle \otimes |\varphi_0 \rangle$ that provides the same initial conditions for the internal pair as the physical initial state \cite{ruggenthaler2014, pellegrini2015}, we then have to solve self-consistently
\begin{align}
&i \frac{\partial}{\partial t}|\psi(t)\rangle =  \left[\frac{\omega_{0}}{2}\hat{\sigma}_{z} + \left(v(t) + v_\text{Mxc}([\sigma_{x},q];t)\right)\hat{\sigma}_{x}\right]|\psi(t)\rangle, 
\\
&\left(\frac{\partial^{2}}{\partial t^{2}} + \omega_{c}^{2}\right)q(t) = -j(t) - \lambda \sigma_{x}(t). \label{eqn7}
\end{align}
Since the photon subsystem is merely a shifted harmonic oscillator we get away with only solving the classical harmonic oscillator equation coupled to the dipole of the matter subsystem. We can then express the coupled response functions of Eq.~(\ref{RabiResp}) in analogy to Eqs.~(\ref{eq:chi_nn})-(\ref{eq:chi_qq}) by the uncoupled auxiliary response functions $\chi_{\sigma_{x},s}^{\sigma_{x}}$ and $\chi_{q,s}^{q}$ as
\begin{eqnarray}
\chi_{\sigma_{x}}^{\sigma_{x}}(t,t') &=& \chi_{\sigma_{x},s}^{\sigma_{x}}(t,t') 
\\
&& + \iint d\tau d\tau' \chi_{\sigma_{x},s}^{\sigma_{x}}(t,\tau) f_{Mxc}^{\sigma_{x}}(\tau,\tau')\chi_{\sigma_{x}}^{\sigma_{x}}(\tau',t') \nonumber
\\
&& +\iint d\tau d\tau' \chi_{\sigma_{x},s}^{\sigma_{x}}(t,\tau) f_{Mxc}^{q}(\tau,\tau')\chi_{\sigma_{x}}^{q}(\tau',t'), \label{N-N} \nonumber
\\
\chi_{q}^{\sigma_{x}}(t,t') &=& \iint d\tau d\tau' \chi_{\sigma_{x},s}^{\sigma_{x}}(t,\tau) f_{Mxc}^{q}(\tau,\tau')\chi_{q}^{q}(\tau',t') 
\\
&&+ \iint d\tau d\tau' \chi_{\sigma_{x},s}^{\sigma_{x}}(t,\tau) f_{Mxc}^{\sigma_{x}}(\tau,\tau')\chi_{q}^{\sigma_{x}}(\tau',t') , \label{N-Q} \nonumber \\
\chi_{\sigma_{x}}^{q}(t,t') &=& \iint d\tau d\tau' \chi_{q,s}^{q}(t,\tau) g_{M}^{\sigma_{x}}(\tau,\tau')\chi_{\sigma_{x}}^{\sigma_{x}}(\tau',t'), \label{Q-N}
\\
\chi_{q}^{q}(t,t') &=& \chi_{q,s}^{q}(t,t') \\ 
&& + \iint d\tau d\tau' \chi_{q,s}^{q}(t,\tau) g_{M}^{\sigma_{x}}(\tau,\tau')\chi_{q}^{\sigma_{x}}(\tau',t'). \label{Q-Q} \nonumber
\end{eqnarray}
The only real difference is that in the Rabi case we do not have a longitudinal interaction and therefore the Mxc contributions come solely from the matter-photon coupling, i.e., $f_{Mxc}^{\sigma_{x}} = f_{pxc}^{\sigma_{x}}$  and $f_{Mxc}^{q} = f_{pxc}^{q}$. This allows us to study exclusively the influence of these new terms and how approximations of them perform. 
\subsection{Matter-photon correlation effect in Maxwell's equations}
Let us follow the previous general section~\ref{sec:Observables} and briefly consider the influence of the matter-photon coupling on the Maxwell's equations in this model system, i.e., Eq.~(\ref{eqn7}). The inhomogeneous Maxwell's equation here accounts for the back-reaction of the matter on the field through the atomic dipole operator $\sigma_{x}([v,j];t)$. If we, for instance, perturb the two-level system directly via a $\delta v(t)$, the response of the Maxwell's equation expressed in terms of the uncoupled problem with the help of Eq.~(\ref{N-N}) becomes
\begin{widetext}
\begin{eqnarray}
\left(\partial_{t}^{2} + \omega_{c}^{2} \right)\delta q(t) &=&  -\lambda \int dt' \chi_{\sigma_{x},s}^{\sigma_{x}}(t,t') \delta v(t')  - \lambda \iiint dt'd\tau d\tau' \chi_{\sigma_{x},s}^{\sigma_{x}}(t,\tau) f_{Mxc}^{\sigma_{x}}(\tau,\tau')\chi_{\sigma_{x}}^{\sigma_{x}}(\tau',t') \delta v(t') \nonumber\\
&& - \lambda\iint d\tau d\tau' \chi_{\sigma_{x},s}^{\sigma_{x}}(t,\tau) f_{Mxc}^{q}(\tau,\tau') \delta q(\tau').  
\end{eqnarray}
\end{widetext}
Having no coupling, i.e., the Mxc terms are zero, merely recovers the usual inhomogeneous Maxwell's equation for a classical external current. The matter system evolves according to the perturbation and we can determine its induced Maxwell field without any back-reaction. The second term describes the matter polarization due to the induced field and leads to an effective self-interaction of the two-level system. If there would be more than one particle this would induce an effective matter-matter interaction as well. The third term then accounts for the field polarization and induces an effective self-interaction in the mode of the light field. That is, the coupling to matter leads to a photon-photon interaction. This can be made more explicit by separating the mean-field contribution $v_\text{M}(t) = \lambda q(t)$ and rewriting the above equation as
\begin{widetext}
\begin{eqnarray}
\left(\partial_{t}^{2} + \omega_{c}^{2} \right)\delta q(t) &=&  -\lambda \int dt' \chi_{\sigma_{x},s}^{\sigma_{x}}(t,t') \delta v(t')  - \lambda \iiint dt'd\tau d\tau' \chi_{\sigma_{x},s}^{\sigma_{x}}(t,\tau) f_{xc}^{\sigma_{x}}(\tau,\tau')\chi_{\sigma_{x}}^{\sigma_{x}}(\tau',t') \delta v(t') \nonumber\\
&& - \lambda^{2}\int d\tau \chi_{\sigma_{x},s}^{\sigma_{x}}(t,\tau)\delta q(\tau) - \lambda\iint d\tau d\tau' \chi_{\sigma_{x},s}^{\sigma_{x}}(t,\tau) f_{xc}^{q}(\tau,\tau') \delta q(\tau').  
\end{eqnarray}
\end{widetext}
The third term on the right-hand side is then the pRPA form of photon-photon response. Similar terms also appear for a perturbation induced by an external current $\delta j(t)$ which can be rewritten with the help of Eq.~(\ref{N-Q}) and the mean-field made explicit as
\begin{widetext}
\begin{eqnarray}
\left(\partial_{t}^{2} + \omega_{c}^{2} \right)\delta q(t) &=&-\delta j(t) -\lambda^{2}\int d\tau \chi_{\sigma_{x},s}^{\sigma_{x}}(t,\tau)\delta q(\tau)  - \lambda \iint d\tau d\tau' \chi_{\sigma_{x},s}^{\sigma_{x}}(t,\tau) f_{xc}^{q}(\tau,\tau')\delta q(\tau') \nonumber\\
&& - \lambda \iiint dt'd\tau d\tau' \chi_{\sigma_{x},s}^{\sigma_{x}}(t,\tau) f_{xc}^{\sigma_{x}}(\tau,\tau')\chi_{q}^{\sigma_{x}}(\tau',t') \delta j(t') .
\end{eqnarray}
\end{widetext}
Here we used that $f_\text{M}^{q}(\tau,\tau') = \lambda \delta(\tau - \tau')$. As is most obvious in the pRPA limit, both types of perturbations lead to the same resonance conditions, i.e., peaks in the responses. They are connected to the combined eigenstates of the matter-photon system. However, the detailed response can differ strongly. That these resonance conditions that we get from the pRPA are indeed connected to the coupled eigenstates we will show next.
\subsection{Application of the pseudo-eigenvalue problem}
As a preparatory step we first rewrite the linear-response problem of the extended Rabi model in terms of the previously introduced pseudo-eigenvalue problem of Eq.~(\ref{eq:casida-deltav1}). In the two-level one-mode case we consider here, this reduces to
\begin{eqnarray}
\left(
\begin{array}{ c c  }
U(\Omega_{q}) &   V(\Omega_{q})     \\
W(\Omega_{q}) &  \omega_{c}^{2}
\end{array}
\right)
\left(
\begin{array}{ c }
\textbf{E}_{1}  \\
\textbf{P}_{1}   
\end{array}	
\right) 
&=& 
\Omega_{q}^{2}
\left(
\begin{array}{ c }
\textbf{E}_{1}  \\
\textbf{P}_{1}  
\end{array}	
\right). \label{example}
\end{eqnarray}
Where the matrices {in the model system reduce to functions of $\Omega_{q}$ as  $U = \omega_{0}^{2} + 2\omega_{0}K(\Omega_{q})$, $V(\Omega_{q}) = 2\omega_{0}^{1/2} M(\Omega_{q})^{1/2} N^{1/2}\omega^{1/2}_{c}$, and $W(\Omega_{q}) = 2\omega^{1/2}_{c} N^{1/2} M(\Omega_{q})^{1/2}\omega_{0}^{1/2} $}. The coupling functions are given explicitly using Eqs.(\ref{52a})-(\ref{52c}) as
\begin{eqnarray}
K(\Omega_{q}) &=&  f_{M}^{\sigma_{x}} + f_{xc}^{\sigma_{x}}(\Omega_{q}) , \nonumber \\
M(\Omega_{q}) &=& f_{M}^{q} + f_{xc}^{q}(\Omega_{q}) , \nonumber \\
N &=& \frac{1}{2\omega_{c}}g_{M}^{\sigma_{x}} , \nonumber
\end{eqnarray}
where the Kohn-Sham states is the dipole matrix element $\varphi_{a}\varphi_{i}^{*}=\langle g|\hat{\sigma}_{x}|e\rangle =1$.
The Mxc kernels can be defined using the inverse of the auxiliary and interacting response functions (see also Eqs.~(\ref{App20})-(\ref{App21}) in the appendix) and are given in frequency space by
\begin{eqnarray}\label{fmxc_kernels}
f_{Mxc}^{\sigma_{x}}(\omega) &=& \left(\chi_{\sigma_{x},s}^{\sigma_{x}}(\omega)\right)^{-1} - \left(\chi_{\sigma_{x}}^{\sigma_{x}}(\omega)\right)^{-1}, \\
f_{Mxc}^{q}(\omega) &=& - \left(\chi_{n}^{q}(\omega)\right)^{-1}. \label{Q-N-RW}
\end{eqnarray}
Here $\left(\chi_{\sigma_{x},s}^{\sigma_{x}}(\omega)\right)^{-1}$, $\left(\chi_{\sigma_{x}}^{\sigma_{x}}(\omega)\right)^{-1}$ and $\left(\chi_{n}^{q}(\omega)\right)^{-1}$ are the inverses of the uncoupled response function of the electronic subsystem, the fully coupled response function of the electronic dipole and of the displacement field of the Rabi model, respectively. With these quantities we then determine spectroscopic observables such as the photoabsorption cross section. To determine this cross section we first note that the linear polarizability $\alpha(\omega)$ induced by the external potential $v(\omega)$ is related to the ``dipole-dipole'' response function as $\alpha(\omega) =\chi_{\sigma_{x}}^{\sigma_{x}}(\omega)$. Using Eq.~(\ref{photo}), we can determine the photoabsorption cross section of the Rabi model (see Fig.~\ref{mixed}~(a) displayed in dotted-red for the numerically exact case).
\begin{equation}
\sigma(\omega) = \frac{4 \pi \omega}{c} \;\Ib m \ \chi_{\sigma_{x}}^{\sigma_{x}}(\omega). \label{photo1}
\end{equation}
Here, the mean of the polarizability was not considered since the Rabi model is a one-dimensional system. Analogously, we define a linear ``field polarizability'' $\beta(\omega)$ due to polarizing the photon mode by an external current. In the same way, we relate the field polarizability to the response function of the photon mode as $\beta(\omega) = \chi_{q}^{q}(\omega)$ and then determine a photonic spectrum from (see Fig.~\ref{mixed}~(b) displayed in dotted-red for the numerically exact case). 
\begin{equation}
\tilde{\sigma}(\omega) \equiv \frac{4 \pi \omega}{c} \;\Ib m \ \chi_{q}^{q}(\omega). \label{photo2}
\end{equation}
Finally, we consider mixed spectroscopic observables where we perturb one subsystem and then consider the response in the other. We analogously employ $ \chi_{\sigma_{x}}^{q}(\omega)$ and $\chi_{q}^{\sigma_{x}}(\omega)$ in Eqs.(\ref{photo1}) and (\ref{photo2}), respectively, to determine a ``mixed polarizability''. If we plot this mixed spectrum (see Fig.~\ref{mixed}~(c) displayed in dotted-red for the numerically exact case), we find that we have positive and negative peaks. Indeed, this highlights that excitations due to external perturbations can be exchanged between subsystems, i.e., energy absorbed in the electronic subsystem can excite the photonic subsystem and vice versa.  
Next, we want to employ the pRPA approximation to the extended Rabi model and try to solve it analytically. The pRPA is equivalent to using the mean-field approximation in the coupled equations, i.e., approximating the electron-photon coupling term as $\hat{\sigma}_{x}\hat{q} \approx \langle\hat{\sigma}_{x}\rangle\hat{q} + \langle\hat{q}\rangle\hat{\sigma}_{x}$. This corresponds then to a coupled Schr\"odinger-Maxwell treatment of the coupled matter-photon problem~\cite{ruggenthaler2017b}. In the {Maxwell} KS equations this leads to approximating the full $v_\text{Mxc}$ by the mean-field potential $v_\text{M} = v_\text{p} = \lambda q$. The mean-field current is known explicitly as $j_\text{M} = \lambda \sigma_{x}.$ In the case of the pseudo-eigenvalue problem this amounts to approximating $K = f_\text{M}^{\sigma_{x}}= 0$, $M = f_\text{M}^{q}= \lambda$ and $N=\frac{1}{2\omega_{c}}g_\text{M}^{\sigma_{x}} = \frac{\lambda}{2\omega_{c}}$. Consequently we have 
\begin{equation*}
U = \omega_{0}^{2}, \quad V = W = 2\lambda\sqrt{\frac{\omega_{0}}{2}}, \quad \omega_{\alpha}^{2} = \omega_{c}^{2} .
\end{equation*}
The resulting nonlinear eigenvalue equation yields the excitation frequencies 
\begin{align}
\Omega_{1}^{2}(-) &=\frac{1}{2}\left(\omega_{0}^2 + \omega_{c}^2\right) - \frac{1}{2}\sqrt{\left(\omega_{0}^2 - \omega_{c}^2 \right)^2 + 8\lambda^{2}\omega_{0}}, 
\\ 
\Omega_{1}^{2}(+) &=\frac{1}{2}\left(\omega_{0}^2 + \omega_{c}^2\right) + \frac{1}{2}\sqrt{\left(\omega_{0}^2 - \omega_{c}^2 \right)^2 + 8\lambda^{2}\omega_{0}}, 
\end{align}
and the corresponding normalized eigenvectors can be given in closed form as
\begin{equation}
\textbf{E}_{1} =
\left(
\begin{array}{ c }
-\sin\theta  \\
\cos\theta
\end{array}	
\right)
, \quad \textrm{and} \quad
\textbf{P}_{1} =
\left(
\begin{array}{ c }
\cos\theta  \\
\sin\theta
\end{array}	
\right)
\end{equation}

The resulting pRPA-approximated spectra are displayed in Fig.~\ref{mixed} in dashed-blue. We will discuss the results in a little more detail at the end of this section. Before we consider a slightly more advanced approximation based on the rotating-wave approximation (RWA).
If we slightly simplify the full Rabi problem by approximating the full coupling as $\hat{\sigma}_x \hat{q} \approx  \frac{1}{\sqrt{2\omega_{c}}}\left(\hat{{\sigma}}_{+}\hat{a} + \hat{{\sigma}}_{-}\hat{a}^{\dagger} \right)$ we end up with the Jaynes-Cumming Hamiltonian~\cite{jaynes1963} given as
\begin{eqnarray}
\hat{H}_{JC}(t) &=& \frac{\omega_{0}}{2}\hat{\sigma}_{z} + \omega_{c}\hat{a}^{\dagger}\hat{a} + \frac{\lambda}{\sqrt{2\omega_{c}}}\left(\hat{{\sigma}}_{+}\hat{a} + \hat{{\sigma}}_{-}\hat{a}^{\dagger} \right) \nonumber\\
&& + j(t)\hat{q} + v(t)\hat{{\sigma}}_{x}.
\end{eqnarray}
Here we used $\hat{{\sigma}}_{\pm} = (\hat{\sigma}_{x} \pm i\hat{\sigma}_{y})/2$. The above approximation is called the RWA because we ignore quickly oscillating terms and thus assume that the excitation of the matter subsystem can only destroy and the de-excitation only create a photon. This approximation is justified (with respect to the full wave function) if we are in the weak coupling regime, i.e., $\lambda\ll\omega_{c}$, and near to resonance, i.e., $\delta=\omega_{0} - \omega_{c} \approx 0$. The ground-state of the Jaynes-Cummings model is the uncoupled tensor product of the matter ground-state and the photon ground-state with ground-state energy of $E_{0}=-\omega_{0}/2$. The excited states of the Jaynes-Cummings Hamiltonian are known analytically and are given by (we only show the lowest lying excited states where a single photon is excited {and for which the matrix elements are non-zero with the ground-state})
\begin{eqnarray}
|-,0\rangle &=& -\sin\theta_{0}|g\rangle|1\rangle + \cos\theta_{0}|e\rangle|0\rangle, \\
|+,0\rangle &=&  \cos\theta_{0}|g\rangle|1\rangle + \sin\theta_{0}|e\rangle|0\rangle. 
\end{eqnarray}

With these eigenstates we find the transition frequencies that correspond to the linear response from the ground state ({due to the approximations involved only} one photon absorbed or emitted) to be
\begin{eqnarray}
\Omega_{-}(0) &=& \frac{1}{2}\left(\omega_{c} + \omega_{0} - \Omega_{0}\right), \label{RWLP} \\ 
\Omega_{+}(0) &=& \frac{1}{2}\left(\omega_{c} + \omega_{0} + \Omega_{0}\right), \label{RWUP}
\end{eqnarray}
where $\Omega_{0} = \sqrt{\delta^{2} + 4\lambda'^{2}}$ where $\lambda'=\frac{\lambda}{\sqrt{2\omega_{c}}}$. 
So we already know where the RWA will generate the poles of the response function. Since we know analytically the eigenfunctions in the RWA, we can construct the RWA response functions analytically. Using the definitions of the Mxc kernels of Eq.~(\ref{fmxc_kernels}) we can then analytically construct the RWA Mxc kernels. These kernels are frequency dependent, therefore the resulting Mxc approximation is non-adiabatic~\cite{ullrich2011}. Substituting them into $M(\Omega_{q})$ we recover the known poles $\Omega_{q} = \Omega_{\pm}(0)$ from Eq.~(\ref{example}). Further, we can then construct the different spectra associated with the RWA. We show them in Fig.~\ref{mixed} in full-orange.

\section{Numerical details}
\label{sec:numerics}
We start by discussing the general setup before considering the specialized situations discussed above. We have implemented the linear-response pseudo-eigenvalue equation of Eq.~(\ref{eq:casida-deltav1}) into the real-space code OCTOPUS~\cite{marques2003,andrade2014}. The absorption spectrum of the benzene molecule has been very successfully studied with TDDFT calculations~\cite{yabana1999,marques2003}. Small organic molecules and benzene in particular are rewarding systems to be studied with TDDFT, since the adiabatic approximation in concert with the local-density approximation (LDA)~\cite{kohn1965,perdew1981} capture the occurring $\Pi$-$\Pi^*$ transition exceptionally well~\cite{yabana1999}. This transition is a characteristic of carbon conjugate compounds~\cite{marques2003} and occurs around 7~eV in the case of a benzene molecule.  To calculate the electronic structure of the benzene molecule, we follow closely the setup of Ref.~\cite{marques2003}. Thus, we use a cylindric real space grid of $8$~{\AA} length with the radius of $6$~{\AA} in the $x$-$y$ plane, and a spacing of $\Delta x = 0.22 \AA$. For the benzene nuclear structure, we use the CC bond length of $1.396$~\AA, and CH bond length of $1.083$~\AA. We explicitly describe the 30 valence electrons, while the core atoms are considered implicitly by LDA Troullier-Martins pseudopotentials~\cite{troullier1993}. In the excited state manifold, we include $500$ unoccupied states in the pseudo-eigenvalue calculation. This number amounts to $30\times500=7500$ pairs of occupied-unoccupied states. Further, to describe the electron-electron interaction in the response functions, we apply the adiabatic LDA (ALDA) kernel, i.e. $f^{n}_{Mxc} \rightarrow f^{n}_\text{Hxc,ALDA} + f^{n}_p$. 
Solving the linear-response pseudo-eigenvalue problem of Eq.~(\ref{eq:casida-deltav1}) provides us with the transition amplitudes, as well as the excitation energies of the correlated electron-photon system. These quantities can be used to calculate photoabsorption spectra by using e.g. Eq.~(\ref{eq:dipole-strength}). In standard calculations, to obtain such spectra and mimic the finite lifetime of the excited state usually a peak-broadening is applied. In our case, {where necessary}, we apply a Lorentzian broadening, i.e. the standard implementation of the OCTOPUS code, that is of the following form
\begin{align}
\Gamma(\omega, \omega_I) = \frac{1}{\pi}\frac{\Delta}{(\omega-\omega_I)^2+\Delta^2},
\label{eq:lolle}
\end{align}
where $\omega_I$ is the excitation frequency, and $\Delta$ the broadening parameter. The actual dipole strength function as defined in Eq.~(\ref{eq:dipole-strength}) is then obtained by
\begin{align}
\label{eq:s-lorentzian}
{S}(\omega) = \sum_I f_I \Gamma(\omega, \omega_I),
\end{align} 
where $f_I$ denotes the oscillator strength as defined in Eq.~(\ref{eq:oscillator-strength}). {We obtain the spectra {for systems not immersed in the photon bath} in this paper, where the peaks have been broadened by applying the broadening as defined in Eq.~(\ref{eq:lolle}) with $\Delta= 0.1361$~eV.\\}

\begin{table}
\begin{center}
\begin{tabular}{ | c | c | c| c | }
\hline
$\lambda_\alpha$ $[$eV$^{1/2}$/nm$]$ &  $E_I$ [eV] & $\langle x_I\rangle$ [A] &  $f_I$ [a.u.]\\\hline \hline
     0 & 6.88  & 0.952 & 0.546\\\hline
     2.77 & 6.69&  0.721&   0.304\\
    2.77  &7.03 & 0.626 &0.241\\\hline
     5.55 & 6.49&  0.791&  0.355\\
    5.55  &7.18 &0.550 & 0.190\\\hline
     8.32 & 6.28&  0.848 & 0.395\\
    8.32  &7.30 &0.482  &0.149\\\hline
    11.09 &    6.06 & 0.896 & 0.426\\
    11.09 & 7.41 &0.420  &0.114 \\\hline
\end{tabular}
\end{center}
\caption{Rabi splitting of the $\Pi-\Pi^*$ transition: electron-photon interaction strength $\lambda_\alpha = |\boldsymbol\lambda_\alpha|$, excitation energy $E_I$, transition dipole moment $x_I$ and the oscillator strength $f_I$.}
\label{tab:casida}
\end{table}

\vspace{10em}

\bibliography{../01_light_matter_coupling} 

\pagebreak
\widetext
\begin{center}
\textbf{\large Supplemental Information:\\{Light-Matter Response in Non-Relativistic Quantum Electrodynamics: Quantum Modifications of Maxwell's Equations}}
\end{center}
\setcounter{section}{0}
\setcounter{equation}{0}
\setcounter{figure}{0}
\setcounter{table}{0}
\setcounter{page}{1}
\makeatletter
\renewcommand{\thesection}{S\arabic{section}}
\renewcommand{\theequation}{S\arabic{equation}}
\renewcommand{\thefigure}{S\arabic{figure}}

\section{Current state of the art for spectroscopic: semi-classical description}
\label{sec:semi-classics}
{To highlight the many differences of the presented framework to the standard linear-response approach we give here a brief recapitulation of the standard (matter-only) theory.}
{The current theoretical description of linear spectroscopic techniques is built on the \textit{semi-classical} approximation~\cite{gilbert2010}. Herein, the many-particle electronic system is treated quantum mechanically while the nuclei are subject to the Born-Oppenheimer approximation and the electromagnetic field appears as an external perturbation. As an external perturbation, the electromagnetic field probes the quantum system, but is not a dynamical variable of the complete system. To arrive at the semi-classical description starting from the full non-relativistic description of the Pauli-Fierz Hamiltonian~\cite{ruggenthaler2015}, several approximations are used to simplify the problem. In the following, we list these approximations explicitly 
\begin{itemize}
\item The mean-field approximation renders the Pauli-Fierz Hamiltonian as a problem of two coupled equations, i.e. the time-dependent Pauli equation and the inhomogeneous Maxwell's equations, and is also know as the Maxwell-Pauli equation~\cite{ruggenthaler2017b}. 
\item The decoupling of these Maxwell-Pauli equations {leads to} the inhomogeneous Maxwell's equation {becoming independent of} the electronic system and all field effects are treated as a classical external field that perturbs the many-electron system.
\item The dipole approximation, which ensures the uniformity of the {external (decoupled)} field over the extend of the electronic system.
\end{itemize}
Based on these approximations the Pauli-Fierz Hamiltonian~\cite{ruggenthaler2017b} reduces to the time-dependent semi-classical Hamiltonian for many-particle systems given as
\begin{align}
\label{eqn:semi-classical} 
\hat{H}_e(t)&=\sum\limits_{i=1}^{N}\left(-\frac{\hbar^2}{2m_e}{\boldsymbol\nabla}_i^2+v(\textbf{r}_i,t)\right)+\frac{e^2}{4 \pi\epsilon_0}\sum\limits_{i>j}^{N}\frac{1}{\left|\textbf{r}_i-\textbf{r}_j\right|},
\end{align}
including the kinetic energy, time-dependent external potential and the longitudinal Coulomb interaction. The time-dependent external potential has two parts $v(\textbf{r},t) = v_0(\textbf{r}) + \delta v(\textbf{r},t)$. Here, $v_0(\textbf{r})$ describes the attractive part of the external potential due to the nuclei and $\delta v(\textbf{r},t) = {e}\textbf{r}\cdot\textbf{E}_{\perp}(t)$ with $\textbf{E}_{\perp}(t)$ being a classical external (transversal) probe field in dipole approximation that couples to the electronic subsystem. In this decoupling limit of light and matter, the many-particle wavefunction is labeled only by the particle coordinate and spin as $\Psi(\vec{r}_{1} \sigma_{1},...,\vec{r}_{N} \sigma_{N})$. In the dipole approximation we can  investigate dipole-related spectroscopic observables such as polarizabiltiy, absorption and emission spectra, etc from linear to all orders in the external perturbation. Consider the particular case of a response of an electronic system to an external weak probe field. In the dipole limit a key observable in the study of electronic and optical excitations in large many-particle systems is the electron density. Formulated within linear-response, the density response to an external perturbation is given as \cite{kubo}:
\begin{align}
\delta n(\textbf{r}t)
&=-\frac{i}{\hbar}\int_{t_{0}}^{t} dt' \int d\textbf{r}' \langle\Psi_{0}|\left[\hat{n}_{I}(\textbf{r}t),\hat{n}_{I}(\textbf{r}'t')\right]|\Psi_{0}\rangle  \nonumber\\
&=\int_{t_{0}}^{t} dt' \int d\textbf{r}' \tilde{\chi}^n_n(\textbf{r}t,\textbf{r}'t') \delta v(\textbf{r}'t') \label{eq:delta_n}.
\end{align}
Here, $\tilde{\chi}^n_n(\textbf{r}t,\textbf{r}'t')$ is the density-density function with respect to the ground-state $\Psi_{0}(\vec{r}_{1} \sigma_{1},...,\vec{r}_{N} \sigma_{N})$. Practical calculations for the response of a many-electron system is a considerable challenge due to the large degrees of freedom. In practice, time-dependent density functional theory (TDDFT) \cite{marques2006,gross1996} is one of the most frequently applied theories to approach this problem. Knowing the electron density in TDDFT we can in principle calculate all observables of interest. Formulated within TDDFT linear-response, the density-density response function of the interacting system can be expressed in terms of non-interacting the density-density response function and an exchange-correlation (xc) kernel that has a form of a Dyson-type equation \cite{petersilka1996}:
\begin{align}
\label{eq:chi_n_n}
\tilde{\chi}^n_n(\textbf{r}t,\textbf{r}'t') 
&=\chi^n_{n, {\rm s}}(\textbf{r}t,\textbf{r}'t') +\iint \text{d}\textbf{x}  \text{d}\tau \iint \text{d}\tau'\text{d}\textbf{y}\chi^n_{n, {\rm s}}(\textbf{r}t,\textbf{x}\tau)f_\text{Hxc}{(\textbf{x}\tau,\textbf{y}\tau')}\tilde{\chi}^n_n{(\textbf{y}\tau',\textbf{r}'t')},
\end{align} 
where $\chi^n_{n, {\rm s}}$ and $f_\text{Hxc} = \left(\chi^{n}_{n, {\rm s}}\right)^{-1} - \left(\tilde{\chi}^n_{n}\right)^{-1}$. One of the most widely employed approaches to TDDFT linear-response is the Casida formalism which can be written in a compact matrix form. The Casida equation obtains the exact excitation energies $\Omega_{q}$ of the many-particle system and requires all occupied and unoccupied Kohn-Sham orbitals and energies including the continuum of states. In practice, the Casida equation is often cast into the following form 
\begin{align}
\label{eq:casida-alternate}
U \textbf{E} = \Omega_{q}^{2} \textbf{E}.
\end{align}
The explicit form of the matrix elements is given as (with $q=(i,a)$)
\begin{align}	
U_{qq'} &= \delta_{qq'}\omega_{q}^{2} + 2\sqrt{\omega_{q}\omega_{q'}}K_{qq'}(\Omega_{q})  ,\\
K_{ai,jb}(\Omega_{q}) &= \iint d\textbf{r} d\textbf{r}'\varphi_i(\textbf{r})\varphi_a^*(\textbf{r})  f_{Hxc}{(\textbf{r},\textbf{r}',\Omega_{q})}\varphi_b(\textbf{r}')\varphi^*_j(\textbf{r}') . \nonumber
\end{align}
{The Casida formalism is well established and has been applied to a variety of systems, see e.g. Refs.~\cite{casida1996, jamorski1996, casida2000, casida2012, yang2013} and references therein.}} 

{The many obvious shortcomings of the approximations that lead to the standard Schr\"odinger equation~\eqref{eqn:semi-classical} are well-known and discussed to some extend in the main part of the paper (for more details see, e.g., Ref.~\cite{ruggenthaler2017b}). We point out that all of the above ubiquitous fundamental equations are modified and the results based on the introduced generalized equations can differ strongly, as discussed in Sec.~\ref{sec:general} of the main article.}

\section{Linear-response in non-relativistic QED}
\label{app:linresp}

{To help the reader with the unfamiliar generalized linear-response framework for coupled light-matter systems, we here} derive the linear-response equations and the ensuing response functions presented in Sec.~\ref{sec:theory}. In the non-relativistic setting of QED, the static and dynamical behavior of the coupled electron-photon systems is given by
\begin{equation}
\hat{H}(t) = \hat{H}_{0} + \hat{H}_{ext}(t). \label{App1}
\end{equation}
Where we define the time-independent electron-photon Hamiltonian as
\begin{eqnarray}
\hat{H}_{0} &=& \hat{T} + \hat{W}_{ee} + \frac{1}{2}\sum_{\alpha=1}^{M}\left[\hat{p}_{\alpha}^{2} + \omega_{\alpha}^{2}\left(\hat{q}_{\alpha} - \frac{\boldsymbol{\lambda}_{\alpha}}{\omega_{\alpha}}\cdot \textbf{R} \right)^{2} \right]  + \sum\limits^{N}_{i=1}v_{0}(\textbf{r}_{i}) + \sum_{\alpha=1}^{M}\frac{j_{\alpha,0}}{\omega_{\alpha}}\hat{q}_{\alpha}, \label{App2}
\end{eqnarray}
where the kinetic energy operator is $\hat{T} =-\frac{\hbar^2}{2m_{e}}\sum^{N}_{i=1}{\boldsymbol\nabla}^2_{i}$, the Coulomb potential is $\hat{W}_{ee} = \frac{e^2}{4\pi\epsilon_0}\sum^{N}_{i< j}\frac{1}{|\textbf{r}_{i}-\textbf{r}_{j}|}$ and the time-dependent external perturbation is given by
\begin{equation}
\hat{H}_{ext}(t) = \hat{V}_{ext}(t) + \hat{J}_{ext}(t).  \label{App3}
\end{equation}
Here, the time-dependent external potential and current are 
\begin{equation}
\hat{V}_{ext}(t) = \sum\limits^{N}_{i=1} v(\textbf{r}_{i},t ), \quad
\hat{J}_{ext}(t) = \sum_{\alpha} \frac{j_{\alpha}(t)}{\omega_{\alpha}}\hat{q}_{\alpha}. \label{App4}
\end{equation}
We now introduce the interaction picture, where a general state vector of the interacting electron-photon system is given by
\begin{equation*}
\Psi_{I}(t) = \hat{U}_{0}^{\dagger}(t)\Psi(t) = e^{i\hat{H}_{0}t/\hbar}\Psi(t),
\end{equation*}
with $\Psi(t)$ as the state vector in the Schr\"{o}dinger picture. Accordingly, an arbitrary operator $\hat{O}$ can be transformed from the Schr\"odinger to the interaction picture by
\begin{align}
\hat{O}_{I}(t)=  \hat{U}_{0}^{\dagger}(t) \hat{O} \hat{U}_{0}(t) .
\end{align}
In the interaction picture, the evolution of the interacting electron-photon system from an initial state $\Psi_0$ is described by the following time-dependent Schr\"odinger equation
\begin{equation}
i\hbar\frac{\partial}{\partial t}\Psi_{I}(t) = \hat{H}_{ext,I}(t)\Psi_{I}(t). \label{App5}
\end{equation}
Through an integration, the above equation can be formally solved to yield 
\begin{equation}
\Psi_{I}(t) = \Psi_{0} - \frac{i}{\hbar}\int_{t_{0}}^{t} dt'\hat{H}_{ext,I}(t')\Psi_{I}(t'). \label{App6}
\end{equation}
If we only keep the first order, we obtain in the Schr\"odinger picture a closed solution
\begin{equation}
\Psi(t) \simeq \hat{U}_{0}(t) \Psi_{0} - \frac{i}{\hbar}\hat{U}_{0}(t) \int_{t_{0}}^{t} dt' \hat{H}_{ext,I}(t')\hat{U}_{0}^{\dagger}(t)\Psi_{0}. \label{App7}
\end{equation}
In our case however, we are not interested in the time evolution of the wave function, but rather in the response of an observable $\hat{O}$ to (small) external perturbations. The change in the expectation value of an arbitrary observable $\hat{O}$ due to the external perturbation $\hat{H}_{ext}(t)$ is given by
\begin{equation}
\delta\langle \hat{O}(t)\rangle = \langle \Psi({t})|\hat{O}|\Psi({t})\rangle - \langle \Psi_{0}|\hat{O}|\Psi_{0}\rangle, \label{App8}
\end{equation}
In linear-response theory, we now assume that the external perturbation in Eq.~(\ref{App4}) is sufficiently small such that Eq.~(\ref{App7}) is a good approximation to Eq.~(\ref{App6}) and that $\Psi_0$ equals the ground-state of Eq.~(\ref{App2}). Thus, if we evaluate Eq.~(\ref{App8}) with Eq.~(\ref{App7}), we obtain
\begin{equation}
\delta\langle \hat{O}(t)\rangle = -\frac{i}{\hbar}\int_{t_{0}}^{t} dt' \langle\Psi_{0}|\left[\hat{O}_{I}(t),\hat{H}_{ext,I}(t')\right]|\Psi_{0}\rangle , \label{App9}
\end{equation}
As a side remark, beyond linear-response solutions can be obtained by higher-order terms in Eq.~(\ref{App6}).
Staying within linear response, we can now use Eq.~(\ref{App9}) to obtain the response of the electron density to $\hat{H}_{ext}(t)$ that is given by
\begin{eqnarray}
\delta n(\textbf{r}t) &=& -\frac{i}{\hbar}\int_{t_{0}}^{t} dt' \int d\textbf{r}' \langle\Psi_{0}|\left[\hat{n}_{I}(\textbf{r}t),\hat{V}_{ext,I}(\textbf{r}'t')\right]|\Psi_{0}\rangle  -\frac{i}{\hbar} \sum_{\alpha}\int_{t_{0}}^{t} dt' \langle\Psi_{0}|\left[\hat{n}_{I}(\textbf{r}t),\hat{J}_{ext,I}(t')\right]|\Psi_{0}\rangle . \nonumber
\end{eqnarray}
Simplifying further, the density response reads
\begin{eqnarray}
\delta n(\textbf{r}t) &= -\frac{i}{\hbar}\int_{t_{0}}^{t} dt' \int d\textbf{r}' \langle\Psi_{0}|\left[\hat{n}_{I}(\textbf{r}t),\hat{n}_{I}(\textbf{r}'t')\right]|\Psi_{0}\rangle \delta v(\textbf{r}'t')  \nonumber\\
-&\frac{i}{\hbar} \sum_{\alpha}\int_{t_{0}}^{t} dt' \frac{1}{{ \omega_{\alpha}}} \langle\Psi_{0}|\left[\hat{n}_{I}(\textbf{r}t),\hat{q}_{\alpha,I}(t')\right]|\Psi_{0}\rangle \delta j_{\alpha}(t'). \nonumber
\end{eqnarray}
The response of the density to the external perturbation $(v(\textbf{r}t),j_{\alpha}(t))$ is 
\begin{eqnarray}
\delta n(\textbf{r}t) &=& \int_{t_{0}}^{\infty} dt' \int d\textbf{r}' \chi_{n}^{n}(\textbf{r}t,\textbf{r}'t') \delta v(\textbf{r}'t')  + \sum_{\alpha}\int_{t_{0}}^{\infty} dt' \chi_{q_{\alpha}}^{n}(\textbf{r}t,t') \delta j_{\alpha}(t') , \nonumber
\end{eqnarray}
where the response functions are
\begin{align}
& \chi_{n}^{n}(\textbf{r}t,\textbf{r}'t') = -\frac{i}{\hbar}\Theta(t-t') \langle\Psi_{0}|\left[\hat{n}_{I}(\textbf{r}t),\hat{n}_{I}(\textbf{r}'t')\right]|\Psi_{0}\rangle , \label{App10}\\
& \chi_{q_{\alpha}}^{n}(\textbf{r}t,t') = -\frac{i}{\hbar}\Theta(t-t')\frac{1}{{ \omega_{\alpha}}} \langle\Psi_{0}|\left[\hat{n}_{I}(\textbf{r}t),\hat{q}_{\alpha,I}(t')\right]|\Psi_{0}\rangle  .\label{App11}
\end{align}

Similarly, the response of the photon coordinate $q_{\alpha}(t)$ to $\hat{H}_{ext}(t)$ is
\begin{eqnarray}
\delta q_{\alpha}(t) &=& -\frac{i}{\hbar}\int_{t_{0}}^{t} dt' \langle\Psi_{0}|\left[\hat{q}_{\alpha,I}(t),\hat{V}_{ext,I}(t')\right]|\Psi_{0}\rangle  -\frac{i}{\hbar}\int_{t_{0}}^{t} dt' \langle\Psi_{0}|\left[\hat{q}_{\alpha,I}(t),\hat{J}_{ext,I}(t')\right]|\Psi_{0}\rangle. \nonumber
\end{eqnarray}
Following similar steps as above, the response of the photon coordinate to the external perturbation $(v(\textbf{r}t),j_{\alpha}(t))$ is
\begin{eqnarray}
\delta q_{\alpha}(t) &=& \int_{t_{0}}^{\infty} dt' \int d\textbf{r}' \chi_{n}^{q_{\alpha}}(t,\textbf{r}'t') \delta v(\textbf{r}'t')  + \sum_{\alpha'}\int_{t_{0}}^{\infty} dt' \chi_{q_{\alpha'}}^{q_{\alpha}}(t,t') \delta j_{\alpha'}(t') , \nonumber
\end{eqnarray}
where the response functions are
\begin{align}
& \chi_{n}^{q_{\alpha}}(t,\textbf{r}'t') = -\frac{i}{\hbar}\Theta(t-t') \langle\Psi_{0}|\left[q_{\alpha,I}(t),\hat{n}_{I}(\textbf{r}'t')\right]|\Psi_{0}\rangle ,  \label{App12} \\
& \chi_{q_{\alpha'}}^{q_{\alpha}}(t,t') = -\frac{i}{\hbar}\Theta(t-t') \frac{1}{{ \omega_{\alpha'}}} \langle\Psi_{0}|\left[q_{\alpha,I}(t),\hat{q}_{\alpha',I}(t')\right]|\Psi_{0}\rangle  . \label{App13}
\end{align}
Alternatively, the response functions of Eqs.(\ref{App10})-(\ref{App13}) can be obtained using the functional dependence of the observables on the external pair $(v(\textbf{r}t),j_{\alpha}(t))$. The wave function of Eq.~(\ref{eqn:TDSE}) has a functional dependence $\Psi([v,j_{\alpha}];t)$ via the Hamiltonian Eq.~(\ref{App1}), i.e., $\hat{H}(t) = \hat{H}([v,j_{\alpha}];t)$. Therefore, through the expectation of electron density and photon displacement coordinate, both have a functional dependence on the external pair as $n([v,j_{\alpha}];\textbf{r}t)$ and $q_{\alpha}([v,j_{\alpha}];t)$, respectively.

Considering the ground-state problem with external potential and current of  $(v_{0}(\textbf{r}),j_{\alpha,0})$, we can perform a functional Taylor expansion of the density $n(\textbf{r}t)$ and photon coordinate $q_{\alpha}(t)$ to first-order as

\begin{eqnarray}
n([v,j_{\alpha}];\textbf{r}t) &=& n([v_{0},j_{\alpha,0}];\textbf{r}) + \iint d\textbf{r}'dt' \frac{\delta n([v_{0},j_{\alpha,0}];\textbf{r}t)}{\delta v(\textbf{r}'t')} \delta v(\textbf{r}'t')  + \sum_{\alpha} \int dt' \frac{\delta n([v_{0},j_{\alpha,0}];\textbf{r}t)}{\delta j_{\alpha}(t')} \delta j_{\alpha}(t') , \nonumber\\
q_{\alpha}([v,j_{\alpha}];t) &=& q_{\alpha}([v_{0},j_{\alpha,0}]) + \iint d\textbf{r}'dt' \frac{\delta q_{\alpha}([v_{0},j_{\alpha,0}];t)}{\delta v(\textbf{r}'t')} \delta v(\textbf{r}'t')  + \sum_{\alpha'} \int dt' \frac{\delta q_{\alpha}([v_{0},j_{\alpha,0}];t)}{\delta j_{\alpha'}(t')} \delta j_{\alpha'}(t') . \nonumber
\end{eqnarray}
This reduces to the response of the electron density and photon coordinate given as
\begin{eqnarray}
\delta n([v,j_{\alpha}];\textbf{r}t) &=&  \iint d\textbf{r}'dt' \chi_{v}^{n}(\textbf{r}t,\textbf{r}'t') \delta v(\textbf{r}'t')   + \sum_{\alpha} \int dt' \chi_{j_{\alpha}}^{n}(\textbf{r}t,t') \delta j_{\alpha}(t') , \nonumber
\end{eqnarray}
and
\begin{eqnarray}
\delta q_{\alpha}([v,j_{\alpha}];t) &=&  \iint d\textbf{r}'dt' \chi_{v}^{q_{\alpha}}(t,\textbf{r}'t') \delta v(\textbf{r}'t')  + \sum_{\alpha'} \int dt' \chi_{j_{\alpha'}}^{q_{\alpha}}(t,t') \delta j_{\alpha'}(t') , \nonumber
\end{eqnarray}
where we define the response functions of the above relation as 
\begin{eqnarray}
\chi_{v}^{n}(\textbf{r}t,\textbf{r}'t') &=& \left.\frac{\delta n([v,j_{\alpha}];\textbf{r}t)}{\delta v(\textbf{r}'t')} \right\vert_{v_{0}(\textbf{r}),j_{\alpha,0}} , \label{App13i}\\
\chi_{j_{\alpha}}^{n}(\textbf{r}t,t') &=&  \left.\frac{\delta n([v,j_{\alpha}];\textbf{r}t)}{\delta j_{\alpha}(t')} \right\vert_{v_{0}(\textbf{r}),j_{\alpha,0}}  , \label{App13ii}\\
\chi_{v}^{q_{\alpha}}(t,\textbf{r}'t') &=& \left.\frac{\delta q_{\alpha}([v,j_{\alpha}];t)}{\delta v(\textbf{r}'t')} \right\vert_{v_{0}(\textbf{r}),j_{\alpha,0}} , \label{App13iii}\\
\chi_{j_{\alpha'}}^{q_{\alpha}}(t,t') &=& \left.\frac{\delta q_{\alpha}([v,j_{\alpha}];t)}{\delta j_{\alpha'}(t')} \right\vert_{v_{0}(\textbf{r}),j_{\alpha,0}} . \label{App13iv}
\end{eqnarray}
These response functions defined in Eqs.(\ref{App10})-(\ref{App13}) and Eqs.(\ref{App13i})-(\ref{App13iv}) are equivalent.

The response functions expressed in the so-called Lehmann representation are given by
\begin{eqnarray}
\chi_{n}^{n}(\textbf{r},\textbf{r}',\omega)  &=& \frac{1}{\hbar}\lim\limits_{\eta\rightarrow 0^{+}}\sum_{k}\left[\frac{f_{k}(\textbf{r})f_{k}^{*}(\textbf{r}')}{\omega - \Omega_{k} + i\eta} - \frac{f_{k}(\textbf{r}')f_{k}^{*}(\textbf{r})}{\omega + \Omega_{k} + i\eta} \right], \nonumber\\
\chi_{q_{\alpha}}^{n}(\textbf{r},\omega)  &=& \frac{1}{\hbar}\lim\limits_{\eta\rightarrow 0^{+}}\sum_{k}\frac{1}{{ \omega_{\alpha}}}\left[\frac{f_{k}(\textbf{r})g^{*}_{\alpha,k}}{\omega - \Omega_{k} + i\eta} - \frac{g_{\alpha,k}f_{k}^{*}(\textbf{r})}{\omega + \Omega_{k} + i\eta} \right], \nonumber\\
\chi_{n}^{q_{\alpha}}(\textbf{r}',\omega)  &=& \frac{1}{\hbar}\lim\limits_{\eta\rightarrow 0^{+}}\sum_{k}\left[\frac{g_{\alpha,k}f_{k}^{*}(\textbf{r}')}{\omega - \Omega_{k} + i\eta} - \frac{f_{k}(\textbf{r}')g_{\alpha,k}^{*}}{\omega + \Omega_{k} + i\eta} \right], \nonumber\\
\chi_{q_{\alpha'}}^{q_{\alpha}}(\omega)  &=& \frac{1}{\hbar}\lim\limits_{\eta\rightarrow 0^{+}}\sum_{k}\frac{1}{{ \omega_{\alpha'}}}\left[\frac{g_{\alpha,k}g^{*}_{\alpha',k}}{\omega - \Omega_{k} + i\eta} - \frac{g_{\alpha',k}g^{*}_{\alpha,k}}{\omega + \Omega_{k} + i\eta} \right], \nonumber
\end{eqnarray}
where $f_{k}(\textbf{r})=  \langle\Psi_{0}|\hat{n}(\textbf{r})|\Psi_{k}\rangle$ and $g_{\alpha,k} = \langle\Psi_{0}|\hat{q}_{\alpha}|\Psi_{k}\rangle$ are the transition matrix elements and $|\Psi_{0}\rangle $ is the correlated electron-photon ground state wave function. The excitation energies $ \Omega_{k} = (E_{k}-E_{0})/\hbar$ of the finite interacting system are the poles of the response functions of the unperturbed system. As a side remark, if we can choose the wave functions $\Psi_0$ and $\Psi_k$ to be real, we find $g_{\alpha,k}=g^*_{\alpha,k}$, and $f_{k}(\textbf{r})=f^*_{k}(\textbf{r})$, thus $\chi_{n}^{q_{\alpha}}(\textbf{r},\omega) = { \omega_{\alpha}} \chi_{q_{\alpha}}^{n}(\textbf{r},\omega)$.

\section{Linear-response within QEDFT }
\label{app:linresp1}

In this section, we present linear-response in QEDFT by employing the maps between interacting and non-interacting system, we express the interacting response functions in terms of two non-interacting response functions and exchange correlation kernels. The responses due to $(v(\textbf{r}t),j_{\alpha}(t))$ are evaluated at the ground-state $(v_{0}(\textbf{r}),j_{\alpha,0})$ and will not be written explicitly.

The non-interacting subsystems moving in an effective potential and current $(v_{s}(\textbf{r}t),j_{\alpha}^{s}(t))$ can be written as a time-dependent problem of the Schr\"{o}dinger 
\begin{equation}
i\hbar\frac{\partial}{\partial t}\Phi(t) = \hat{H}_{\textrm{KS}}(t) \Phi(t). \label{App14}
\end{equation}
Here, $\Phi(t)$ is the wave function of the auxiliary non-interacting system and the non-interacting effective Hamiltonian $\hat{H}_{\textrm{KS}}(t) = \hat{H}_{\textrm{KS}}^{(0)} + \hat{H}_{\textrm{KS}}^{(ext)}(t)$ that is meant to reproduce the exact density and displacement field, is given explicitly as
\begin{eqnarray}
\hat{H}_{\textrm{KS}}^{(0)} &=& \hat{T} + \hat{H}_{pt} + \left(v_{0}(\textbf{r}) + v_{Mxc}^{(0)}([n, q_{\alpha}];\textbf{r}) \right)+ \sum_{\alpha}\frac{1}{\omega_{\alpha}} \left(j_{\alpha,0} + j_{\alpha,Mxc}^{(0)}[n, q_{\alpha}]\right)\hat{q}_{\alpha}, \nonumber
\end{eqnarray}
and 
\begin{eqnarray}
\hat{H}_{\textrm{KS}}^{(ext)}(t) &=& \left(v(\textbf{r}t) + v_{Mxc}([n,q_{\alpha}];\textbf{r}t) \right)  + \sum_{\alpha}\frac{1}{\omega_{\alpha}} \left(j_{\alpha}(t) + j_{\alpha,Mxc}\left([n,q_{\alpha}];t\right)\right)\hat{q}_{\alpha}. \nonumber
\end{eqnarray} 
Here $\hat{H}_{pt} = \frac{1}{2}\sum_{\alpha=1}^{M}\left[\hat{p}_{\alpha}^{2} + \omega_{\alpha}^{2}\hat{q}_{\alpha}^{2} \right]$ is the oscillator for the photon mode and the mean-field xc potential and current are defined as
\begin{eqnarray}
v_{Mxc}([n, q_{\alpha}];\textbf{r}t) &:=& v_{s}([n];\textbf{r}t) - v([n,q_{\alpha}];\textbf{r}t) , \label{App15} \\
j_{\alpha,Mxc}([n, q_{\alpha}];t) &:=& j_{\alpha}^{s}([q_{\alpha}];t) - j_{\alpha}([n,q_{\alpha}];t)  . \label{App16}
\end{eqnarray}
In the above definitions of $v_{Mxc}([n, q_{\alpha}];\textbf{r}t)$ and $j_{\alpha,Mxc}([n, q_{\alpha}];t)$, the initial state dependence of the interacting $\Psi_{0}$ and non-interacting $\Phi_{0}$ system has been dropped. For completeness, the definition of $j_{\alpha,Mxc}([n, q_{\alpha}];t)$ accounts for a functional dependence on $q_{\alpha}$ but this term can be calculated explicitly since it has no xc part as seen in Eq.~(\ref{Max0}). The simplified form of $j_{\alpha,Mxc}$ is shown in Eq.~(\ref{m_current}).

Through similar steps as in Eqs.(\ref{App5})-(\ref{App7}), in first-order the solution of the Schr\"{o}dinger-Kohn-Sham equation reads
\begin{equation}
\Phi(t) \simeq \hat{U}_{\textrm{KS},0}(t) \Phi_{0} - \frac{i}{\hbar}\hat{U}_{\textrm{KS},0}(t) \int_{t_{0}}^{t} dt' \hat{H}_{\textrm{KS},I}^{(ext)}(t')\hat{U}_{\textrm{KS},0}^{\dagger}(t)\Phi_{0}. \label{App17}
\end{equation}
where $\hat{U}_{\textrm{KS},0} =  e^{-i\hat{H}_{\textrm{KS}}^{(0)}t/\hbar}$. Next, the bijective mapping between the interacting and non-interacting system that yields the same density and photon coordinate is given as
\begin{equation}
(v(\textbf{r}t),j_{\alpha}(t)) \xleftrightarrow[\Psi_{0}]{1:1} (n(\textbf{r}t), q_{\alpha}(t)) \xleftrightarrow[\Phi_{0}]{1:1}  (v_{s}(\textbf{r}t),j_{\alpha}^{s}(t)), \label{App18}
\end{equation}
which can be inverted as $(v_{s}([v,j_{\alpha}];\textbf{r}'t'),j_{\alpha}^{s}([v,j_{\alpha}];t'))$. The response of the electronic subsystem due to the perturbations with the external pair $(v(\textbf{r}t),j_{\alpha}(t))$ is
\begin{eqnarray}
\delta n(\textbf{r}t) &=& -\frac{i}{\hbar}\iint d\tau d\textbf{x} \iint dt'  d\textbf{r}'\langle\Phi_{0}|\left[\hat{n}_{I}(\textbf{r}t),\hat{n}_{I}(\textbf{x}\tau)\right]|\Phi_{0}\rangle   \frac{ \delta v_{s}([v,j_{\alpha}];\textbf{x}\tau)}{\delta v(\textbf{r}'t')}\delta v(\textbf{r}'t') \nonumber \\
&& -\frac{i}{\hbar}\iint d\tau  d\textbf{x} {\sum_\alpha}  \int dt'\langle\Phi_{0}|\left[\hat{n}_{I}(\textbf{r}t),\hat{n}_{I}(\textbf{x}\tau)\right]|\Phi_{0}\rangle  \frac{ \delta v_{s}([v,j_{\alpha}];\textbf{x}\tau)}{\delta j_{\alpha}(t')}\delta j_{\alpha}(t') .  \nonumber
\end{eqnarray}
Where $\langle\Phi_{0}|\left[\hat{n}_{I}(\textbf{r}t),\hat{q}_{\alpha,I}(\tau)\right]|\Phi_{0}\rangle=0$ since both, electronic and photonic subsystems, are independent in the non-interacting system. From Eq.~(\ref{App18}), we have $(v_{s}([n];\textbf{r}t),j_{\alpha}^{s}([q_{\alpha}];t))$ such that the above equation becomes
\begin{align}
\delta n(\textbf{r}t) &= \iint d\tau d\textbf{x} \iint dt'd\textbf{r}'\iint  d\tau'd\textbf{y}\chi_{n,s}^{n}(\textbf{r}t,\textbf{x}\tau)   \frac{ \delta v_{s}([n];\textbf{x}\tau)}{\delta n(\textbf{y}\tau')} \frac{ \delta n([v,j_{\alpha}];\textbf{y}\tau')}{\delta v(\textbf{r}'t')}\delta v(\textbf{r}'t')  \nonumber \\
&+ \iint d\tau d\textbf{x}  \sum_{\alpha}\int dt'  \iint  d\tau'd\textbf{y}\chi_{n,s}^{n}(\textbf{r}t,\textbf{x}\tau)\frac{ \delta v_{s}([n];\textbf{x}\tau)}{\delta n(\textbf{y}\tau')} \frac{ \delta n([v,j_{\alpha}];\textbf{y}\tau')}{\delta j_{\alpha}(t')} \delta j_{\alpha}(t'), \label{App19}
\end{align}
where $ \chi_{n,s}^{n}(\textbf{r}t,\textbf{x}\tau) = (-i/\hbar)\Theta(t-\tau) \langle\Phi_{0}|\left[\hat{n}_{I}(\textbf{r}t),\hat{n}_{I}(\textbf{x}\tau)\right]|\Phi_{0}\rangle$ is the non-interacting density-density response function. For clarity, the above density response is $\delta n(\textbf{r}t) = \delta n_{v}(\textbf{r}t)  + \delta n_j(\textbf{r}t) $, where $(\delta n_{v}(\textbf{r}t),\delta n_j(\textbf{r}t))$ is the density response to the external pair $(v(\textbf{r}t),j_{\alpha}(t))$, respectively.

Using Eqs.(\ref{App15}) and (\ref{App16}), we define the mean-field xc kernels as:
\begin{align}
& f_{Mxc}^{n}([n,q_{\alpha}];\textbf{r}t,\textbf{r}'t') = \frac{\delta v_{s}([n];\textbf{r}t)}{\delta n(\textbf{r}'t')}  - \frac{\delta v([n,q_{\alpha}];\textbf{r}t)}{\delta n(\textbf{r}'t')} , \label{App20} \\
& f_{Mxc}^{q_{\alpha}}([n,q_{\alpha}];\textbf{r}t,t') =  - \frac{\delta v([n,q_{\alpha}];\textbf{r}t)}{\delta q_{\alpha}(t')} , \label{App21} \\
& g_{Mxc}^{n}([n,q_{\alpha}];t,\textbf{r}'t') =  - \frac{\delta j_{\alpha}([n,q_{\alpha}];t)}{\delta n(\textbf{r}'t')}  , \label{App22} \\
& g_{Mxc}^{q_{\alpha'}}([n,q_{\alpha}];t,t') = \frac{\delta j_{\alpha}^{s}([q_{\alpha}];t)}{\delta q_{\alpha'}(t')} -\frac{\delta j_{\alpha}([n,q_{\alpha}];t)}{\delta q_{\alpha'}(t')} , \label{App23} 
\end{align}
where $\frac{\delta v_{s}([n];\textbf{r}t)}{\delta q_{\alpha}(t')} = 0 = \frac{\delta j_{\alpha}^{s}([q_{\alpha}];t)}{\delta n(\textbf{r}'t')}$. These kernels are the respective inverse of the interacting and non-interacting response functions.

From Eq.~(\ref{App19}), density response to $\delta v(\textbf{r}t)$ can be written in terms of the density-density response function given by
\begin{eqnarray}
\chi_{n}^{n}(\textbf{r}t,\textbf{r}'t') &=& \iint d\tau d\textbf{x} \chi_{n,s}^{n}(\textbf{r}t,\textbf{x}\tau)  \iint d\tau'  d\textbf{y}f_{Mxc}^{n}([n,q_{\alpha}];\textbf{x}\tau,\textbf{y}\tau') \frac{ \delta n([v,j_{\alpha}];\textbf{y}\tau')}{\delta v(\textbf{r}t')} \nonumber  \\
&& + \iint d\tau d\textbf{x} \chi_{n,s}^{n}(\textbf{r}t,\textbf{x}\tau) \iint d\tau' d\textbf{y} \frac{\delta v([n,q_{\alpha}];\textbf{x}\tau)}{\delta n(\textbf{y}\tau')}  \frac{ \delta n([v,j_{\alpha}];\textbf{y}\tau')}{\delta v(\textbf{r}'t')} . \nonumber  
\end{eqnarray}
Making the following substitution in the above equation
\begin{align}
&\iint d\textbf{y}d\tau' \frac{\delta v([n,q_{\alpha}];\textbf{x}\tau)}{\delta n(\textbf{y}\tau')}\frac{\delta n([v,j_{\alpha}];\textbf{y}\tau')}{\delta v(\textbf{r}'t')} = \delta(\textbf{x}-\textbf{r}')\delta(\tau-t')- \sum_{\alpha}\int d\tau' \frac{\delta v([n,q_{\alpha}];\textbf{x}\tau)}{\delta q_{\alpha}(\tau')}\frac{\delta q_{\alpha}([v,j_{\alpha}];\tau')}{\delta v(\textbf{r}'t')} , \nonumber
\end{align}
where $\delta v([n,q_{\alpha}];\textbf{x}\tau)/\delta v(\textbf{r}'t') = \delta(\textbf{x}-\textbf{r}')\delta(\tau-t')$, we obtain the relation
\begin{eqnarray}
\chi_{n}^{n}(\textbf{r}t,\textbf{r}'t') &=& \chi_{n,s}^{n}(\textbf{r}t,\textbf{r}'t') + \iiiint d\tau d\textbf{x} d\tau' d\textbf{y} \chi_{n,s}^{n}(\textbf{r}t,\textbf{x}\tau)   f_{Mxc}^{n}(\textbf{x}\tau,\textbf{y}\tau')  \chi_{n}^{n}(\textbf{y}\tau',\textbf{r}'t')  \nonumber  \\
&& + \sum_{\alpha}\iiint d\tau d\textbf{x}d\tau' \chi_{n,s}^{n}(\textbf{r}t,\textbf{x}\tau)      f_{Mxc}^{q_{\alpha}}(\textbf{x}\tau,\tau')  \chi_{n}^{q_{\alpha}}(\tau',\textbf{r}'t').  \label{App24}   
\end{eqnarray}
Next, the density response to $\delta j_{\alpha}(t)$ in Eq.~(\ref{App19}) is expressed in terms of the response function as
\begin{eqnarray}
\chi_{q_{\alpha}}^{n}(\textbf{r}t,t') &=& \iint d\tau d\textbf{x} \chi_{n,s}^{n}(\textbf{r}t,\textbf{x}\tau)  \iint  d\tau'  d\textbf{y}f_{Mxc}^{n}(\textbf{x}\tau,\textbf{y}\tau')  \frac{ \delta n([v,j_{\alpha}];\textbf{y}\tau')}{\delta j_{\alpha}(t')}  \nonumber  \\
&& + \iint d\tau d\textbf{x} \chi_{n,s}^{n}(\textbf{r}t,\textbf{x}\tau)  \iint  d\tau' d\textbf{y}\frac{\delta v([n,q_{\alpha}];\textbf{x}\tau)}{\delta n(\textbf{y}\tau')}  \frac{ \delta n([v,j_{\alpha}];\textbf{y}\tau')}{\delta j_{\alpha}(t')} .  \nonumber  
\end{eqnarray}
Using the relation (obtained from $\delta v([n,q_{\alpha}];\textbf{x}\tau)/ \delta j_{\alpha}(t')$)  
\begin{align}
& \iint d\textbf{y}d\tau' \frac{\delta v([n,q_{\alpha}];\textbf{x}\tau)}{\delta n(\textbf{y}\tau')}\frac{\delta n([v,j_{\alpha}];\textbf{y}\tau')}{\delta j_{\alpha}(t')}  = - \sum_{\alpha'} \int d\tau'   \frac{\delta v([n,q_{\alpha}];\textbf{x}\tau)}{\delta q_{\alpha'}(\tau')}\frac{\delta q_{\alpha'}([v,j_{\alpha}];\tau')}{\delta j_{\alpha}(t')} , \nonumber
\end{align}
the response function is given as
\begin{eqnarray}
\chi_{q_{\alpha}}^{n}(\textbf{r}t,t') &=& \iiiint d\tau d\textbf{x} d\tau'  d\textbf{y} \chi_{n,s}^{n}(\textbf{r}t,\textbf{x}\tau)  f_{Mxc}^{n}(\textbf{x}\tau,\textbf{y}\tau')  \chi_{q_{\alpha}}^{n}(\textbf{y}\tau',t')  \nonumber  \\
&& + \sum_{\alpha'}\iiint d\tau d\textbf{x} d\tau' \chi_{n,s}^{n}(\textbf{r}t,\textbf{x}\tau)   f_{Mxc}^{q_{\alpha'}}(\textbf{x}\tau,\tau') \chi_{q_{\alpha}}^{q_{\alpha'}}(\tau',t')  . \label{App25}
\end{eqnarray}

Similarly, the response to the photonic subsystem to linear perturbations from the external pair $(v(\textbf{r}t),j_{\alpha}(t))$ is
\begin{eqnarray}
\delta q_{\alpha}(t) &=& -\frac{i}{\hbar} \sum_{\beta}\int_{t_{0}}^{t} d\tau \frac{1}{{ \omega_{\beta}}} \langle\Phi_{0}|\left[q_{\alpha,I}(t),q_{\beta,I}(\tau)\right]|\Phi_{0}\rangle  \iint dt' d\textbf{r}' \frac{ \delta j_{\beta}^{s}([v,j_{\alpha}];\tau)}{\delta v(\textbf{r}'t')}\delta v(\textbf{r}'t') \nonumber \\
&& -\frac{i}{\hbar}\sum_{\beta}\int_{t_{0}}^{t} d\tau \frac{1}{{ \omega_{\beta}}}  \langle\Phi_{0}|\left[q_{\alpha,I}(t),q_{\beta,I}(\tau)\right]|\Phi_{0}\rangle  \sum_{\alpha'}  \int dt' \frac{ \delta j_{\beta}^{s}([v,j_{\alpha}];\tau)}{\delta j_{\alpha'}(t')} \delta j_{\alpha'}(t') ,  \nonumber
\end{eqnarray}
where $\langle\Phi_{0}|\left[\hat{q}_{\alpha,I}(t),\hat{n}_{I}(\textbf{x}\tau)\right]|\Phi_{0}\rangle=0$ in the non-interacting system. By defining the non-interacting photon-photon response function as $\chi_{q_{\beta,s}}^{q_{\alpha}}(t,\tau) = (-i/\hbar)\Theta(t-\tau)(1/{ \omega_{\beta}})\langle\Phi_{0}|\left[q_{\alpha,I}(t),q_{\beta,I}(\tau)\right]|\Phi_{0}\rangle$ and using  Eq.~(\ref{App18}), where we have $(v_{s}([n];\textbf{r}t),j_{\alpha}^{s}([q_{\alpha}];t))$, the response can be written as
\begin{eqnarray}
\delta q_{\alpha}(t) &=&  \sum_{\beta}\int d\tau \chi_{q_{\beta,s}}^{q_{\alpha}}(t,\tau) \sum_{\beta'} \iiint dt'd\textbf{r}'d\tau'  \frac{ \delta j_{\beta}^{s}([q_{\alpha}];\tau)}{\delta q_{\beta'}(\tau')} \frac{ \delta q_{\beta'}([v,j_{\alpha}];\tau')}{\delta v(\textbf{r}'t')} \delta v(\textbf{r}'t') \nonumber \\
&& +  \sum_{\beta}\int  d\tau  \chi_{q_{\beta,s}}^{q_{\alpha}}(t,\tau) \sum_{\alpha',\beta'}  \iint dt' d\tau'  \frac{ \delta j_{\beta}^{s}([q_{\alpha}];\tau)}{\delta q_{\beta'}(\tau')} \frac{ \delta q_{\beta'}([v,j_{\alpha}];\tau')}{\delta j_{\alpha'}(t')}\delta j_{\alpha'}(t') . \label{App26}
\end{eqnarray} 
The above response of the displacement field is $\delta q_{\alpha}(t) = \delta q_{\alpha,v}(t)  + \delta q_{{\alpha},j}(t) $, where $(\delta q_{\alpha,v}(t),\delta q_{{\alpha},j}(t) )$ is the response to the external pair $(v(\textbf{r}t),j_{\alpha}(t))$, respectively.

From Eq.~(\ref{App26}), the field response to $\delta v(\textbf{r}t)$ can be written in terms of the photon-density response function as
\begin{align*}
\chi_{n}^{q_{\alpha}}(t,\textbf{r}'t') =&  \sum_{\beta}\int d\tau \chi_{q_{\beta,s}}^{q_{\alpha}}(t,\tau) \sum_{\beta'} \int d\tau'  g_{Mxc}^{q_{\beta'}}(\tau,\tau') \chi_{n}^{q_{\beta'}} (\tau',\textbf{r}'t') \nonumber \\
&  +  \sum_{\beta}\int d\tau \chi_{q_{\beta,s}}^{q_{\alpha}}(t,\tau)\sum_{\beta'}  \int d\tau' \frac{\delta j_{\beta}([n,q_{\alpha}];\tau)}{\delta q_{\beta'}(\tau')} \frac{ \delta q_{\beta'}([v,j_{\alpha}];\tau')}{\delta v(\textbf{r}'t')} .  \nonumber 
\end{align*}
Using the relation (obtained from $\delta j_{\beta}([n,q_{\alpha}];\tau)/ \delta v(\textbf{r}'t')$)
\begin{align}
& \sum_{\beta'} \int d\tau'\frac{\delta j_{\beta}([n,q_{\alpha}];\tau)}{\delta q_{\beta'}(\tau')}\frac{\delta q_{\beta'}([v,j_{\alpha}];\tau')}{\delta v(\textbf{r}t')}   =  -  \iint d\tau' d\textbf{y}   \frac{\delta j_{\beta}([n,q_{\alpha}];\tau)}{\delta n(\textbf{y}\tau')}\frac{\delta n([v,j_{\alpha}];\textbf{y}\tau')}{\delta v(\textbf{r}'t')}  , \nonumber
\end{align}
the response function is given as
\begin{align}
\chi_{n}^{q_{\alpha}}(t,\textbf{r}'t') &=  \sum_{\beta}\int d\tau \iint d\tau' d\textbf{y}\chi_{q_{\beta,s}}^{q_{\alpha}}(t,\tau)  g_{Mxc}^{n_{\beta}}(\tau,\textbf{y}\tau') \chi_{n}^{n}(\textbf{y}\tau',\textbf{r}'t')  , \label{App27}
\end{align}
where $g_{Mxc}^{n_{\beta}}=g_{M}^{n_{\beta}}$ and $g_{Mxc}^{q_{\alpha}}=0$ as determined from the equation of motion for the displacement field. Also, from Eq.~(\ref{App26}), field response to $\delta j_{\alpha}$ can be written in terms of the photon-photon response function as 
\begin{eqnarray}
\chi_{q_{\alpha'}}^{q_{\alpha}}(t,t') &=&  \sum_{\beta} \int d\tau  \chi_{q_{\beta,s}}^{q_{\alpha}}(t,\tau)  \sum_{\beta'}\int d\tau' g_{Mxc}^{q_{\beta'}}(\tau,\tau')\chi_{q_{\alpha'}}^{q_{\beta'}} (\tau',t') \nonumber\\
&&+  \sum_{\beta}\int d\tau \chi_{q_{\beta,s}}^{q_{\alpha}}(t,\tau) \sum_{\beta'}\int d\tau' \frac{\delta j_{\beta}([n,q_{\alpha}];\tau)}{\delta q_{\beta'}(\tau')}\frac{\delta q_{\beta'}([n,q_{\alpha}];\tau')}{\delta j_{\alpha'}(t')} .  \nonumber
\end{eqnarray}
Making the following substitution (where  $\delta j_{\beta}([n,q_{\alpha}];\tau)/\delta j_{\alpha'}(t') = \delta(\tau-t')\delta_{\beta,\alpha'}$) in the above equation
\begin{align}
&\sum_{\beta'}\int d\tau' \frac{\delta j_{\beta}([n,q_{\alpha}];\tau)}{\delta q_{\beta'}(\tau')}\frac{\delta q_{\beta'}([v,j_{\alpha}];\tau')}{\delta j_{\alpha'}(t')}  = \delta(\tau-t')\delta_{\beta,\alpha'} - \iint d\tau' d\textbf{x}\frac{\delta j_{\beta}([n,q_{\alpha}];\tau)}{\delta n(\textbf{x}\tau')}\frac{\delta n([v,j_{\alpha}];\textbf{x}\tau')}{\delta j_{\alpha'}(t')}, \nonumber
\end{align} 
yields the photon-photon response function
\begin{align}
\chi_{q_{\alpha'}}^{q_{\alpha}}(t,t') &= \chi_{q_{\alpha',s}}^{q_{\alpha}}(t,t') + \sum_{\beta}\iiint d\tau d\tau'd\textbf{x}  \chi_{q_{\beta,s}}^{q_{\alpha}}(t,\tau) g_{Mxc}^{n_{\beta}}(\tau,\textbf{x}\tau')\chi_{q_{\alpha'}}^{n}(\textbf{x}\tau',t')  , \label{App28}
\end{align}
where $g_{Mxc}^{q_{\beta'}}=0$ since $j_{\alpha,M}$ in Eq.~(\ref{m_current}) has no functional dependency on $q_\alpha$.

\section{Matrix formulation of QEDFT response equations}
\label{app:linresp2}

In this section we present a matrix formulation of non-relativistic QEDFT response equations which in {the no-coupling} limit reduces to Casida equation. Through a Fourier transform of Eqs.(\ref{App24})-(\ref{App25}) and Eqs.(\ref{App27})-(\ref{App28}) and making a substitution into Eqs.(\ref{App13a})-(\ref{App13d}), we express the responses in the following form:
\begin{align}
\delta n_v(\textbf{r},\omega) &= \sum_{i,a} \left[\varphi_a(\textbf{r})\varphi^{*}_{i}(\textbf{r}) \textbf{P}^{(1)}_{ai,v}(\omega) +  \varphi_i(\textbf{r})\varphi^*_a(\textbf{r}) \textbf{P}^{(1)}_{ia,v}(\omega)\right], \label{App29} \\
\delta n_j(\textbf{r},\omega) &= \sum_{i,a} \left[\varphi_a(\textbf{r})\varphi^*_i(\textbf{r}) \textbf{P}^{(1)}_{ai,j}(\omega) +  \varphi_i(\textbf{r})\varphi^*_a(\textbf{r}) \textbf{P}^{(1)}_{ia,j}(\omega)\right], \label{App30}\\
\delta q_{\alpha,v}(\omega) & = \textbf{L}^{(1)}_{\alpha,v,-}(\omega) + \textbf{L}^{(1)}_{\alpha,v,+}(\omega), \label{App31}\\
\delta q_{\alpha,j}(\omega) & = \textbf{L}^{(1)}_{\alpha,j,-}(\omega) + \textbf{L}^{(1)}_{\alpha,j,+}(\omega). \label{App32}
\end{align}
Here, the subscripts $(v,j)$ on the first-order responses $\textbf{P}^{(1)}_{ia,v}$, $\textbf{P}^{(1)}_{ia,j}$, $\textbf{P}^{(1)}_{ai,v}$, $\textbf{P}^{(1)}_{ai,j}$, $\textbf{L}^{(1)}_{\alpha,v,\pm}$ and $\textbf{L}^{(1)}_{\alpha,j,\pm}$ shows to what external perturbations $(\delta v(\textbf{r},t), \delta j_{\alpha}(t))$ is being considered to induce the coupled responses. In defining Eqs.(\ref{App29})-(\ref{App32}), we used the static KS orbitals in the Lehmann spectral representation of $ \chi_{n,s}^{n}(\textbf{r},\textbf{r}',\omega)$ and photon-photon response function $\chi_{q_{\alpha,s}}^{q_{\alpha}}(\omega)$ for a single-photon in Fock number basis are given as
\begin{align}
\begin{split}
\chi_{n,s}^{n}(\textbf{r},\textbf{r}',\omega) 
&= \sum_{i,a} \left( \frac{\psi_{a}(\textbf{r})\psi_{i}(\textbf{r}')\psi_{i}^{*}(\textbf{r})\psi_{a}^{*}(\textbf{r}')}{\omega - (\epsilon_{a} - \epsilon_{i}) + i\eta} -\frac{\psi_{i}(\textbf{r})\psi_{a}(\textbf{r}')\psi_{a}^{*}(\textbf{r})\psi_{i}^{*}(\textbf{r}')}{\omega + (\epsilon_{a} - \epsilon_{i}) + i\eta} \right)  ,  \\
\chi_{q_{\alpha,s}}^{q_{\alpha}}(\omega) 
&= \frac{1}{2\omega_{\alpha}^{2}} \left(\frac{1}{\omega - \omega_{\alpha} + i\eta} - \frac{1}{\omega + \omega_{\alpha} + i\eta} \right) . \nonumber
\end{split}
\end{align}
where the summations over occupied and unoccupied Kohn-Sham orbitals are performed according to $\sum_{i} = \sum_{i=1}^{N}$ and $\sum_{a} = \sum_{a=N+1}^{\infty}$ and from here on $\lim_{\eta\rightarrow 0^{+}}$ is implied. The first-order responses $\textbf{P}^{(1)}_{ia,v}$, $\textbf{P}^{(1)}_{ia,j}$, $\textbf{P}^{(1)}_{ai,v}$, $\textbf{P}^{(1)}_{ai,j}$, $\textbf{L}^{(1)}_{\alpha,v,\pm}$ and $\textbf{L}^{(1)}_{\alpha,j,\pm}$ are given by
\begin{align}
\left[\omega - \omega_{ai} \right] \textbf{P}^{(1)}_{ai,v}(\omega) &= \int d\textbf{r} \varphi_i(\textbf{r})\varphi^*_a(\textbf{r}) \delta v_{\textrm{KS},v}^{(1)}(\textbf{r},\omega), \label{App33}\\
\left[\omega + \omega_{ai} \right] \textbf{P}^{(1)}_{ia,v}(\omega) &= -\int d\textbf{r}  \varphi_a(\textbf{r})\varphi^*_i(\textbf{r}) \delta v_{\textrm{KS},v}^{(1)}(\textbf{r},\omega), \label{App34}\\
\left[\omega - \omega_{ai} \right] \textbf{P}^{(1)}_{ai,j}(\omega) &= \int d\textbf{r}  \varphi_i(\textbf{r})\varphi^*_a(\textbf{r}) \delta v_{\textrm{KS},j}^{(1)}(\textbf{r},\omega), \label{App35}\\
\left[\omega + \omega_{ai} \right] \textbf{P}^{(1)}_{ia,j}(\omega) &= -\int d\textbf{r}  \varphi_a(\textbf{r})\varphi^*_i(\textbf{r}) \delta v_{\textrm{KS},j}^{(1)}(\textbf{r},\omega), \label{App36}\\
\left[\omega - \omega_{\alpha} \right] \textbf{L}^{(1)}_{\alpha,v,-}(\omega) &= \frac{1}{2\omega_{\alpha}^{2}} \delta j_{\alpha,\textrm{KS},v}^{(1)}(\omega), \label{App37}\\
\left[\omega + \omega_{\alpha} \right] \textbf{L}^{(1)}_{\alpha,v,+}(\omega) &= -\frac{1}{2\omega_{\alpha}^{2}} \delta j_{\alpha,\textrm{KS},v}^{(1)}(\omega), \label{App38}\\
\left[\omega - \omega_{\alpha} \right] \textbf{L}^{(1)}_{\alpha,j,-}(\omega) &= \frac{1}{2\omega_{\alpha}^{2}} \delta j_{\alpha,\textrm{KS},j}^{(1)}(\omega), \label{App39}\\
\left[\omega + \omega_{\alpha} \right] \textbf{L}^{(1)}_{\alpha,j,+}(\omega) &= -\frac{1}{2\omega_{\alpha}^{2}} \delta j_{\alpha,\textrm{KS},j}^{(1)}(\omega) ,\label{App40}
\end{align}
where $\omega_{ai}=(\epsilon_{a} - \epsilon_{i} )$ and the respective effective potentials and currents $(\delta v_{s,\nu}(\textbf{r},\omega),j_{\alpha,\nu}^{s}(\omega))$ as
\begin{eqnarray}
\delta v_{\textrm{KS},v}^{(1)}(\textbf{r},\omega) &=& \delta v(\textbf{r},\omega) + \int d\textbf{r}'f_{Mxc}^{n}(\textbf{r},\textbf{r}',\omega) \delta n_{v}(\textbf{r}',\omega)  + \sum_{\alpha} f_{Mxc}^{q_{\alpha}}(\textbf{r},\omega) \delta q_{\alpha,v}(\omega), \label{App41} \\
\delta v_{\textrm{KS},j}^{(1)}(\textbf{r},\omega) &=&  \int d\textbf{r}'f_{Mxc}^{n}(\textbf{r},\textbf{r}',\omega) \delta n_j(\textbf{r}',\omega)  + \sum_{\alpha} f_{Mxc}^{q_{\alpha}}(\textbf{r},\omega) \delta q_{\alpha,j}(\omega), \label{App42} \\
\delta j_{\alpha,\textrm{KS},v}^{(1)}(\omega) &=&  \int d\textbf{r}g_{M}^{n_{\alpha}}(\textbf{r}) \delta n_{v}(\textbf{r},\omega) , \label{App43} \\ 
\delta j_{\alpha,\textrm{KS},j}^{(1)}(\omega) &=&  \delta j_{\alpha}(\omega) +   \int d\textbf{r}g_{M}^{n_{\alpha}}(\textbf{r}) \delta n_{j}(\textbf{r},\omega) . \label{App44} 
\end{eqnarray}
The mean-field kernel is given by $g_{M}^{n_{\alpha}}(\textbf{r}) = - \omega_{\alpha}^{2}\boldsymbol{\lambda}_{\alpha}\cdot \textbf{r}$. As stated above, the subscripts $(v,j)$ on the responses, KS potentials and currents signifies as to what external perturbations $(\delta v(\textbf{r},t), \delta j_{\alpha}(t))$ is being considered. The Kohn-Sham scheme of QEDFT decouples the interacting system such that the responses are paired as $(\delta n_v(\textbf{r},\omega), \delta q_{\alpha,v}(\omega))$ due to $\delta v(\textbf{r},\omega)$ and $(\delta n_j(\textbf{r},\omega), \delta q_{\alpha,j_{\alpha}}(\omega))$ due to  $\delta j_{\alpha}(\omega)$. Therefore, substituting Eqs.(\ref{App41}) and (\ref{App43}) into Eqs.(\ref{App33})-(\ref{App34}) and Eqs.(\ref{App37})-(\ref{App38}) and after some simplification, we obtain
\begin{align}
&\sum_{j,b}\left[\delta_{ab}\delta_{ij}\left(\omega_{ai} - \omega \right) +K_{ai,jb}(\omega) \right]\textbf{P}^{(1)}_{bj,v}(\omega) +   K_{ai,bj}(\omega) \textbf{P}^{(1)}_{jb,v}(\omega) + \sum_\alpha \delta_{ab}\delta_{ij}M_{\alpha,bj}(\omega)\left(\textbf{L}^{(1)}_{\alpha,v,-}(\omega) + \textbf{L}^{(1)}_{\alpha,v,+}(\omega) \right) \nonumber\\
&= -v_{ai}(\omega),  \label{App45}   \\
&\sum_{j,b}\left[\delta_{ab}\delta_{ij} \left(\omega_{ai} +\omega \right) + K_{ia,bj}(\omega)\right]\textbf{P}^{(1)}_{jb,v}(\omega) + K_{ia,jb}(\omega) \textbf{P}^{(1)}_{bj,v}(\omega) + \sum_\alpha\delta_{ab}\delta_{ij} M_{\alpha,jb}(\omega)\left(\textbf{L}^{(1)}_{\alpha,v,-}(\omega) + \textbf{L}^{(1)}_{\alpha,v,+}(\omega) \right)  \nonumber\\
&= -v_{ia}(\omega), \label{App46}   \\
&\left[\omega_\alpha -\omega \right] \textbf{L}^{(1)}_{\alpha,v,-}(\omega) +  \sum_{jb} \left[N_{\alpha,jb}\textbf{P}^{(1)}_{bj,v}(\omega) +  N_{\alpha,bj} \textbf{P}^{(1)}_{jb,v}(\omega)\right] =0,  \label{App47}  \\
&\left[\omega_\alpha +\omega \right] \textbf{L}^{(1)}_{\alpha,v,+}(\omega) + \sum_{jb} \left[N_{\alpha,jb} \textbf{P}^{(1)}_{bj,v}(\omega) +  N_{\alpha,bj} \textbf{P}^{(1)}_{jb,v}(\omega)\right] = 0, \label{App48} 
\end{align}
Also, substituting Eqs.(\ref{App42}) and (\ref{App44}) into Eqs.(\ref{App35})-(\ref{App36}) and Eqs.(\ref{App39})-(\ref{App40}) and after some simplification, we obtain
\begin{align}
&\sum_{j,b}\delta_{ab}\delta_{ij}\left[\left(\left(\omega_{ai} -\omega \right) + {K_{ai,jb}(\omega)}\right) \textbf{P}^{(1)}_{bj,j}(\omega) + K_{ai,bj}(\omega) \textbf{P}^{(1)}_{jb,j}(\omega) + \sum_{\alpha} M_{\alpha,bj}(\omega)\left[\textbf{L}^{(1)}_{\alpha,j,-}(\omega) + \textbf{L}^{(1)}_{\alpha,j,+}(\omega)\right]\right]=0  , \label{App49}  \\
&\sum_{j,b}\delta_{ab}\delta_{ij}\left[(\left( \omega_{ai} + \omega \right) + K_{ia,bj}(\omega)) \textbf{P}^{(1)}_{jb,j}(\omega) + K_{ia,jb}(\omega) \textbf{P}^{(1)}_{bj,j}(\omega) + \sum_{\alpha} M_{\alpha,jb}(\omega)\left[\textbf{L}^{(1)}_{\alpha,j,-}(\omega) + \textbf{L}^{(1)}_{\alpha,j,+}(\omega)\right]\right]=0  , \label{App50}  \\
&\left[\omega_\alpha -\omega \right] \textbf{L}^{(1)}_{\alpha,j,-}(\omega) + \sum_{jb} \left[N_{\alpha,jb} \textbf{P}^{(1)}_{bj,j}(\omega) +  N_{\alpha,bj} \textbf{P}^{(1)}_{jb,j}(\omega)\right] = -\frac{1}{2\omega_{\alpha}^{2}}  \delta j_{\alpha}(\omega) , \label{App51}  \\
&\left[\omega +\omega_\alpha \right] \textbf{L}^{(1)}_{\alpha,j,+}(\omega) +  \sum_{jb} \left[N_{\alpha,jb} \textbf{P}^{(1)}_{bj,j}(\omega) +  N_{\alpha,bj} \textbf{P}^{(1)}_{jb,j}(\omega)\right] = -\frac{1}{2\omega_{\alpha}^{2}}  \delta j_{\alpha}(\omega) , \label{App52} 
\end{align}
where we defined the coupling matrices 
\begin{align}
K_{ai,jb}(\omega) &= \iint d\textbf{r} d\textbf{y}\varphi_i(\textbf{r})\varphi_a^*(\textbf{r})  f^n_{Mxc}{(\textbf{r},\textbf{y},\omega)}\varphi_b(\textbf{y})\varphi^*_j(\textbf{y}) , \label{App52a}\\
M_{\alpha,ai}(\omega)  &=\int d\textbf{r} \varphi_i(\textbf{r})\varphi_a^*(\textbf{r}) f^{q_\alpha}_{Mxc}{ (\textbf{r},\omega)} , \label{App53b} \\
N_{\alpha,ia} &= \frac{1}{2\omega_{\alpha}^{2}} \int d\textbf{r} \varphi^*_i(\textbf{r})\varphi_a(\textbf{r}) g^{n_{\alpha}}_{M}(\textbf{r}) , \label{App53c} 
\end{align}
and
\begin{align}
v_{ia}(\omega) &= \int d\textbf{r}\varphi^*_i(\textbf{r})\delta v(\textbf{r},\omega)\varphi_a(\textbf{r}) .\label{App52d}
\end{align}
The coupling matrix $N_{\alpha,ia}$ has no frequency dependence since this is just the mean-field kernel of the photon modes. We now introduce the following abbreviations $L(\omega)=\delta_{ab}\delta_{ij}\left(\epsilon_a -\epsilon_i \right) + K_{ai,jb}(\omega)$, ${K}(\omega) = K_{ai,jb}(\omega)$,  $M(\omega) = M_{\alpha,bj}(\omega)$, $N = N_{\alpha,bj}$, $\textbf{X}_{1}(\omega) = \textbf{P}^{(1)}_{bj,v}(\omega)$, $\textbf{Y}_{1}(\omega) = \textbf{P}^{(1)}_{jb,v}(\omega)$, $\textbf{X}_{2}(\omega) = \textbf{P}^{(1)}_{bj,j}(\omega)$, $\textbf{Y}_{2}(\omega) = \textbf{P}^{(1)}_{jb,j}(\omega)$, $\textbf{A}_{1}(\omega) = \textbf{L}^{(1)}_{\alpha,v,-}(\omega)$, $\textbf{B}_{1}(\omega) = \textbf{L}^{(1)}_{\alpha,v,+}(\omega)$, $\textbf{A}_{2}(\omega) = \textbf{L}^{(1)}_{\alpha,j,-}(\omega)$, $\textbf{B}_{2}(\omega) = \textbf{L}^{(1)}_{\alpha,j,+}(\omega)$, $V(\omega) = -v_{ai}(\omega)$, $J_{\alpha}(\omega) = -\frac{\delta j_{\alpha}(\omega)}{2\omega_{\alpha}^{2}} $. 

Using these notations, we cast Eqs.(\ref{App45})-(\ref{App48}) and Eqs.(\ref{App49})-(\ref{App52}) into two matrix equations given by
\begin{align}
&\left[
\begin{pmatrix}
L(\omega) & K(\omega) &  M(\omega) &M(\omega)  \\
K^*(\omega) & L(\omega) &  M^*(\omega) &M^*(\omega)  \\
N  & N^*  &\omega_\alpha & 0  \\
N  & N^*  & 0 & \omega_\alpha 
\end{pmatrix} 
+\omega
\begin{pmatrix}
-1 & 0 & 0 & 0\\
0 & 1 & 0 & 0  \\
0 & 0 & -1 & 0 \\
0 & 0 & 0 & 1 
\end{pmatrix}
\right]
\begin{pmatrix}
\textbf{X}_{1}(\omega) \\
\textbf{Y}_{1}(\omega) \\
\textbf{A}_{1}(\omega) \\
\textbf{B}_{1}(\omega)
\end{pmatrix}=\begin{pmatrix}
V(\omega) \\
V^*(\omega) \\
0 \\
0 
\end{pmatrix} \label{App53}\\
&\left[
\begin{pmatrix}
L(\omega) & K(\omega) & M(\omega) & M(\omega)\\
K^*(\omega) &L(\omega)  & M^{*}(\omega) & M^{*}(\omega)\\
N  & N^* & \omega_\alpha & 0  \\
N  & N^* & 0 &\omega_\alpha 
\end{pmatrix} 
+\omega
\begin{pmatrix}
-1& 0 & 0 & 0 \\
0 & 1 & 0 & 0 \\
0 & 0 &-1 & 0 \\
0 & 0 & 0 & 1 
\end{pmatrix}
\right]
\begin{pmatrix}
\textbf{X}_{2}(\omega) \\
\textbf{Y}_{2}(\omega) \\
\textbf{A}_{2}(\omega) \\
\textbf{B}_{2}(\omega)
\end{pmatrix}=\begin{pmatrix}
0 \\
0 \\
J_{\alpha}(\omega) \\
J_{\alpha}(\omega)
\end{pmatrix} \label{App54}
\end{align}
Next, we argue that the right hand side of the above matrices remains finite as the frequency $\omega$ approaches the exact excitation frequencies $\omega\rightarrow \Omega_{q}$ of the interacting system while the density and displacement field responses on the left hand side has poles at the true excitation frequencies $\Omega_{q}$. This allows us to cast Eq.~(\ref{App53}) and Eq.~(\ref{App54}) into an eigenvalue problem
\begin{align}
&\left.
\begin{pmatrix}
L(\Omega_q) & K(\Omega_q) &  M(\Omega_q) &M(\Omega_q)  \\
K^*(\Omega_q) & L(\Omega_q) &  M^*(\Omega_q) &M^*(\Omega_q)  \\
N & N^* &\omega_\alpha & 0  \\
N & N^* &0 & \omega_\alpha 
\end{pmatrix}\begin{pmatrix}
\textbf{X}_{1}(\Omega_q) \\
\textbf{Y}_{1}(\Omega_q) \\
\textbf{A}_{1}(\Omega_q) \\
\textbf{B}_{1}(\Omega_q) 
\end{pmatrix}
=\Omega_q
\begin{pmatrix}
1 & 0 & 0 & 0\\
0 & -1 & 0 & 0  \\
0 & 0 & 1 & 0 \\
0 & 0 & 0 & -1 
\end{pmatrix}
\right.
\begin{pmatrix}
\textbf{X}_{1}(\Omega_q) \\
\textbf{Y}_{1}(\Omega_q) \\
\textbf{A}_{1}(\Omega_q) \\
\textbf{B}_{1}(\Omega_q) 
\end{pmatrix}  \label{App55}
\end{align}
\begin{align}
&\left.
\begin{pmatrix}
(\Omega_q) & K(\Omega_q) &  M(\Omega_q) & M(\Omega_q)  \\
K^*(\Omega_q) & L(\Omega_q) &  M^*(\Omega_q) & M^*(\Omega_q)  \\
N & N^* &\omega_\alpha & 0  \\
N & N^* &0 & \omega_\alpha 
\end{pmatrix}\begin{pmatrix}
\textbf{X}_{2}(\Omega_q) \\
\textbf{Y}_{2}(\Omega_q) \\
\textbf{A}_{2}(\Omega_q) \\
\textbf{B}_{2}(\Omega_q) 
\end{pmatrix}
=\Omega_q
\begin{pmatrix}
1 & 0 & 0 & 0\\
0 & -1 & 0 & 0  \\
0 & 0 & 1 & 0 \\
0 & 0 & 0 & -1 
\end{pmatrix}
\right.
\begin{pmatrix}
\textbf{X}_{2}(\Omega_q) \\
\textbf{Y}_{2}(\Omega_q) \\
\textbf{A}_{2}(\Omega_q) \\
\textbf{B}_{2}(\Omega_q) 
\end{pmatrix} \label{App56}
\end{align}
It is convenient to cast Eqs.(\ref{App55}) and (\ref{App56}) into a Hermitian eigenvalue problem which is given by
\begin{eqnarray}
\left(
\begin{array}{ c c  }
U &   V     \\
W &  \omega_{\alpha}^{2}
\end{array}
\right)
\left(
\begin{array}{ c }
\textbf{E}_{1} \\
\textbf{P}_{1}  
\end{array}	
\right) 
&=& 
\Omega_{q}^{2}
\left(
\begin{array}{ c }
\textbf{E}_{1} \\
\textbf{P}_{1} 
\end{array}	
\right), \label{App57}\\
\left(
\begin{array}{ c c  }
U &   V    \\
W &  \omega_{\alpha}^{2}
\end{array}
\right)
\left(
\begin{array}{ c }
\textbf{E}_{2}  \\
\textbf{P}_{2}  
\end{array}	
\right) 
&=& 
\Omega_{q}^{2}
\left(
\begin{array}{ c }
\textbf{E}_{2} \\
\textbf{P}_{2} 
\end{array}	
\right), \label{App58}
\end{eqnarray}
where we assumed real-valued orbitals, i.e., $K=K^{*}$, $M=M^{*}$ and $N=N^{*}$, and the matrices are given by $U = (L-K)^{1/2}(L + K)(L-K)^{1/2}$, $V = 2(L-K)^{1/2}M^{1/2}N^{1/2}\omega_{\alpha}^{1/2}$, $W = 2\omega_{\alpha}^{1/2}N^{1/2}M^{1/2}(L-K)^{1/2}$, and the eigenvectors are $\textbf{E}_{1} = N^{1/2}(L-K)^{-1/2}(\textbf{X}_{1} + \textbf{Y}_{1})$ and $\textbf{P}_{1} = M^{1/2}\omega_{\alpha}^{-1/2}(\textbf{A}_{1} + \textbf{B}_{1})$. 

The pseudo-eigenvalue problem of Eqs.(\ref{App57}) and (\ref{App58}) is the final form of QEDFT matrix equation for obtaining exact excitation frequencies and oscillator strengths.

\section{Oscillator Strengths}
\label{app:linresp3}

In this section, we derive the oscillator strengths resulting from the eigenvectors of the pseudo-eigenvalue problem of Eqs.(\ref{App57}) and (\ref{App58}).  Multiplying out Eq.~(\ref{App53}), we write the matrix equation in the form
\begin{eqnarray}
(L + K)(\textbf{X}_{1} +\textbf{Y}_{1}) + 2M(\textbf{A}_{1} + \textbf{B}_{1}) - \omega (\textbf{X}_{1} - \textbf{Y}_{1}) &=& -2\boldsymbol{v}, \nonumber\\
(L - K)(\textbf{X}_{1} - \textbf{Y}_{1}) - \omega (\textbf{X}_{1} + \textbf{Y}_{1}) &=& 0, \nonumber\\
2N(\textbf{X}_{1} + \textbf{Y}_{1}) + \omega_{\alpha}(\textbf{A}_{1} + \textbf{B}_{1}) - \omega (\textbf{A}_{1} -\textbf{B}_{1} ) &=& 0, \nonumber\\
\omega_{\alpha} (\textbf{A}_{1} - \textbf{B}_{1}) - \omega (\textbf{A}_{1} +\textbf{B}_{1} ) &=& 0 .\nonumber
\end{eqnarray}
From here on we set $S=(L-K)$, the above pair of equations now becomes
\begin{align}
& S(L + K)\textbf{E}_{1} + 2SM\textbf{P}_{1} - \omega^{2}\textbf{E}_{1} = -2S\boldsymbol{v} , \nonumber\\
& 2\omega_{\alpha} N \textbf{E}_{1} + \omega_{\alpha}^{2} \textbf{P}_{1} - \omega^{2} \textbf{P}_{1} = 0 . \nonumber
\end{align}
This can be written in matrix form as
\begin{align}
&\left[
\left(
\begin{array}{ c c  }
S(L+K) &   2SM     \\
2\omega_{\alpha}N   &  \omega_{\alpha}^{2}
\end{array}
\right)
-
\omega^{2}
\left(
\begin{array}{ c c  }
1 &   0     \\
0   & 1
\end{array}
\right)
\right]
\left(
\begin{array}{ c }
\textbf{E}_{1}  \\
\textbf{P}_{1} 
\end{array}	
\right) = -
\left(
\begin{array}{ c }
2S\boldsymbol{v}  \\
0 
\end{array}	
\right),  \label{os1}
\end{align}
where $\textbf{E}_{1}=\textbf{X}_{1} + \textbf{Y}_{1}$ and $\textbf{P}_{1}=\textbf{A}_{1} + \textbf{B}_{1}$. We perform the same steps as above to make the nonlinear eigenvalue problem Hermitian and obtain
\begin{align}
& \left[C - \omega^{2} \mathbb{1} \right]
\left(
\begin{array}{ c }
N^{1/2}S^{-1/2}\textbf{E}_{1}  \\
M^{1/2}\omega_{\alpha}^{-1/2}\textbf{P}_{1} 
\end{array}	
\right) = -
\left(
\begin{array}{ c }
2N^{1/2}S^{1/2}\boldsymbol{v}  \\
0 
\end{array}	
\right),  \label{os2}
\end{align}
where $C =
\left(
\begin{array}{ c c  }
U  &   V     \\
W   &  \omega_{\alpha}^{2}
\end{array}
\right)$. We determine the vectors given as
\begin{align}
&\textbf{E}_{1} = -2S^{1/2}\left[C - \omega^{2} \mathbb{1} \right]^{-1} S^{1/2}\boldsymbol{v} , \label{os3} \\
&\textbf{P}_{1} = -2\omega_{\alpha}^{1/2}M^{-1/2}\left[C - \omega^{2} \mathbb{1} \right]^{-1} N^{1/2}S^{1/2}\boldsymbol{v} .  \label{os4}
\end{align}
When $\textbf{Z}_{I}$ is normalized, we can use the spectral expansion to get
\begin{equation}
\left[C - \omega^{2} \mathbb{1}\right]^{-1} = \sum_{I}\frac{\textbf{Z}_{I}\textbf{Z}_{I}^{\dagger} }{\Omega_{I}^{2} - \omega^{2}},  \label{os5}
\end{equation}
where $\textbf{Z}_{I} =
\left(
\begin{array}{ c }
\textbf{E}_{1I}  \\
\textbf{P}_{1I}  
\end{array}
\right)$. The oscillator strength for the density-density response function which is related to the dynamic polarizability is given in Eq.(\ref{eq:oscillator-strength}).

\subsection{Oscillator strength for the photon-matter response function}

Next, we substitute the expression of the spectral expansion Eq.~(\ref{os5}) in Eq.~(\ref{os4}) and by substituting $\textbf{P}_{1}$ in Eq.~(\ref{App30}) yields
\begin{eqnarray}
\delta q_{\alpha,v}(\omega) &=& \sum_{I} \left\{ \frac{2\omega_{\alpha}^{1/2}M^{-1/2}\textbf{Z}_{I}\textbf{Z}_{I}^{\dagger} N^{1/2}S^{1/2}}{ \omega^{2} - \Omega^{2}_{I}}   \right\}  v(\omega). \nonumber 
\end{eqnarray}
The oscillator strength is given by 
\begin{eqnarray}
f_{I,\alpha}^{pn} = 2\omega_{\alpha}^{1/2}M^{-1/2}\textbf{Z}_{I}\textbf{Z}_{I}^{\dagger}N^{1/2}S^{1/2}. \label{os11}
\end{eqnarray}
Also, from Eq.(\ref{App13b}) and using the Lehmann representation of the response function $\chi_{n}^{q_{\alpha}}(\textbf{r}',\omega) $ the response $\delta q_{\alpha,v}(\omega)$ is given by
\begin{eqnarray}
\delta q_{\alpha,v}(\omega)  &=& \int d \textbf{r}' \sum_{k}\left[\frac{2\Omega_{k}\langle\Psi_{0}|\hat{q}_{\alpha}|\Psi_{k}\rangle\langle\Psi_{k}|\hat{n}(\textbf{r}')|\Psi_{0}\rangle}{\omega^{2} - \Omega_{k}^{2} } \right] \delta v(\textbf{r}',\omega), \nonumber
\end{eqnarray}
The oscillator strength of Eq.(\ref{os11}) can be expressed as matrix elements of the internal pair $\left(\hat{n}(\textbf{r}),\hat{q}_{\alpha} \right)$ as
\begin{equation}
f_{\alpha,k}(\textbf{r}')= 2 \Omega_{k} \langle\Psi_{0}|\hat{q}_{\alpha}|\Psi_{k}\rangle\langle\Psi_{k}|\hat{n}(\textbf{r}')|\Psi_{0}\rangle \equiv f_{I,\alpha}^{pn}.
\end{equation}

\subsection{Oscillator strength for the matter-photon response function}

Following similar steps as above with Eq.~(\ref{App54}) we obtain
\begin{eqnarray}
\textbf{E}_{2} &=& -2S^{1/2}N^{-1/2}\left[C - \omega^{2}\mathbb{1} \right]^{-1}M^{1/2}\omega_{\alpha}^{1/2} J'_{\alpha} , \label{os12}\\
\textbf{P}_{2} &=& -2\omega_{\alpha}^{1/2}\left[C - \omega^{2}\mathbb{1} \right]^{-1}\omega_{\alpha}^{1/2} J'_{\alpha} . \label{os13} 
\end{eqnarray}
where $J'_{\alpha}(\omega) = \frac{j_{\alpha}(\omega)}{2\omega_{\alpha}^{2}} $ and $J_{\alpha}(\omega) = -J'_{\alpha}(\omega)$. By substituting the spectral expansion Eq.~(\ref{os5}) in $\textbf{E}_{2}$ and further substituting in Eq.~(\ref{App31}) yields
\begin{eqnarray}
\delta n_{j}(\textbf{r},\omega) = -2\sum_{ia,I} \frac{\Phi_{ia}S^{1/2}N^{-1/2}\textbf{Z}_{I}\textbf{Z}_{I}^{\dagger} M^{1/2}\omega_{\alpha}^{1/2} \Phi_{ai} }{\left(\Omega_{I}^{2} - \omega^{2}\right)} J'_{\alpha}(\omega). \nonumber
\end{eqnarray}
Following a similar procedure as above, we express the density response to the external charge current as
\begin{eqnarray}
\delta n_{j}(\textbf{r},\omega) &=& \sum_{I}\left\{ \frac{\Phi_{ia} S^{1/2}N^{-1/2}\textbf{Z}_{I}\textbf{Z}_{I}^{\dagger} M^{1/2}\omega_{\alpha}^{1/2}\Phi_{ia}}{ \omega^{2} - \Omega^{2}_{I}}  \right\} \frac{j_{\alpha}(\omega)}{\omega_{\alpha}^{2}}, \nonumber 
\end{eqnarray}
where $\Phi_{ia}(\textbf{r}) = \varphi_{i}^{*}(\textbf{r}) \varphi_{a}(\textbf{r})$ and the oscillator strength is given by
\begin{eqnarray}
f_{I,\alpha}^{np} = \frac{1}{\omega_{\alpha}}\Phi_{ia} S^{1/2}N^{-1/2}\textbf{Z}_{I}\textbf{Z}_{I}^{\dagger}M^{1/2}\omega_{\alpha}^{1/2}\Phi_{ia}. \label{os5a} 
\end{eqnarray}
From Eq.(\ref{App13c}) and using the Lehmann representation of the response function $\chi_{q_{\alpha}}^{n}(\textbf{r},\omega) $, the response $\delta n_{j}(\textbf{r},\omega)$ is given by
\begin{eqnarray}
\delta n_{j}(\textbf{r},\omega)  &=& \sum_{\alpha,k} \left[\frac{2\Omega_{k}\langle\Psi_{0}|\hat{n}(\textbf{r})|\Psi_{k}\rangle\langle\Psi_{k}|\hat{q}_{\alpha}|\Psi_{0}\rangle}{\omega^{2} - \Omega_{k}^{2} } \right] \frac{\delta j_{\alpha}(\omega)}{\omega_{\alpha}}, \nonumber
\end{eqnarray}
The oscillator strength of Eq.(\ref{os5a}) can be expressed as matrix elements of the internal pair $\left(\hat{n}(\textbf{r}),\hat{q}_{\alpha} \right)$ as
\begin{equation}
f_{k,\alpha}(\textbf{r})= 2 \Omega_{k}\langle\Psi_{0}|\hat{n}(\textbf{r})|\Psi_{k}\rangle \langle\Psi_{k}|\hat{q}_{\alpha}|\Psi_{0}\rangle \equiv f_{I,\alpha}^{np} .
\end{equation}

\subsection{Oscillator strength for the photon-photon response function}
We define a collective photon coordinate for the $\alpha$ modes $Q = \sum_{\alpha} q_{\alpha}$ (in analogy with $\textbf{R} = \sum_{i} e \textbf{r}_{i}$ ). By perturbing the photon field through the photon coordinate with an external charge current $j_{\alpha}(\omega)$, we induce a polarization of the field of mode $\alpha$ which we denote as $Q(\omega) = \sum_{\alpha} \beta_{\alpha}(\omega) j_{\alpha}(\omega)$. Where $\beta_{\alpha}(\omega)$ is the polarizability of field of the $\alpha$ mode. To first-order, the collective coordinate is given by
\begin{equation}
\delta Q(t) = \sum_{\alpha}  \delta q_{\alpha}(t). \label{OS11}
\end{equation}
The  field polarizability in frequency space can be written as
\begin{equation}
\beta_{\alpha}(\omega)  = \sum_{\alpha'} \frac{\delta q_{\alpha}(\omega)}{\delta j_{\alpha'}(\omega)}.\label{OS12}
\end{equation}
By substituting Eq.~(\ref{os13}) in Eq.~(\ref{App32}) and using the spectral expansion yields
\begin{eqnarray}
\delta q_{\alpha,j}(\omega) = -\sum_{I}  \frac{2\omega_{\alpha}^{1/2}\textbf{Z}_{I}\textbf{Z}_{I}^{\dagger} \omega_{\alpha}^{1/2} }{\Omega_{I}^{2} - \omega^{2}} J'_{\alpha} . \nonumber
\end{eqnarray}
By substituting the above relation in Eq.~(\ref{OS12}) we obtain
\begin{align*}
\beta_{\alpha}(\omega) = -\sum_{\alpha'}\sum_{I}  \frac{2\omega_{\alpha}^{1/2}\textbf{Z}_{I}\textbf{Z}_{I}^{\dagger} \omega_{\alpha}^{1/2} }{\Omega_{I}^{2} - \omega^{2}} \frac{\delta j_{\alpha}(\omega)/2\omega_{\alpha}^{2}}{\delta j_{\alpha'}(\omega)} ,
\end{align*}
which simplifies to
\begin{align}
\beta_{\alpha}(\omega) = -\sum_{I} \frac{1}{\omega_{\alpha}^{2}} \frac{\omega_{\alpha}^{1/2}\textbf{Z}_{I}\textbf{Z}_{I}^{\dagger} \omega_{\alpha}^{1/2} }{\Omega_{I}^{2} - \omega^{2}} . \label{OS13}
\end{align}
Eq.~(\ref{OS13}) is the field polarizability analogous to the atomic polarizability tensor of Eq.~(\ref{atom-pol}). As in Eq.(\ref{photo}) in which the molecular isotropic polarizability, $\alpha(\omega)$ is defined as the mean value of three diagonal elements of the polarizability tensor, i.e., $\alpha(\omega) = 1/3\left(\alpha_{xx}(\omega) + \alpha_{yy}(\omega) + \alpha_{zz}(\omega)\right)$, we analogously define an absorption cross section of the field given by
\begin{align}
\tilde{\sigma}_{\alpha}(\omega) \equiv \frac{4 \pi \omega}{c} \Ib m \ \text{Tr}\beta_{\alpha}(\omega)/3 . \label{photo1a}
\end{align} 
For the oscillator strength, from Eq.(\ref{App13d}) and using the Lehmann representation of the response function $\chi_{q_{\alpha'}}^{q_{\alpha}}(\omega) $ the response $\delta q_{\alpha,j}(\omega)$ is given by
\begin{align}
\delta q_{\alpha,j}(\omega)  &=  \sum_{\alpha',k}\left[\frac{2\Omega_{k}\langle\Psi_{0}|\hat{q}_{\alpha}|\Psi_{k}\rangle\langle\Psi_{k}|\hat{q}_{\alpha'}|\Psi_{0}\rangle}{\omega^{2} - \Omega_{k}^{2} } \right] \frac{\delta j_{\alpha'}(\omega)}{\omega_{\alpha'}}. \nonumber
\end{align}
We find the oscillator strength
\begin{align}
f_{I,\alpha}^{pp} = \frac{1}{3\omega_{\alpha}^{2}}\left|\textbf{Z}^\dagger_{I} \omega_{\alpha}^{1/2}\right|^2 = \frac{2}{3}\Omega_{I}\sum_{\alpha'}\frac{1}{\omega_{\alpha'}} \langle\Psi_{0}|\hat{q}_{\alpha}|\Psi_{I}\rangle\langle \Psi_{I}|\hat{q}_{\alpha'}|\Psi_{0}\rangle .
\end{align}

\end{document}